\documentclass{emulateapj}
\usepackage{apjfonts}
\usepackage[]{natbib}
\usepackage{graphics}
\usepackage{color}
     
\newcommand{\etal}{et al.}  
\newcommand{\per}{\ensuremath{^{-1}}}
\newcommand{\persq}{\ensuremath{^{-2}}}
\newcommand{\hal}{H\ensuremath{\alpha}}
\newcommand{\hbeta}{H\ensuremath{\beta}} 
\newcommand{\hgamma}{H\ensuremath{\gamma}} 
\newcommand{\hdelta}{H\ensuremath{\delta}} 
\newcommand{\heii}{\ion{He}{2}}
\newcommand{\feii}{\ion{Fe}{2}}

\newcommand{\hst}{\emph{HST}}

\newcommand{\kms}{km~s\ensuremath{^{-1}}}

\newcommand{\mbh}{\ensuremath{M_\mathrm{BH}}}

\newcommand{\msigma}{\ensuremath{\mbh-\sigma}}
\newcommand{\rblr}{\ensuremath{r_{\mathrm{BLR}}}}

\newcommand{\fvar}{\ensuremath{F_\mathrm{var}}}
\newcommand{\maxratio}{\ensuremath{R_\mathrm{max}}}
\newcommand{\sigmaline}{\ensuremath{\sigma_\mathrm{line}}}

\begin{document} 

\title{The Lick AGN Monitoring Project 2011: \\  Spectroscopic
  Campaign and Emission-Line Light Curves}

\author{
  Aaron J. Barth\altaffilmark{1}, 
  Vardha N. Bennert\altaffilmark{2},  
  Gabriela Canalizo\altaffilmark{3},  
  Alexei V. Filippenko\altaffilmark{4}, 
  Elinor L. Gates\altaffilmark{5}, 
  Jenny E. Greene\altaffilmark{6},  
  Weidong Li\altaffilmark{4,26}, 
  Matthew A. Malkan\altaffilmark{7}, 
  Anna Pancoast\altaffilmark{8}, 
  David J. Sand\altaffilmark{9},   
  Daniel Stern\altaffilmark{10},  
  Tommaso Treu\altaffilmark{7},  
  Jong-Hak Woo\altaffilmark{11}, 
  Roberto J. Assef\altaffilmark{12},  
  Hyun-Jin Bae\altaffilmark{13}, 
  Brendon J. Brewer\altaffilmark{14},  
  S. Bradley Cenko\altaffilmark{15,16}, 
  Kelsey I. Clubb\altaffilmark{4},  
  Michael C. Cooper\altaffilmark{1}, 
  Aleksandar M. Diamond-Stanic\altaffilmark{17}, 
  Kyle D. Hiner\altaffilmark{18}, 
  Sebastian F. H\"{o}nig\altaffilmark{19}, 
  Eric Hsiao\altaffilmark{20}, 
  Michael T. Kandrashoff\altaffilmark{4}, 
  Mariana S. Lazarova\altaffilmark{21}, 
  A. M. Nierenberg\altaffilmark{22}, 
  Jacob Rex\altaffilmark{4},
  Jeffrey M. Silverman\altaffilmark{23}, 
  Erik J. Tollerud\altaffilmark{24,27}, and 
  Jonelle L. Walsh\altaffilmark{25} 
}

\altaffiltext{1}{Department of Physics and Astronomy, 4129 Frederick
  Reines Hall, University of California, Irvine, CA, 92697-4575, USA;
  barth@uci.edu}

\altaffiltext{2}{Physics Department, California Polytechnic State
University, San Luis Obispo, CA 93407, USA}

\altaffiltext{3}{Department of Physics and Astronomy, University of
  California, Riverside, CA 92521, USA}

\altaffiltext{4}{Department of Astronomy, University of California,
Berkeley, CA 94720-3411, USA}

\altaffiltext{5}{Lick Observatory, P.O. Box 85, Mount Hamilton, CA
95140, USA}

\altaffiltext{6}{Department of Astrophysical Sciences, Princeton
  University, Princeton, NJ 08544, USA}

\altaffiltext{7}{Department of Physics and Astronomy, University of
California, Los Angeles, CA 90095-1547, USA}

\altaffiltext{8}{Department of Physics, University of California,
Santa Barbara, CA 93106, USA}

\altaffiltext{9}{Texas Tech University, Physics Department, Box
  41051, Lubbock, TX 79409-1051, USA}

\altaffiltext{10}{Jet Propulsion Laboratory, California Institute of
Technology, 4800 Oak Grove Boulevard, Pasadena, CA 91109, USA}

\altaffiltext{11}{Astronomy Program, Department of Physics and
Astronomy, Seoul National University, Seoul 151-742, Republic of
Korea}

\altaffiltext{12}{N\'ucleo de Astronom\'ia de la Facultad de
  Ingenier\'ia, Universidad Diego Portales, Av. Ej\'ercito Libertador
  441, Santiago, Chile}

\altaffiltext{13}{Department of Astronomy and Center for Galaxy
Evolution Research, Yonsei University, Seoul 120-749, Republic of
Korea}

\altaffiltext{14}{Department of Statistics, The University of
  Auckland, Private Bag 92019, Auckland 1142, New Zealand}

\altaffiltext{15}{Astrophysics Science Division, NASA Goddard Space
  Flight Center, MC 661, Greenbelt, MD 20771, USA}

\altaffiltext{16}{Joint Space-Science Institute, University of
  Maryland, College Park, MD 20742, USA}

\altaffiltext{17}{Department of Astronomy, University of Wisconsin,
Madison, WI 53706, USA}

\altaffiltext{18}{FONDECYT Postdoctoral Fellow, Departamento de
  Astronom\'{i}a, Universidad de Concepci\'{o}n, Chile}

\altaffiltext{19}{School of Physics \& Astronomy, University of
  Southampton, Southampton SO17 1BJ, UK}

\altaffiltext{20}{Department of Physics and Astronomy, Aarhus
  University, Ny Munkegade, DK-8000 Aarhus C, Denmark}

\altaffiltext{21}{Dept.\ of Physics and Physical Science, University
  of Nebraska at Kearney, Kearney, NE 68849, USA}

\altaffiltext{22}{Center for Cosmology and Astro-Particle Physics, The
  Ohio State University, Columbus OH 43210, USA}

\altaffiltext{23}{Department of Astronomy, The University of Texas, 
Austin, TX 78712-0259, USA}

\altaffiltext{24}{Astronomy Department, Yale University, New Haven, CT
  06510, USA}

\altaffiltext{25}{George P. and Cynthia Woods Mitchell Institute for
  Fundamental Physics and Astronomy, Department of Physics and
  Astronomy, Texas A\&M University, College Station, TX 77843-4242,
  USA}

\altaffiltext{26}{Deceased 12 December 2011}

\altaffiltext{27}{Hubble Fellow}

\begin{abstract}
In the Spring of 2011 we carried out a 2.5 month reverberation mapping
campaign using the 3 m Shane telescope at Lick Observatory, monitoring
15 low-redshift Seyfert 1 galaxies.  This paper describes the
observations, reductions and measurements, and data products from the
spectroscopic campaign.  The reduced spectra were fitted with a
multicomponent model in order to isolate the contributions of various
continuum and emission-line components.  We present light curves of
broad emission lines and the AGN continuum, and measurements of the
broad \hbeta\ line widths in mean and root-mean square (rms) spectra.
For the most highly variable AGNs we also measured broad \hbeta\ line
widths and velocity centroids from the nightly spectra.  In four AGNs
exhibiting the highest variability amplitudes, we detect
anticorrelations between broad \hbeta\ width and luminosity,
demonstrating that the broad-line region ``breathes'' on short
timescales of days to weeks in response to continuum variations.  We
also find that broad \hbeta\ velocity centroids can undergo
substantial changes in response to continuum variations; in NGC 4593
the broad \hbeta\ velocity shifted by $\sim250$ \kms\ over a one-month
duration.  This reverberation-induced velocity shift effect is likely
to contribute a significant source of confusion noise to binary black
hole searches that use multi-epoch quasar spectroscopy to detect
binary orbital motion.  We also present results from simulations that
examine biases that can occur in measurement of broad-line widths from
rms spectra due to the contributions of continuum variations and
photon-counting noise.
\end{abstract}


\section{Introduction}

Broad emission lines are a hallmark property of active galactic nuclei
(AGNs), first highlighted in pioneering work by \citet{seyfert1943}.
Originating from the broad-line region (BLR), they serve as important
diagnostics of physical conditions within the AGN as well as dynamical
tracers providing information about the mass of the supermassive black
hole at the core of the AGN central engine. Unfortunately, even for
the nearest AGNs the BLR is too small to be directly resolved by any
current observational capability.  For example, in the well-studied
nearby Seyfert 1 galaxy NGC 5548, the radius of the \hbeta-emitting
portion of the BLR is $\sim4-20$ light-days \citep{bentz2007},
corresponding to an angular radius of just 9--45 microarcseconds.

Although the spatial structure of the BLR cannot be directly resolved,
the time-variable nature of AGN emission makes it possible to resolve
the BLR's size and structure in the time domain via reverberation
mapping \citep{blandfordmckee, peterson2001}.  Ionizing photons from
the AGN's central engine are reprocessed by BLR gas into emission-line
photons. From the vantage point of a distant observer, random
variations in the flux of the AGN continuum will be followed by
corresponding variations in the fluxes of emission lines from the BLR,
with a time delay that depends on the light-travel time across the
BLR.  Cross-correlation of emission-line light curves against light
curves of the AGN continuum gives a measure of the BLR radius \rblr,
under the assumption that the size of the continuum emission region is
much smaller than \rblr.  Reverberation mapping thus requires
measurement of light curves of the variable AGN continuum and emission
lines, with a cadence sufficient to resolve the temporal fluctuations
adequately.  When \rblr\ is combined with a velocity parameter
measured from the width of the broad line, a so-called virial mass
estimate can be derived for the central black hole. Virial masses have
been determined through reverberation mapping for several dozen
low-redshift AGNs, and these measurements form the foundation of
secondary ``single-epoch'' virial mass methods, which provide nearly
all of the empirical information currently available on the
cosmological growth history and redshift evolution of supermassive
black holes \citep[for a comprehensive review, see][]{shen2013review}.

A major goal for reverberation mapping observations is the detection
of velocity-resolved emission-line variations --- that is, variations
in the shapes of broad emission-line profiles, not merely variations
in total emission-line flux. Line profile variability conveys
information about the kinematics of BLR gas.  Measurements of profile
variability can provide a means to discriminate between rotationally
dominated, inflowing, or outflowing BLR kinematics
\citep[e.g.,][]{welshhorne}, while high-cadence and high
signal-to-noise ratio (S/N) observations can potentially reveal a
wealth of detail about BLR structure and kinematics \citep{horne2004}.

It is a nontrivial challenge to obtain observations with sufficiently
high cadence and high S/N suitable for measurement of broad-line
profile variations.  Consequently, many of the largest
reverberation-mapping programs have focused largely on measurement of
flux variations and derivation of black hole virial masses
\citep[e.g.,][and references therein]{kaspi2000, peterson2004}.  Over
the past decade or so, various observing programs have succeeded in
measuring velocity-resolved variability signatures in broad \hbeta; an
early example was a sparsely sampled program at the Hobby-Eberly
telescope targeting Mrk 110 \citep{kollatschny2003}.  More recently,
densely sampled observing campaigns have led to rapid progress.  Using
data from the 2008 Lick AGN Monitoring Project campaign, a 64-night
program carried out at the Lick 3 m telescope, \citet{bentz2008,
  bentz2009lamp} detected a strong gradient in the reverberation lag
across the \hbeta\ line profile in Arp 151 (Mrk 40), and the campaign
provided some degree of velocity resolution of the
\hbeta\ reverberation lag in five additional AGNs.  Similarly,
intensive monitoring campaigns carried out at the MDM Observatory have
recently yielded impressive detections of velocity-resolved
reverberation \citep{denney2009vr, grier2013a}. The collective results
of these programs suggest a surprising diversity in the kinematic
properties of AGNs, with different objects exhibiting behavior
consistent with rotation, infall, or outflow as the dominant kinematic
behavior in the BLR.

These high-cadence monitoring programs at Lick and MDM have spurred
the development of new methods for analysis of reverberation data.
The traditional approach of deriving black hole masses by applying a
simple recipe to derive the ``virial product'' [VP $ = \rblr (\Delta
  V)^2/G$, where \rblr\ is determined from the reverberation lag and
  $\Delta V$ is a measure of the broad-line width] suffers from
several shortcomings. In particular, the ``BLR radius'' measured from
the reverberation lag gives a highly oversimplified measure of BLR
structure, since the BLR actually spans a very broad range in
radius. Similarly, a single measure of BLR velocity derived from an
emission-line width is a gross oversimplification of the complex
velocity structure of a real BLR.  More worrying is the possibility
that the BLR dynamics might be affected by outflows or other
nongravitational motions, in which case simple virial mass estimates
might be entirely misleading \citep{krolik2001}.  \citet{pancoast2011}
proposed a different approach to reverberation data analysis, in which
black hole masses can be determined via a forward-modeling procedure
that computes the response to continuum variations of a physically
motivated dynamical model for BLR clouds. By exploring a broad
parameter space of BLR dynamical models and comparing the model
predictions directly with the observed time series of broad
emission-line profiles, direct constraints can be placed on the black
hole mass, BLR inclination and opening angle, radial emissivity
profile, and other parameters.  This approach has been applied to
several AGNs from our high-cadence Lick programs \citep{brewer2011,
  pancoast2012, pancoast2014b}.  \citet{li2013} presented applications
of an independent implementation of a similar method to archival data
from several observing campaigns.

Although reverberation mapping has seen dramatic progress over the
past several years stemming from these high-cadence monitoring
campaigns, there are still only a handful of AGNs having time-series
data with sufficient sampling and S/N for these dynamical modeling
methods to be optimally successful.  In order to obtain more
velocity-resolved reverberation data suitable for dynamical modeling,
we embarked on a new reverberation mapping program in the spring of
2011.  Spectroscopic observations were conducted at the Lick 3 m Shane
telescope, and $V$-band photometric monitoring was carried out using
several queue-scheduled and robotic telescopes.  This paper describes
the spectroscopic campaign and presents measurements from the Lick
data.

A new aspect of our data analysis for this 2011 program is the use of
spectral fitting and decomposition techniques to isolate different
portions of the variable spectra. In the past, nearly all
emission-line light curves for reverberation mapping were measured by
subtracting a local, linear continuum model underlying strong emission
lines such as \hbeta\ in order to isolate the emission-line flux.
This method is generally effective for measuring integrated light
curves of strong emission lines having high variability
amplitude. However, simple linear continuum subtraction is unable to
remove host-galaxy starlight features that can affect the broad-line
profiles and widths, and it cannot disentangle blended emission lines.
The alternative approach of fitting models to deblend components such
as \ion{Fe}{2} from \hbeta\ had seldom been applied to nightly spectra
from reverberation campaigns \citep[e.g.,][]{bian2010}. Our initial
experiments with spectral decomposition using observations of Mrk 50
from this program established that multicomponent fits made it
possible to measure light curves for \ion{He}{2} $\lambda4686$ even
when the line was heavily diluted by starlight, and to detect
variability and measure reverberation lags of the optical \ion{Fe}{2}
blends \citep{barth2011:mrk50, barth2013}.  Using a similar fitting
method, \citet{park2012a} carried out decompositions of spectra from
the 2008 Lick AGN Monitoring Project campaign \citep{bentz2009lamp}.
This work demonstrated clear improvements in determination of the mean
and root-mean square (rms) line profiles for \hbeta\ after removal of
the host-galaxy starlight and other blended line components from the
spectra.  In this paper, we present a complete description of our
fitting method and results from applying the spectroscopic
decomposition procedure to our full 2011 monitoring sample.

The outline of this paper is as follows.  Section
\ref{sec:sampleselection} describes the sample selection, and \S
\ref{sec:observations} describes the spectroscopic observations
carried out at Lick.  Details of the data reduction, calibrations, and
the fitting procedure applied to the blue-side spectra are given in \S
\ref{sec:dataprocessing}.  Sections \ref{sec:lightcurves} and
\ref{sec:linewidths} present the measurements of emission-line light
curves, broad \hbeta\ line widths, and velocity centroid variations.
Section \ref{sec:widthluminosity} examines the anticorrelation between
broad \hbeta\ width and luminosity in the single-epoch spectra of
highly variable sources.  In \S \ref{sec:velocityshifts} we describe
how broad-line velocity shifts induced by asymmetric emission-line
reverberation can potentially mimic the observational signature of
orbital motion in a binary black hole system, and we consider the
implications of broad-line reverberation for spectroscopic binary
black hole searches.  Section \ref{sec:conclusions} presents a summary
of our results and conclusions.  In Appendix
\ref{appendix:discardedtargets}, we display spectra of additional AGNs
that were observed during the first few nights of our program as we
finalized our target list for monitoring.  Appendices
\ref{appendix:widthsimulations} and \ref{appendix:biases} present
simulations designed to clarify the relationship between line widths
measured from mean and rms spectra, and to demonstrate potential
sources of bias that can afflict broad-line widths measured from rms
spectra.

\section{Sample Selection}
\label{sec:sampleselection}

The major science goal for this program was to measure
velocity-resolved reverberation signals in \hbeta\ for bright Seyfert
1 galaxies, and we primarily selected targets having strong broad
\hbeta\ emission from the Sloan Digital Sky Survey (SDSS) Data Release
7 archive \citep{abazajian2009} and other AGN catalogs in the
literature.  Our reverberation campaign began in late March and
continued through the middle of June. This set a requirement for the
sample to contain AGNs distributed over a broad range of right
ascension such that the entire sample could be observed over nearly
the full duration of the program.  Targets were selected to have
apparent magnitudes of $V \lesssim 17$ so that high-S/N spectra could
be obtained in short exposures (10--40 min).  We chose AGNs for which
estimated \hbeta\ reverberation lags were $<25$ days based on either
the radius-luminosity relation from \citet{bentz2009rl} or previous
reverberation measurements.  Finally, the standard configuration of
the Kast spectrograph employs a 600-line grism in the blue camera and
a dichroic with cutoff wavelength 5500 \AA, and the blue-camera
throughput drops steeply at $\lambda>5400$ \AA.  This imposes a
practical redshift limit of $z\lesssim0.08$ in order for [\ion{O}{3}]
$\lambda5007$ to fall on the high-sensitivity portion of the blue
camera. This line is used as the reference for flux calibration of the
spectral time series.

Two targets, Mrk 817 and Zw 229-015, were included in the sample in
order to coordinate with a \emph{Spitzer} infrared monitoring program
taking place in 2011 (\emph{Spitzer} program ID 70119; PI V. Gorjian).
Zw 229-015 was also a \emph{Kepler} monitoring target during this time
\citep{mushotzky2011, carini2012, edelson2014}.

We assembled an initial set of 15 high-priority targets plus several
alternates which included narrow-line Seyfert 1 galaxies, AGNs that
were somewhat fainter than our primary targets, and candidate AGNs for
which no high-quality recent spectra were available in the literature.
During the first few usable nights of the campaign, we found that
three of our top-priority targets selected from SDSS had transformed
into Type 1.9 Seyferts with essentially no broad \hbeta\ emission
remaining (see Appendix \ref{appendix:discardedtargets} for details).
Several alternate targets were then observed and we chose the most
promising ones to fill the gaps in the sample's right ascension
range.  Our final monitoring sample is listed in Table
\ref{sampleproperties}, and Appendix \ref{appendix:discardedtargets}
presents spectra of the alternate and discarded targets.

Some of these AGNs have been observed in previous reverberation
programs, including Mrk 40 \citep[Arp
  151;][]{bentz2008,bentz2009lamp}, Mrk 279 \citep{maoz1990,
  santos2001}, Mrk 817 \citep{peterson1998, denney2010}, NGC 4593
\citep{dietrich1994, onken2003, denney2006}, and Zw 229-015
\citep{barth:zw229}.  Revised lag measurements of Mrk 279, Mrk 817,
and NGC 4593 were also given in the compilation of
\citet{peterson2004}.  Recently, \citet{wang2014} presented results
from a reverberation campaign that included Mrk 486 and Mrk 493 as
well. New observations of previously well-observed targets such as Mrk
40 can potentially test whether velocity-resolved reverberation lags
remain constant or vary over durations comparable to the dynamical
timescale of the BLR.

Owing to the combination of criteria given above, the observed sample
is not complete or unbiased in terms of black hole mass, luminosity,
broad-line width, or any other fundamental or observed property.  This
is a recurring issue for all AGN reverberation mapping campaigns to
date, and should be kept in mind when extrapolating reverberation
results to derive single-epoch masses for AGNs at high luminosities or
Eddington ratios not probed by the reverberation-mapped sample. This
point has been emphasized by \citet{richards2011} and
\citet{shen2013review} in the context of examining biases inherent in
single-epoch masses derived from the \ion{C}{4} line.

\begin{deluxetable*}{llcccccccc}
\tablecaption{Sample Properties and Observation Parameters}
\tablehead{
  \colhead{Galaxy} &
  \colhead{Alt.\ Name} &
  \colhead{$z$} &
  \colhead{$D_L$} &
  \colhead{$A_V$} & 
  \colhead{Slit PA} & 
  \colhead{$N_\mathrm{obs}$} &
  \colhead{Mean Sampling} & 
  \colhead{Median} & 
  \colhead{Median} \\
    \colhead{} &
  \colhead{} &
  \colhead{} &
  \colhead{(Mpc)} &
  \colhead{(mag)} & 
  \colhead{(deg)} & 
  \colhead{} &
  \colhead{(days)} &
  \colhead{Airmass} & 
  \colhead{S/N}
}
\startdata
Mrk 40      & Arp 151     & 0.0211 & 92.3  & 0.039 & 90   & 39 & 1.97 & 1.07 & 72 \\
Mrk 50      &             & 0.0234 & 102.5 & 0.044 & 180  & 55\tablenotemark{a} & 1.81\tablenotemark{c} & 1.29 & 77 \\ 
Mrk 141     &             & 0.0417 & 185.2 & 0.028 & 90   & 36  & 1.79 & 1.16 & 107 \\
Mrk 279     & PG 1351+695 & 0.0305 & 134.3 & 0.044 & 120  & 34  & 2.30 & 1.19 & 106 \\
Mrk 486     & PG 1535+547 & 0.0389 & 172.4 & 0.040 & 90   & 27  & 2.28 & 1.08 & 94 \\
Mrk 493     &             & 0.0313 & 137.9 & 0.068 & 70   & 32  & 2.39 & 1.08 & 88 \\
Mrk 504     & PG 1659+294 & 0.0359 & 158.7 & 0.135 & 60   & 36  & 2.18 & 1.05 & 89 \\
Mrk 704     &             & 0.0292 & 128.5 & 0.079 & 45   & 38  & 1.68 & 1.15 & 70 \\
Mrk 817     & PG 1434+590 & 0.0315 & 138.8 & 0.019 & 90   & 27  & 2.42 & 1.08 & 84 \\
Mrk 841     & PG 1501+106 & 0.0364 & 161.0 & 0.082 & 45   & 35  & 1.82 & 1.14 & 94 \\
Mrk 1392    &             & 0.0361 & 160.1 & 0.125 & 45   & 38  & 2.02 & 1.24 & 98 \\
Mrk 1511    & NGC 5940    & 0.0339 & 149.7 & 0.112 & 45   & 40  & 2.00 & 1.21 & 72 \\
NGC 4593    & Mrk 1330    & 0.0090 & 39.0  & 0.068 & 45   & 43  & 1.85 & 1.44 & 121 \\
PG 1310-108 & II SZ 10    & 0.0343 & 151.5 & 0.143 & 180  & 35  & 2.30 & 1.58 & 50 \\
Zw 229-015  &             & 0.0279 & 122.6 & 0.198 & 56.5 & 29\tablenotemark{b} &  2.72\tablenotemark{c} & 1.07 & 49 \\
\enddata
\tablecomments{Redshifts are taken from NED.  Luminosity distances
  ($D_L$) are calculated from redshifts assuming a WMAP9 cosmology
  with $H_0 = 69.7$ km s\per\ Mpc\per, $\Omega_M = 0.281$, and
  $\Omega_\Lambda = 0.7185$ \citep{hinshaw:wmap}, using the
  \citet{wright:cosmo} calculator.  Galactic extinctions ($A_V$) are
  from NED and are based on \citet{schlafly2011}, assuming
  $R_V=3.1$. $N_\mathrm{obs}$ is the total number of spectroscopic
  observations of each AGN. Mean sampling is the mean time interval
  between successive observations, in days. (The median sampling
  interval is 1.0 days for all AGNs.)  The median S/N gives the median
  S/N per pixel between rest wavelengths 4600 and 4700 \AA\ in the
  extracted blue-side spectra, for the full series of observations of
  each AGN.  \tablenotetext{a}{Twelve of the 55 spectra of Mrk 50 were
    obtained before the start of our main campaign during nights
    assigned to other projects.}  \tablenotetext{b}{Three of the 29
    spectra of Zw 229-015 were obtained after the end of our main
    campaign during nights assigned to other projects.}
  \tablenotetext{c}{Mean sampling for Mrk 50 and Zw 229-015 was
    calculated based on observations taken during the main campaign.}
\label{sampleproperties}}
\end{deluxetable*}

\section{Observations}
\label{sec:observations}

This project was allocated 69 nights at the Lick 3 m Shane telescope,
distributed between 2011 March 27 and June 13 (all dates are UT).
Interspersed with our nights were occasional gaps, mostly during or
close to full moon, for other projects such as exoplanet search
programs.

Observations were conducted using the Kast double spectrograph
\citep{millerstone}.  On the blue side (spatial scale 0\farcs43
pixel\per), we used a 600 lines mm\per\ grism covering 3440--5515
\AA\ at 1.02 \AA\ pixel\per.  On the red side (spatial scale 0\farcs78
pixel\per), we used a 600 lines mm\per\ grating blazed at 7500 \AA,
giving coverage of 5410--8200 \AA\ at 2.35 \AA\ pixel\per.  The d55
dichroic was employed to split the light between the blue and red
cameras.\footnote{In our 2008 Lick campaign \citep{bentz2009lamp}, we
  used only the red arm of the Kast spectrograph because the blue-side
  CCD had recently failed and was replaced at the time with a
  temporary, lower-efficiency detector.} A slit width of 4\arcsec\ was
adopted in order to mitigate slit losses while also preserving
moderate spectral resolution. Calibration frames taken each afternoon
included bias exposures, dome flats, and wavelength-calibration lamps
including He, Ne, Ar, Hg, and Cd.  The red-side CCD exhibits a steep
change in bias level at the short-wavelength end, while the blue-side
detector has extremely clean and flat bias structure with no
discernible persistent features.

The red-side spectra are affected by strong fringing at $\lambda
\gtrsim 7000$ \AA, but it can be removed fairly well by flat-field
division if a dome-flat exposure is taken at the same telescope
position and spectrograph rotation angle as the object.  While the
dome-flat exposures are very short (1--2 s), this requires additional
overhead time for the dome to be rotated or even capped (if the target
is close to the zenith).  In order to maximize the on-source exposure
time during the nights, we opted to use only afternoon dome-flat
exposures rather than observing dome flats at the position of each AGN
during the nights.  This trade-off compromises the data quality of the
red-side exposures at wavelengths longer than 7000 \AA, but it does
not affect the blue-side data which are the main focus of our
reverberation measurements.

Exposure times for each AGN were between 300 and 2400 s and were
adjusted during the campaign as the nights grew shorter.  Longer
integrations were usually split into two individual exposures to
facilitate cosmic-ray rejection.  

Each AGN was observed at a fixed slit position angle (PA) throughout
the program in order to maintain consistency in the amount of
spatially extended host-galaxy emission within the slit (Table
\ref{sampleproperties}).  Additionally, the flux-scaling method
applied to the spectra relies on the assumption of a constant
[\ion{O}{3}] flux, and observing at a fixed PA ensures consistency in
the portion of the narrow-line region subtended by the slit, which can
be important for low-redshift AGNs having spatially extended emission.
The PA for each object was chosen to be close to the parallactic angle
\citep{filippenko1982} during the portion of the campaign when the
object would be observed at relatively higher airmass.  Since the
observations were not carried out with the slit at the parallactic
angle, wavelength-dependent slit losses have an impact on the quality
of the relative flux calibration, although those losses are somewhat
mitigated by the use of the 4\arcsec-wide slit.  Slit losses
contribute spurious scatter to the continuum light curves measured
from the blue end of the spectra and to the \hal\ light curves
measured from the red-side data.  For Zw 229-015, the slit PA was set
to 56\fdg5 in order to include a foreground star in the slit and for
consistency with previous measurements \citep{barth:zw229}.

On clear nights we attempted to observe each AGN in the sample, but Zw
229-015 (at right ascension 19 hours) was only occasionally observed
during the first month of the program.  As the campaign progressed and
nights became shorter, exposure times were shortened for most objects.
We also ended monitoring for Mrk 141, Mrk 486, Mrk 817, and Mrk 841
about two weeks before the end of the campaign in order to conserve
observing time for higher-priority targets.  The weather at Lick was
worse than average during the spring of 2011.  Of the 69 nights, 20
were entirely lost to bad weather, and another 12 nights suffered
$\gtrsim50\%$ losses of observing time to clouds or high
humidity. Only 23 nights were fully usable with no significant
weather-related losses, while the remaining 14 nights were mildly to
moderately impacted by clouds.  Consequently, the number of
observations per AGN was relatively small, and for most targets we
obtained between 30 and 40 observations during the 69-night program.
Table \ref{sampleproperties} lists the number of exposures, median
airmass, and median S/N at 4600--4700 \AA\ for each object.  During
partially usable nights, the objects observed at highest priority
included Mrk 40, Mrk 50, Mrk 1511, and NGC 4593.

In order to extend our light curves for two AGNs, we also requested
additional observations from other observers using the Kast
spectrograph: Mrk 50 from January through March of 2011, and Zw
229-015 in June and July.  The Mrk 50 data (including observations by
other observers) were previously presented by \citet{barth2011:mrk50}.
Unfortunately, the 12 observations taken prior to the start of our
campaign suffered from large sampling gaps owing to winter weather,
and do not add much additional information to the lag measurements.
For Zw 229-015, three additional observations were taken 20--23 days
after the end of our main campaign.

During each night, time was also set aside for one spectroscopic
observation of a bright supernova or other transient.  Observations
taken during our program have been included in publications examining
the spectral evolution of SN 2010jl \citep{smith2012:sn2010jl}, SN
2011ay \citep{foley2013}, SN 2011by \citep{silverman:latetime}, and
the young stellar object PTF 10nvg \citep{hillenbrand2013}, and have
been described in Central Bureau Electronic Telegrams (CBETs) 2681 (SN
2011ay), 2699 (SN 2011bp), 2701 (SN 2011bq), 2702 (SN 2011br), 2712
(SN 2011cc), 2716 (SN 2011cf), 2721 (SN 2011cj), and 2752 (SN
2011dt).\footnote{CBETs are distributed via
  \\ \texttt{http://www.cbat.eps.harvard.edu/index.html }.}

\begin{figure*}
\begin{center}
\scalebox{0.80}{\includegraphics{{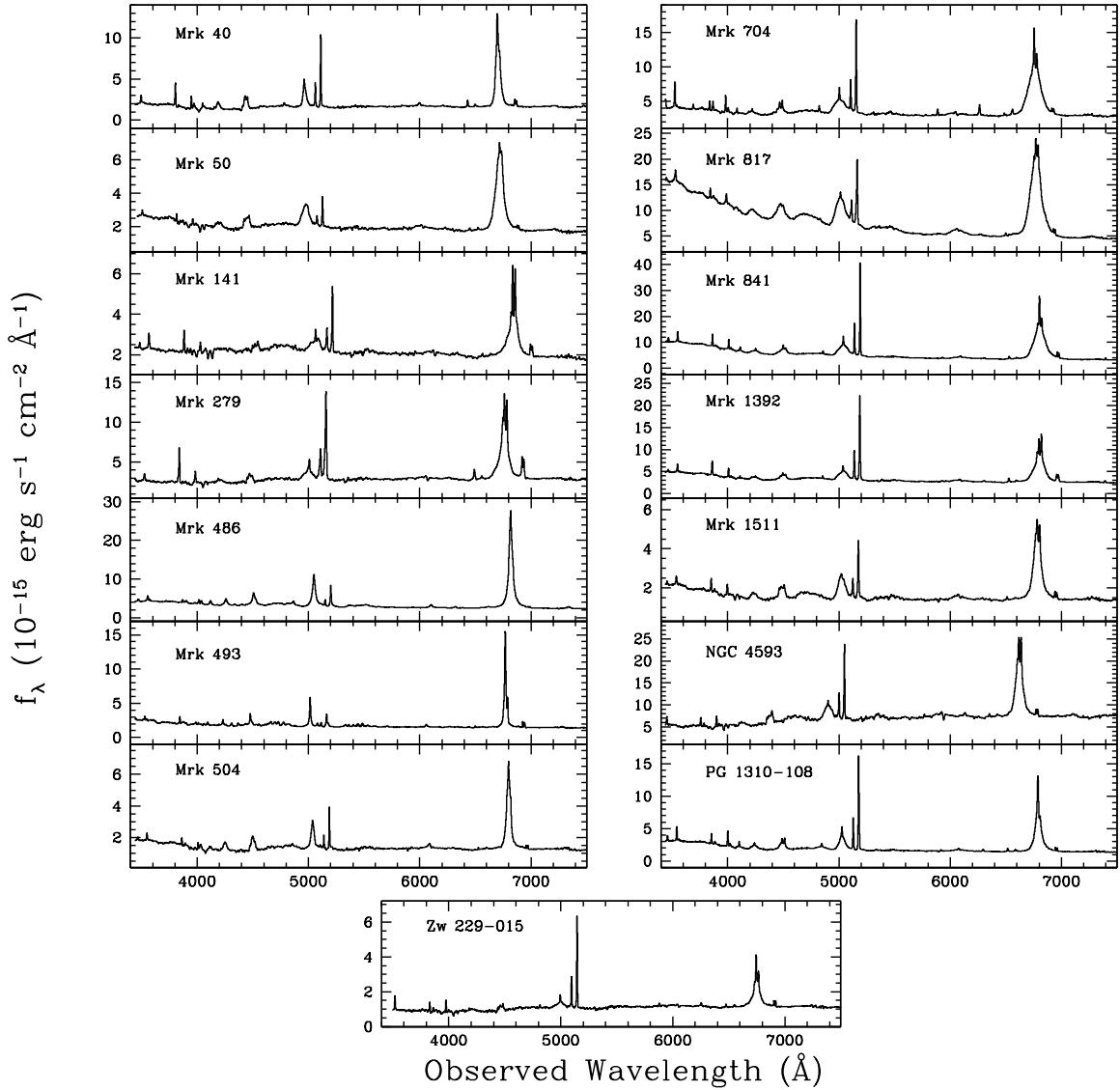}}}
\end{center}
\caption{Mean spectra of the sample including data from the blue-side
  and red-side cameras.  The displayed wavelength range is truncated
  at 7500 \AA.
\label{meanspectra}}
\end{figure*}

\section{Data Processing}
\label{sec:dataprocessing}

\subsection{Spectroscopic Data Reduction}

The initial data reduction followed standard procedures including
overscan subtraction, two-dimensional bias subtraction (for the
red-side CCD only), flat-fielding, and cosmic-ray cleaning using the
LA-Cosmic routine \citep{vandokkum2001} implemented as an
IRAF\footnote{IRAF is distributed by the National Optical Astronomy
  Observatories, which are operated by the Association of Universities
  for Research in Astronomy, Inc., under cooperative agreement with
  the National Science Foundation.} script.  The IRAF
\texttt{response} task was used to produce normalized flat-field
frames having pixel values close to unity.  Spectral extractions were
done with the IRAF \texttt{apall} task.  An unweighted extraction of
width 10\farcs3 was adopted in order to minimize the impact of nightly
seeing variations.  The wavelength scale was established by applying a
polynomial fit to the line-lamp spectra observed each afternoon, and
then applied to the extracted spectra.  For each wavelength-calibrated
spectrum, a small final linear shift to the wavelength scale was
applied based on offsets measured from strong emission lines in the
corresponding night-sky spectrum.  Flux calibration and removal of
telluric absorption features \citep[following the procedure
  of][]{wadehorne} were carried out using IDL routines as described by
\citet{matheson2000}.  The calibrated spectra were rebinned to a
linear wavelength scale of 1.0 \AA\ pixel\per\ on the blue side and
2.0 \AA\ pixel\per\ on the red side.  When two exposures of an AGN
were taken on the same night, they were combined to produce a final
weighted average spectrum.

Error spectra produced by the \texttt{apall} task were propagated
through all steps of the calibration pipeline.  For spectra of these
spatially extended AGNs, it is necessary to use the unweighted
extractions because optimal extractions \citep{horne1986} have a
tendency to truncate the peaks of strong emission lines such as
[\ion{O}{3}] when they have a different spatial extent than the
surrounding continuum flux.  However, the \texttt{apall} error
spectrum actually gives the pixel uncertainties (owing to photon
counting and readout noise) on the optimally weighted extraction, not
on the unweighted extraction.  For our high-S/N observations
(typically with S/N $\approx$ 50--100 pixel\per), the differences
between the unweighted and optimally weighted extractions are very
small, and the error spectra will only slightly underestimate the
actual photon-counting uncertainties.  As described below,
systematic uncertainties in flux calibration are often a major or
dominant source of noise in the spectroscopic light curves. This
additional source of error is not accounted for in the extracted error
spectra and must be added separately to the error budget after the
spectral rescaling procedure is applied.

\subsection{Photometric Scaling}
\label{photscale}

Conditions at Lick are rarely photometric during the spring season,
and the flux-calibrated spectra do not share a consistent overall flux
scale.  The wavelength scales of the calibrated spectra also exhibit
small offsets from night to night, caused by random miscentering of
the AGN in the spectrograph slit.  Before carrying out the
emission-line measurements, the wavelength and flux scales of the
spectra must be aligned.

The blue-side data were treated as follows for each AGN individually.
First, wavelength shifts between the nightly spectra were measured
using a cross-correlation procedure, sampling in steps of 0.1
pixel. The wavelength shifts determined from cross-correlation were
removed from the wavelength vector of each spectrum, and the aligned
spectra were then averaged to produce an initial mean reference
spectrum.

In order to place the reference spectrum on an approximate absolute
photometric scale, we examined the nightly observing logs and selected
nights listed by the observers as being either photometric or very
clear; these candidate clear nights numbered only 14 during the
69-night campaign.  For each AGN, the [\ion{O}{3}] $\lambda5007$ flux
was measured from each candidate clear night's spectrum by integrating
the line flux above a straight-line fit to adjacent continuum regions
on either side of the line.  Then, the standard deviation of these
line fluxes was calculated, and $>2\sigma$ outliers were discarded
from the list of candidate photometric observations.  This process was
then repeated until there were no remaining $2\sigma$ outliers in
[\ion{O}{3}] flux, and the mean [\ion{O}{3}] flux from the remaining
spectra was taken to be the best estimate of the true line flux.  The
number of final ``clear-night'' spectra varied for each AGN and ranged
between 6 and 9.  Table \ref{oiiitable} lists the [\ion{O}{3}] fluxes
based on this procedure; the quoted uncertainty in the table is the
standard deviation of the fluxes measured on candidate photometric
nights after excluding outliers.  The reference spectrum was then
multiplied by a scaling factor to normalize its [\ion{O}{3}] flux to
this photometric value. We caution that these [\ion{O}{3}] fluxes
should be considered only as estimates of the true fluxes, given the
scarcity of truly photometric conditions at Lick.

\begin{deluxetable}{lccc}
\tablecaption{[\ion{O}{3}] and Featureless Continuum Fluxes}
\tablehead{
  \colhead{Object} &
  \colhead{$f$([\ion{O}{3}])} &
  \colhead{$\sigma_\mathrm{nx}$([\ion{O}{3}])} & 
  \colhead{$f_{\lambda}$(5100~\AA)} \\
}
\startdata
Mrk 40      & $68.8\pm6.2$   & 0.007 & $0.56\pm0.08$ \\
Mrk 50      & $16.3\pm2.1$   & 0.033 & $1.20\pm0.20$ \\
Mrk 141     & $34.1\pm3.7$   & 0.016 & $1.09\pm0.16$ \\
Mrk 279     & $157.6\pm7.7$  & 0.006 & $0.93\pm0.10$ \\
Mrk 486     & $65.4\pm4.3$   & 0.019 & $2.15\pm0.26$ \\
Mrk 493     & $25.8\pm1.6$   & 0.023 & $1.43\pm0.17$ \\
Mrk 504     & $17.3\pm0.5$   & 0.015 & $0.57\pm0.06$ \\
Mrk 704     & $114.7\pm14.5$ & 0.005 & $2.32\pm0.37$ \\
Mrk 817     & $143.7\pm18.7$ & 0.013 & $5.69\pm0.92$ \\
Mrk 841     & $270.4\pm19.1$ & 0.009 & $3.63\pm0.45$ \\
Mrk 1392    & $180.0\pm28.8$ & 0.006 & $2.41\pm0.45$ \\
Mrk 1511    & $31.3\pm0.5$   & 0.015 & $0.86\pm0.09$ \\
NGC 4593    & $144.4\pm5.8$  & 0.023 & $2.64\pm0.29$ \\
PG 1310-108 & $128.2\pm2.7$  & 0.007 & $1.66\pm0.17$ \\
Zw 229-015  & $37.4\pm1.3$   & 0.007 & $0.56\pm0.06$ \\
\enddata
\tablecomments{[\ion{O}{3}] line fluxes are given in the observed
  frame in units of $10^{-15}$ erg cm\persq\
  s\per, and uncertainties represent the standard deviation of
  [\ion{O}{3}] fluxes measured on candidate photometric nights.  
  The quantity $\sigma_\mathrm{nx}$ is the normalized excess
  scatter in the [\ion{O}{3}] light curve, as defined in \S
  \ref{photscale}. For Mrk 50, data points taken before the start of the
  main campaign were excluded from the computation of
  $\sigma_\mathrm{nx}$, because the earlier observations were taken
  with heterogeneous spectrograph setups and had a much larger flux
  scaling scatter.  The AGN featureless continuum flux density is given
  in the AGN rest frame at 5100~\AA\ in  units of 
  $10^{-15}$ erg cm\persq\ s\per\ \AA\per, and is 
  measured from the power-law component of the mean spectrum
  decomposition and corrected for Galactic extinction.}
\label{oiiitable}
\end{deluxetable}

Then, the procedure described by \citet{vgw1992} was applied to the
data to align the wavelength and flux scales under the assumption that
the [\ion{O}{3}] flux is intrinsically constant during the duration of
the monitoring campaign.  Long-duration monitoring of NGC 5548
confirms that this is a reasonable assumption, although [\ion{O}{3}]
flux variations have been detected on timescales of years
\citep{peterson2013}.  The \citet{vgw1992} procedure applies a scaling
factor, a linear wavelength shift, and a Gaussian broadening to the
individual night's spectrum, and finds the parameter values which
minimize the value of $\chi^2$ when fitting a low-order polynomial to
the difference between the reference spectrum and the adjusted
individual spectrum, over a small wavelength range including
[\ion{O}{3}] and some redward continuum.  Since the spectral focus and
line-spread function can vary from night to night, and either the
individual spectrum or the reference spectrum might have a narrower
line-spread function, the procedure is repeated with the Gaussian
broadening kernel applied to the reference spectrum instead of the
individual spectrum, and the final scaled spectrum is taken to be the
better of the two options in terms of the $\chi^2$ value obtained.
Instead of using a grid search to find the best-fit values of the
three free parameters as described by \citet{vgw1992}, our method uses
downhill simplex optimization, implemented with the IDL
\texttt{amoeba} procedure, to minimize $\chi^2$.  We found that over
the full time series of spectra for a given AGN, the shifts in
wavelength scale needed to align the spectra were at the level of
$\sim1$ \AA\ rms.  The final blue-side spectra are rebinned to a
uniform wavelength scale of 1.0 \AA\ pixel\per\ with integer
wavelength values in \AA.

To test the effectiveness of the spectral scaling, we can examine the
scatter in [\ion{O}{3}] flux values for each AGN over the course of
the campaign.  Since the scaling procedure essentially matches the
[\ion{O}{3}] profile for each nightly spectrum to the reference
spectrum, but does not explicitly force all [\ion{O}{3}] fluxes to be
identical, the dispersion among [\ion{O}{3}] fluxes for a given AGN
gives a measure of the final random uncertainties in the flux
calibration after the scaling procedure has been applied.  We use the
normalized excess variance $\sigma_\mathrm{nx}^2$ to quantify the
light-curve variance over and above the amount expected from the
propagated measurement uncertainties:
\begin{equation}
\sigma_\mathrm{nx}^2 = \frac{1}{N\langle f \rangle^2}
\sum_{i=1}^{N}[(f_i - \langle f \rangle)^2 - \delta_i^2],
\end{equation}
where $N$ is the number of observations in the light curve, $\langle f
\rangle$ is the mean flux, and $f_i$ and $\delta_i$ are the individual
flux measurements and their uncertainties from photon counting and
detector read noise.\footnote{For clarity, throughout this paper we use
  the letter $\delta$ to denote measurement uncertainties, and
  $\sigma$ to denote dispersions of a sample of measurements or
  dispersions (widths) of line profiles.}  Then, $\sigma_\mathrm{nx}$
is a measure of the fractional rms scatter in the light curve over and
above the amount expected from the propagated uncertainties; we
refer to this quantity as the ``normalized excess scatter.''  We
measured $\sigma_\mathrm{nx}$ from the [\ion{O}{3}] light curves for
each AGN, and the values are listed in Table \ref{oiiitable}.  The
values of $\sigma_\mathrm{nx}$ range from 0.005 to 0.033 (i.e., the
excess scatter in the light curves ranges from 0.5\% to 3.3\% of the
mean [\ion{O}{3}] fluxes).  The objects with the highest [\ion{O}{3}]
photometric scatter tend to be AGNs having low [\ion{O}{3}] equivalent
width, such as Mrk 50 and Mrk 493, for which the \citet{vgw1992}
scaling method does not work optimally.

We interpret $\sigma_\mathrm{nx}$ as a measure of the level of random
error in the final flux normalization of the spectra.  For
measurements of reverberation lags or other quantities from the light
curve, this source of error can be incorporated into the measurement
error budget by adding $\sigma_\mathrm{nx} \times \langle f \rangle$
in quadrature to the photon-counting uncertainties on each light-curve
point $f$.  Since the spectral scaling is optimized for the wavelength
of the [\ion{O}{3}] $\lambda5007$ line, the relative scaling at much
longer or shorter wavelengths is expected to be worse than the amount
indicated by $\sigma_\mathrm{nx}$, as a result of atmospheric
dispersion and wavelength-dependent slit losses.

After applying the scaling procedure, mean and rms spectra are
constructed following \citet{peterson2004}.  The mean spectrum is a
simple average of the individual nightly spectra.  The rms spectrum is
calculated by taking the standard deviation of flux values at each
pixel over the time series, and gives an illustration of the relative
variability amplitude of different spectral regions. Other
prescriptions for calculating rms spectra have been explored by
\citet{park2012a}.  Since the rms spectrum includes the contribution
of photon-counting noise and other sources of noise or random error in
the data, in addition to genuine AGN variability, the amplitude of the
rms spectrum is higher than it would be for the ideal case of
noise-free data. Simulations presented in Appendix \ref{appendix:biases}
demonstrate that photon-counting noise will bias the widths of broad
lines in rms spectra to values that are typically several percent
lower than they would be in the absence of noise.

The red-side data were generally flux-calibrated using standard-star
observations obtained simultaneously with the blue-side standard-star
observations. In the reduced data, the region of wavelength overlap
between the blue-side and red-side spectra is 5420--5500 \AA.  For
each AGN observation, we normalized the red-side flux scale by
applying a scaling factor to match it to the scaled blue-side spectrum
over this wavelength range, and then produced mean and rms spectra
from these rescaled red-side data.  Figure \ref{meanspectra} displays
the mean spectra of each AGN from the blue and red camera data
combined.

For a few of the AGNs, we found that \hal\ light curves measured from
these rescaled spectra had noticeably larger scatter in flux values
than the \hbeta\ light curves.  This large excess scatter for the
red-side data probably arises from a combination of the nonparallactic
slit orientation, the seeing-dependent ratio of AGN to host-galaxy
continuum flux, and the fact that the spectral normalization is
optimized for wavelengths close to [\ion{O}{3}].  Large excess scatter
was particularly noticeable in the \hal\ light curves of Mrk 40 and
Mrk 279.  For both of these objects, we applied the \cite{vgw1992}
scaling method separately to the red-side spectra, using the
[\ion{O}{1}] $\lambda6300$ line as a reference.  This significantly
improved the quality of the \hal\ light curves for these two objects,
although a few extreme outlier points were still present in the data.
For other objects, we found that either rescaling using [\ion{O}{1}]
did not noticeably improve the \hal\ light curves, or, in some cases
such as Mrk 50, [\ion{O}{1}] was too weak to use as a reference line
in the spectral scaling.  We experimented with using the [\ion{S}{2}]
doublet as the reference line, but this produced poor results because
[\ion{S}{2}] is very weak in most of these AGNs and is usually blended
with the red wing of \hal.

\subsection{Spectral Fitting}

Traditionally, measurements of emission-line fluxes for reverberation
mapping have been carried out by selecting two regions nearly free of
emission lines on either side of the emission line of interest, then
fitting and subtracting a straight line to model and remove the
continuum underlying the emission line.  While this procedure is
simple to implement and can generally produce accurate light curves
for strong emission lines, there are several drawbacks associated with
it.  A straight-line fit does not adequately represent the shape of
the starlight continuum, which often makes a significant contribution
to the spectra of Seyfert 1 nuclei (particularly when observed through
a wide slit). Stellar absorption features in the host-galaxy starlight
produce wiggles and bumps that become superposed on the broad
\hbeta\ line profile (including stellar \hbeta\ absorption), and these
contaminating features are not removed when a straight-line continuum
is assumed.  Furthermore, while there are some regions around
\hbeta\ where the flux from other emission lines is relatively low,
there may still be contributions from weak features such as
\ion{Fe}{2} or \ion{He}{2} in the selected continuum fitting regions,
and it is effectively impossible to define a true continuum region
that is entirely devoid of AGN emission features.  These issues can
compromise measurements of emission-line light curves and broad-line
widths or profile shapes in the mean and rms spectra.  Measurements of
velocity-resolved lags in the \hbeta\ line can be compromised by
blending with \ion{He}{2}, \ion{He}{1}, and \ion{Fe}{2} if these
features are not removed before measurement of the reverberation
signals, and biased determination of the true continuum level owing to
contamination by blended lines or starlight could affect measurements
of the lag in the faint high-velocity wings of \hbeta.  In addition to
strong lines such as \hbeta, it is also of interest to measure light
curves for weaker features including \ion{Fe}{2} or \ion{He}{2}, but
these lines can be severely blended with each other as well as with
starlight and other spectral components, and a straight-line continuum
fit is usually not adequate for isolating the contributions of weaker
broad lines to the spectrum.

In order to address these problems, we have implemented a spectral
fitting method which, when applied to each nightly spectrum from a
reverberation campaign, decomposes the data into independent spectral
components.  Fitting of AGN spectra to deblend line and continuum
components (including the broad \ion{Fe}{2} blends) has a long history
with many different implementations in the literature
\citep[e.g.,][]{wills1985, dietrich2002, woo2006, gh07, shen2008,
  kovacevic2010}, but only recently have such methods been applied to
time-series spectroscopic data from reverberation campaigns.  The
procedure described below is an updated version of the method applied
to our data in earlier papers in this series \citep{barth2011:mrk50,
  barth2013}, with several minor modifications and improvements.  The
method described by \cite{park2012a} uses a different fitting
procedure and a different set of model components, but we find that
measurements of \hbeta\ light curves and other parameters are very
similar when carried out using the results of the two independent
fitting codes when applied to the same set of input spectra.

The primary goal of our procedure is to provide an accurate fit to the
region surrounding \hbeta\ so that the broad \hbeta\ profile can be
isolated from other features.  Consequently, accurate fitting over a
small wavelength region is a higher priority than fitting the entire
available wavelength range of the observations.  Fits to a larger
wavelength region \citep[e.g.,][]{denney2009se} would include several
additional emission lines that are of little direct interest for the
reverberation measurement, and extending the fitting region blueward
could compromise the accuracy of the fit in the \hbeta\ region owing to
the complexity of the spectrum at bluer wavelengths.  We fit only the
blue-side spectra, and the redward limit of the fits is set by the red
end of the blue-side CCD data.  Fits were carried out over a rest
wavelength range of approximately 4200--5300 \AA, with slight
differences for each AGN depending on redshift, and the fitting
procedure is applied to the data after scaling by the \citet{vgw1992}
procedure.  The Levenberg-Marquardt routines in the IDL \texttt{mpfit}
package \citep{markwardt2009} are employed for $\chi^2$ minimization
to optimize the fits, using the propagated error spectra for
determination of $\chi^2$.  The model components and parameters are as
follows.

\emph{Starlight}: We used an 11~Gyr, solar metallicity, single-burst
spectrum from \citet{bc03} to model the host-galaxy starlight. We
additionally carried out tests in which a younger (290 Myr) population
component was added. Its contribution was generally small and not well
constrained, and it was omitted in the final fits.  During the fitting
process, the starlight spectrum was rebinned to a log-linear wavelenth
scale, broadened by convolution with a Gaussian kernel, and then
rebinned again to match the linear wavelength scale of the data.  Free
parameters include the flux normalization of the starlight spectrum,
the redshift, and the velocity width of the broadening kernel.  Since
the spectral resolution of the \citet{bc03} spectra differs from the
Lick data, the convolution kernel width does not directly correspond
to the actual stellar velocity dispersion in the host galaxy.  The
velocity-broadening dispersion parameter was limited to a maximum
value of 350 \kms\ in order to prevent it from increasing to
unphysically large values; this limit was reached only for Mrk 486, in
which stellar absorption features are essentially invisible over the
wavelength range of the fits.

\emph{AGN featureless continuum}: The AGN continuum was modeled as a
power law, with free parameters including the flux normalization at
rest wavelength 5100 \AA\ and the power-law spectral index.  Over the
wavelength range that was fitted, Balmer continuum emission does not
contribute to the spectrum.  We do not ascribe any physical meaning to
night-to-night changes in the best-fitting spectral index, since these
can be caused by wavelength-dependent slit losses resulting from
differential atmospheric refraction and variable seeing.

\emph{[\ion{O}{3}]}: A fourth-order Gauss-Hermite function
\citep{vdm93} was used to model the [\ion{O}{3}] $\lambda5007$
profile.  Free parameters included the amplitude, centroid wavelength,
line dispersion, and the higher-order moments $h_3$ and $h_4$.  The
4959 \AA\ line was modeled with the same velocity profile and an
amplitude fixed to 1/3 of that the 5007 \AA\ line, and a fixed
rest-frame wavelength separation of 47.9 \AA.  A fourth-order
Gauss-Hermite function proved to be adequate for modeling the
[\ion{O}{3}] profiles well in all but one of the AGNs; as described
below, we modified the set of fit components for Mrk 279.

\emph{\hbeta}: For the broad component of \hbeta, a fourth-order
Gauss-Hermite function was used. This provides sufficient flexibility
to fit the broad-line profiles for nearly all of the AGNs well, with
the exception of objects such as Mrk 141 having distinct shoulders or
humps on their profiles.  Free parameters include the broad component
centroid, dispersion, $h_3$, $h_4$, and amplitude.  The narrow
component was set to have a profile identical to [\ion{O}{3}], but a
small shift was allowed in the wavelength centroid of the narrow
component as another free parameter.

\emph{\ion{He}{2}}: The \ion{He}{2} $\lambda4686$ line was modeled as
the sum of two Gaussians representing the broad and narrow components.
For each component, free parameters included the centroid, dispersion,
and amplitude.  Initial trial fits indicated that the broad and narrow
components sometimes had offset centroid wavelengths, and for some
AGNs the fits failed noticeably when the two components were forced to
have the same centroid.  In some cases this might be the result of
broad \ion{He}{1} $\lambda4713$ being present in the data but not
included as a fit component.  If the broad \ion{He}{2} model component
incorporated some \ion{He}{1} $\lambda4713$ flux then this would
induce a spurious but small redward shift relative to its true
wavelength.  \citet{veron2002} illustrate examples of fits that
include both \ion{He}{2} $\lambda4686$ and \ion{He}{1} $\lambda4713$,
but since these features are badly degenerate in broad-lined AGNs we
do not attempt to include the \ion{He}{1} $\lambda4713$ line in our
fits as a separate component.

\emph{\ion{He}{1}}: We included the possible contributions of
\ion{He}{1} broad lines at 4471, 4922, and 5016 \AA\ rest wavelength.
In order to keep the number of free parameters to a minimum for these
relatively weak lines, we used Gaussian profiles and followed
\citet{vp05} in setting the 4922 and 5016 \AA\ lines to have the same
amplitude, but we did not force any relationship between the
amplitudes of these lines and the 4471 \AA\ line.  All three
\ion{He}{1} lines were assumed to have identical velocity widths, and
their velocity centroids were set to be identical to that of \hbeta.
The \ion{He}{1} components thus required only three additional free
parameters to represent the amplitudes and widths of these features.
The \ion{He}{1} 4922 and 5016 \AA\ lines are in the ``red shelf'' of
\hbeta, and are blended with the red wing of \hbeta, the [\ion{O}{3}]
lines, and with \ion{Fe}{2} features \citep{veron2002}.  In fact the
4922 and 5016 \AA\ lines are nearly degenerate with two \ion{Fe}{2}
features that appear in the iron templates. The amplitude of these two
features in the fits depends strongly on which \ion{Fe}{2} template is
chosen.  While the sum of the \ion{Fe}{2} and \ion{He}{1} fluxes in
this spectral region is fairly well determined by the fits, we do not
attempt to extract \ion{He}{1} light curves because of the ambiguity
in separating these features from the \ion{Fe}{2} emission.  For most
AGNs, the best-fitting amplitude of the \ion{He}{1} $\lambda4471$ line
is either zero or negligibly small, regardless of \ion{Fe}{2} template
choice.

\begin{figure}[t!]
\scalebox{0.4}{\includegraphics{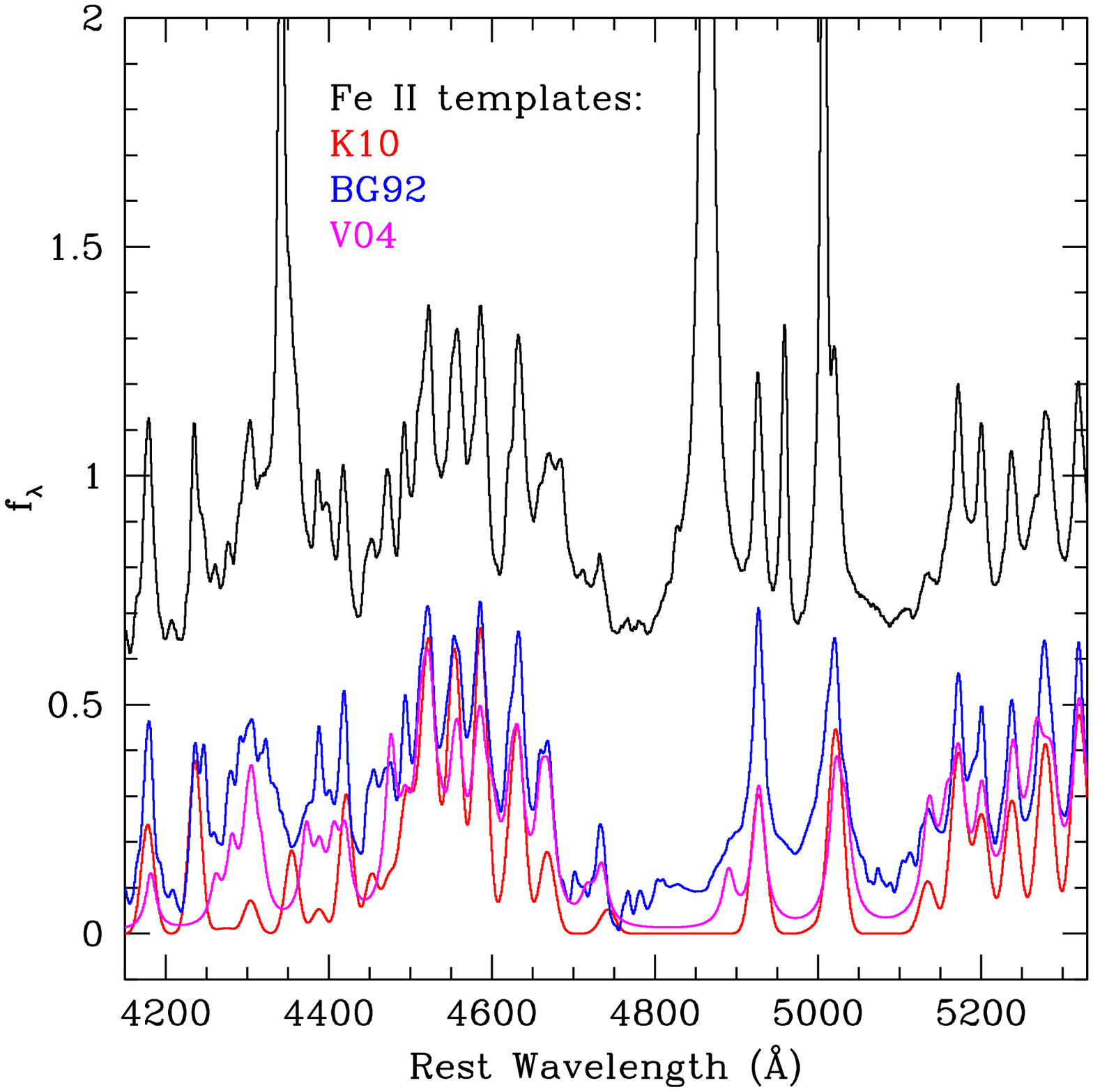}}
\caption{Illustration of the differences between the \ion{Fe}{2}
  templates of \citet{bg92} (blue), \citet{veron2004} (magenta), and
  \citet{kovacevic2010} (red) when fitted to the spectrum of Mrk 493.
  The Mrk 493 mean spectrum, after subtraction of the AGN featureless
  continuum and starlight components, is shown in black, and has been
  shifted upward by 0.7 units for clarity.  The colored curves show
  the \ion{Fe}{2} templates with velocity broadening and normalization
  as determined from the best fit to the mean spectrum.}
\label{fetemplates}
\end{figure}

\begin{figure*}[t!]
\scalebox{0.92}{\includegraphics{{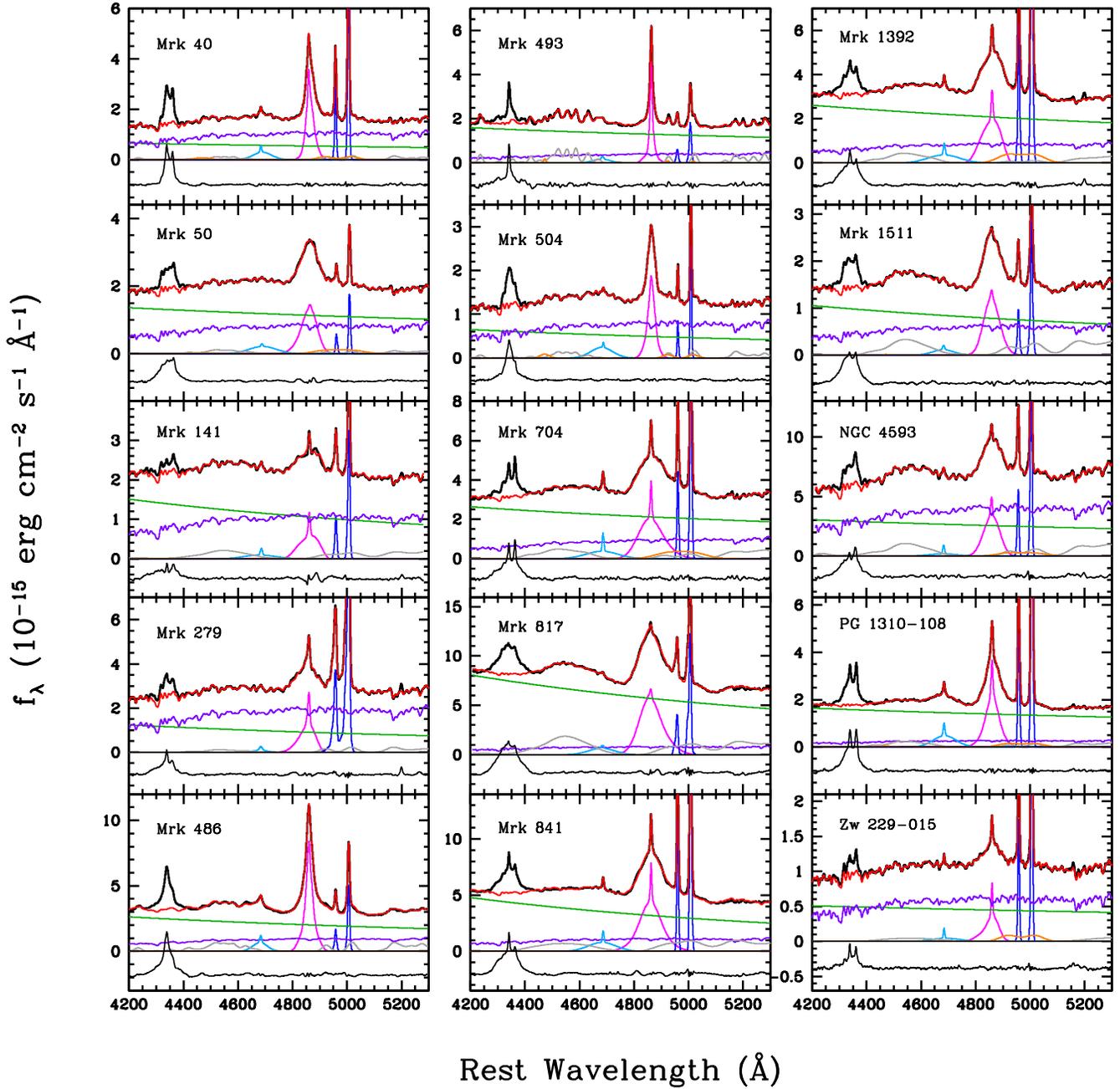}}}
\caption{Fits to the blue-side mean spectra.  The observed mean
  spectra are shown in black.  Model components include starlight
  (purple), the AGN featureless continuum (green), \hbeta\ (magenta),
  \ion{He}{2} (cyan), \ion{He}{1} (orange), \citet{kovacevic2010}
  \ion{Fe}{2} template (grey), and [\ion{O}{3}] (blue).  The sum of
  all model components is shown in red, superposed on the observed
  spectrum.  The residual spectrum (data $-$ total model) is shown in
  black and is offset downward to negative values for clarity.  This
  residual contains emission features including \hgamma, [\ion{O}{3}]
  $\lambda4363$, and [\ion{N}{1}] $\lambda5200$ that are not
  incorporated as components in the model. }
\label{spectralfits}
\end{figure*}

\emph{\ion{Fe}{2}}: The blends of broad \ion{Fe}{2} emission were
modeled using template spectra, including those described by
\citet{bg92}, \citet{veron2004}, and \citet{kovacevic2010}.  We
carried out fits separately with each of these templates in order to
examine systematic differences in the results.  Template spectra were
convolved with a Gaussian kernel in velocity space, similar to the
broadening of the starlight spectrum described above. In each case,
free parameters included the \ion{Fe}{2} velocity shift relative to
broad \hbeta, the broadening kernel width, and the normalization of
the broadened template spectrum.  The \citet{bg92} and
\citet{veron2004} templates are monolithic and only one normalization
parameter is required.  The \citet{kovacevic2010} template \citep[see
  also][]{shapovalova2012} is composed of five separate templates
representing different emission multiplet groups, and a normalization
parameter is required for each one.  In a given spectral fit using the
\citet{kovacevic2010} templates, each of the five templates was
assumed to have the same velocity broadening and velocity shift.  The
\citet{kovacevic2010} templates are distributed with a discrete
sequence of velocity broadenings, and we used the narrowest templates
(corresponding to 700 \kms\ broadening) and then applied a Gaussian
velocity broadening during the fitting process as described above.
Examination of the fits for a very narrow-lined AGN such as Mrk 493
clearly illustrate the contribution of \ion{Fe}{2} in the red wings of
both \hbeta\ and [\ion{O}{3}] $\lambda5007$, while in more typical
broad-lined AGNs these \ion{Fe}{2} features are more difficult to
identify individually and their contribution to this spectral region
may be less obvious unless a multicomponent fit is carried out.

\begin{figure}
\scalebox{0.44}{\includegraphics{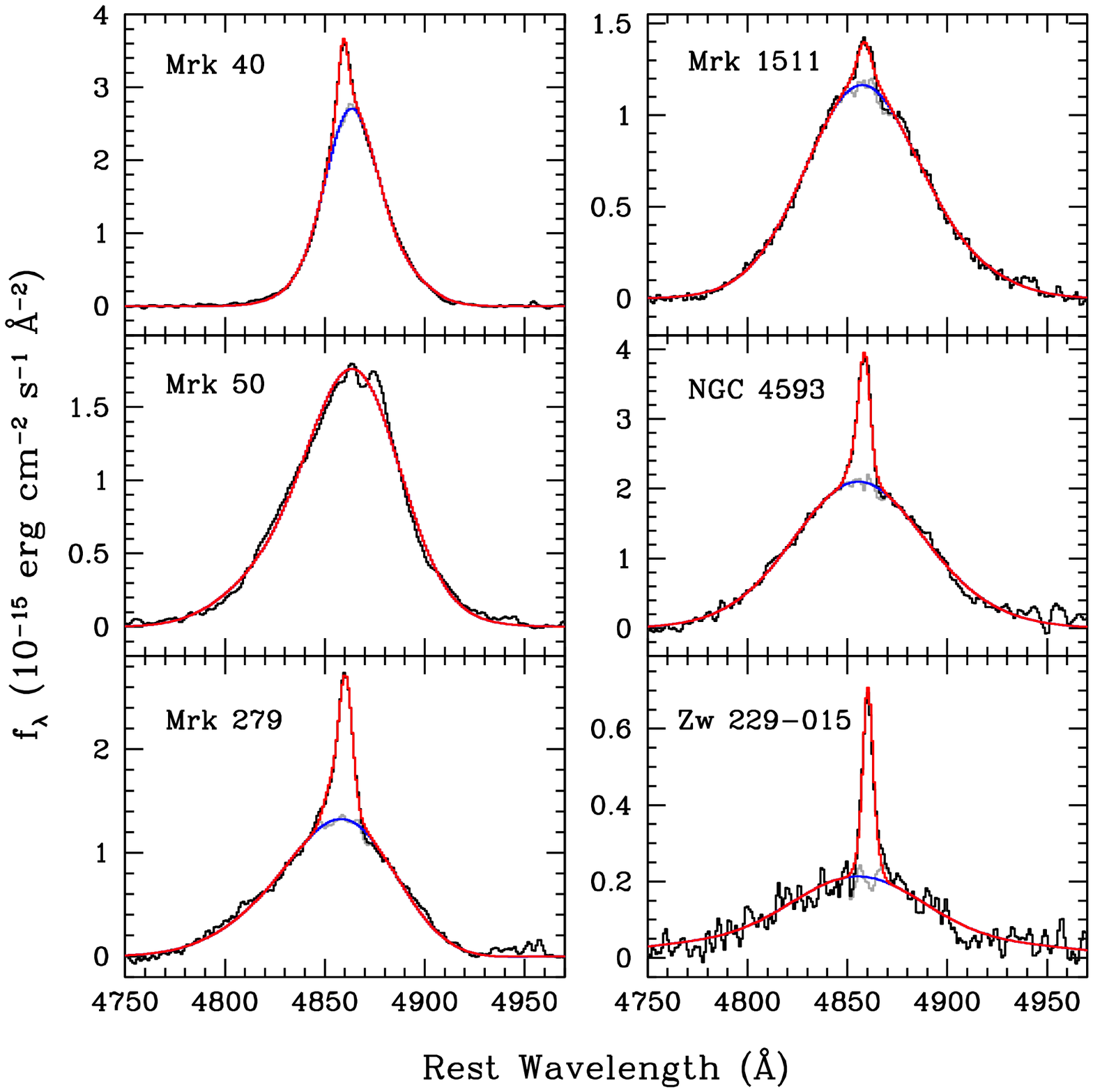}}
\caption{Examples of fits to the \hbeta\ line in single-epoch
  observations.  The displayed data are from observations taken on
  2011-04-27.  In each panel, the decomposed \hbeta\ profile obtained
  by subtracting all model components other than \hbeta\ from the
  spectrum is shown in black.  This includes both the narrow and broad
  components of \hbeta.  The broad component spectrum after
  additionally subtracting the narrow \hbeta\ model component is shown
  in grey.  Blue and red curves show the best-fitting broad
  \hbeta\ and total \hbeta\ (broad + narrow) model components. }
\label{hbetacomps}
\end{figure}

\emph{Reddening}: All model components were reddened by a
\citet{ccm1989} reddening law, with $E(B-V)$ as a free parameter in
the fit.  This parameter accounts for possible reddening within the AGN
host galaxy as well as Galactic reddening, and also accounts for
possible errors in relative flux calibration resulting from the
nonparallactic slit orientation or other wavelength-dependent flux
calibration errors.  For some AGNs, the best-fitting values of
$E(B-V)$ varied significantly from night to night, and we do not
interpret the fitted parameter values as actual measurements of the
reddening toward the AGNs. In principle, the AGN continuum and BLR,
NLR, and host-galaxy light could have different line-of-sight
reddening, but the data would not provide sufficient constraints to
obtain meaningful values of reddening for these components
independently.

In the \hbeta\ and [\ion{O}{3}] models, the wings of Gauss-Hermite
functions can extend to negative values for certain values of $h_3$
and $h_4$. Negative values were clipped and replaced with zero. This
only affects the distant wings of line profiles and has a very minor
effect on the fits.

Other emission lines are present in the spectrum that could be fitted
by adding components and free parameters to the model, but we chose to
restrict the number of components to those listed above in order to
avoid either slowing down the fitting procedure or introducing
additional fitting degeneracies.  Instead, other emission features
were accounted for simply by setting a wavelength region around those
features to have zero weight in the fits.  The largest feature that
was masked out was the blend of \hgamma\ and [\ion{O}{3}]
$\lambda4363$.  This blend typically has a complex shape that, if
included in the fit, would require three overlapping Gauss-Hermite or
Gaussian components and potentially nine or more additional free
parameters.  The masked region corresponded to 4280--4400 \AA\ (rest
wavelength) but was adjusted in a few cases for objects having very
broad lines.  We also masked out a small region around the narrow
[\ion{N}{1}] lines at 5199 and 5201 \AA.

Our procedure fits all of these spectral components to the data
simultaneously, in contrast to other methods which fit and remove
continuum and \ion{Fe}{2} components prior to fitting the
\hbeta\ profile \cite[e.g.,][]{shen2008,park2012a}. Fits were carried out
separately using each of the three \ion{Fe}{2} templates.  The total
number of free parameters in the fits is 29 when the monolithic
\ion{Fe}{2} templates are used, or 33 when the multicomponent
\citet{kovacevic2010} templates are used.  For each AGN, the mean
spectrum was fitted first, and then the best-fit parameters determined
for the high S/N mean spectrum were used as the initial parameters for
the fits to each nightly spectrum.  In most cases, fits using the
\citet{kovacevic2010} templates returned the best $\chi^2_\nu$ values,
and we use these templates for the final fit results presented in this
paper.  Figure \ref{fetemplates} illustrates the systematic
differences between the three templates when fitted to the spectrum of
Mrk 493, which has the strongest and narrowest \ion{Fe}{2} lines of
the AGNs in our sample.  The template differences are most pronounced
at wavelengths underlying the broad \hgamma\ and \hbeta\ lines and the
\hbeta\ red shelf region. 

The fitting procedure was modified for Mrk 279.  Its [\ion{O}{3}]
lines are unusually broad with strong asymmetric blue wings which
contribute substantially to the ``red shelf'' region redward of
\hbeta, and the 4th-order Gauss-Hermite model gave a poor fit to the
      [\ion{O}{3}] doublet leaving very strong residuals in this
      region.  For this object, we removed the \ion{He}{1} components
      from the model (since they did not appear to be necessary in
      this case) and replaced them with two additional Gaussian
      components representing the blue wings of the [\ion{O}{3}]
      $\lambda\lambda4959, 5007$ lines.  The \hbeta\ model also
      required modifications.  In order to prevent the broad
      \hbeta\ width from blowing up to arbitrarily high values due to
      fitting degeneracy in the red wing, we restricted the broad
      component model to $\sigma < 30$ \AA. We also found that the
      narrow component of \hbeta\ had a significantly different
      profile from [\ion{O}{3}] in Mrk 279, and allowed the width of
      narrow \hbeta\ to vary rather than tying it to the [\ion{O}{3}]
      width.  This resulted in
      $\sigma$(\hbeta$_\mathrm{n}$)$\approx0.75\sigma$([\ion{O}{3}]).

The best fits to the mean spectrum of each AGN, using the
\cite{kovacevic2010} template, are shown in Figure \ref{spectralfits}.

For each spectrum, the best-fitting model components are saved.
Additionally, we save the decomposed \hbeta\ profile, made by
subtracting all model components other than \hbeta\ from the original
spectrum. This spectrum also contains the \hgamma+[\ion{O}{3}] blend
as part of the fitting residual.  We save two versions of this
spectrum, one including and one excluding the narrow
\hbeta\ component.  These decomposed spectra (rather than the
noise-free models) are used for measurement of the \hbeta\ light
curves and line widths.  Figure \ref{hbetacomps} illustrates examples
of the decomposed \hbeta\ profiles and component models for six AGNs
from a single night's data.

\begin{figure*}[t!]
\scalebox{0.92}{\includegraphics{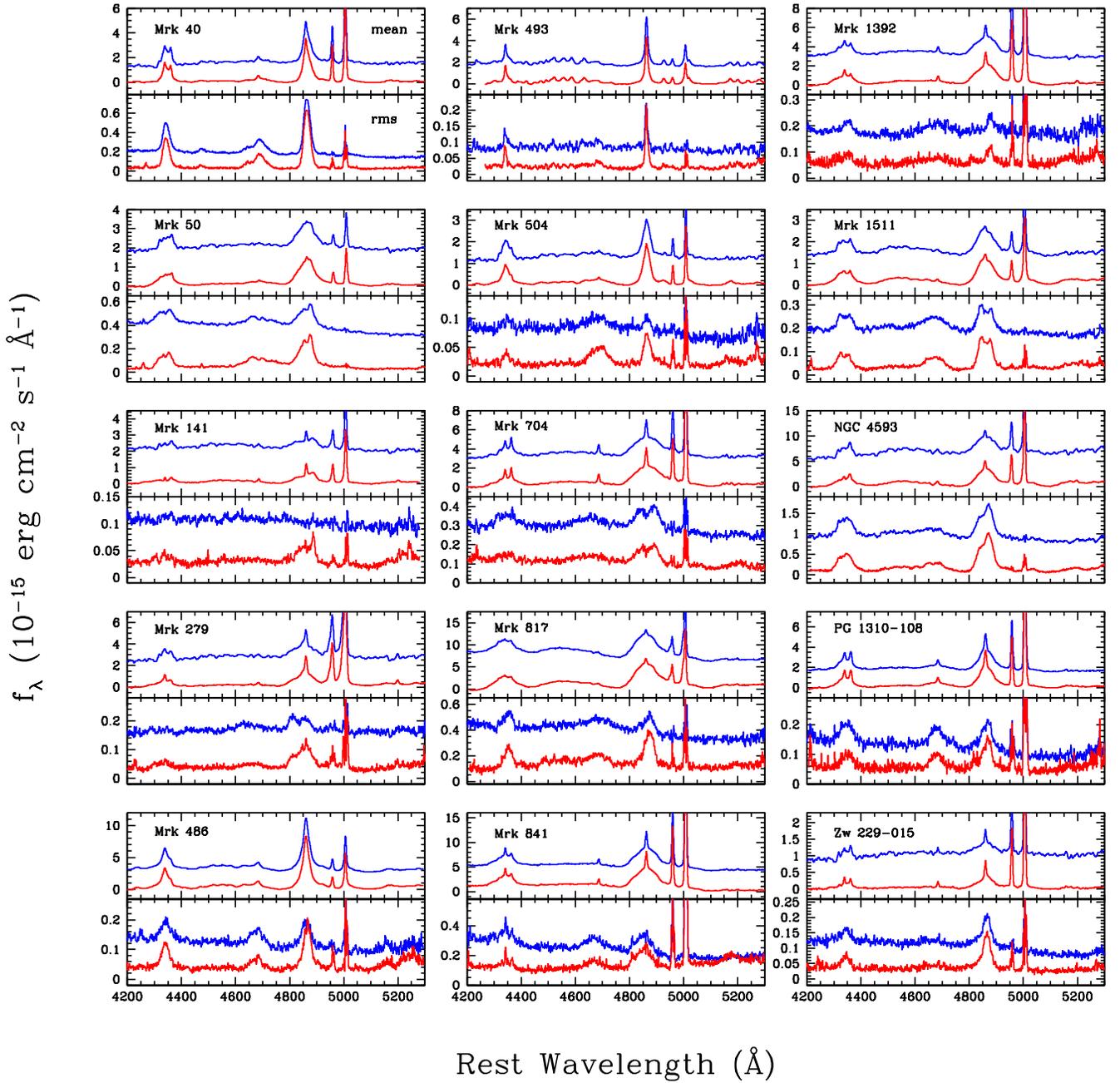}}
\caption{Mean and rms spectra for the blue-side fitting region. In
  each panel, the upper subpanel shows shows the standard mean
  spectrum (blue), and a ``line-only'' mean spectrum constructed after
  subtraction of the AGN featureless continuum and starlight
  components from each nightly spectrum (red).  The lower subpanel
  shows the standard rms spectrum (blue), and an rms spectrum
  constructed after subtraction of the AGN featureless continuum and
  starlight components from each nightly spectrum (red). }
\label{meanrmsspectra}
\end{figure*}

\begin{figure}
\scalebox{0.43}{\includegraphics{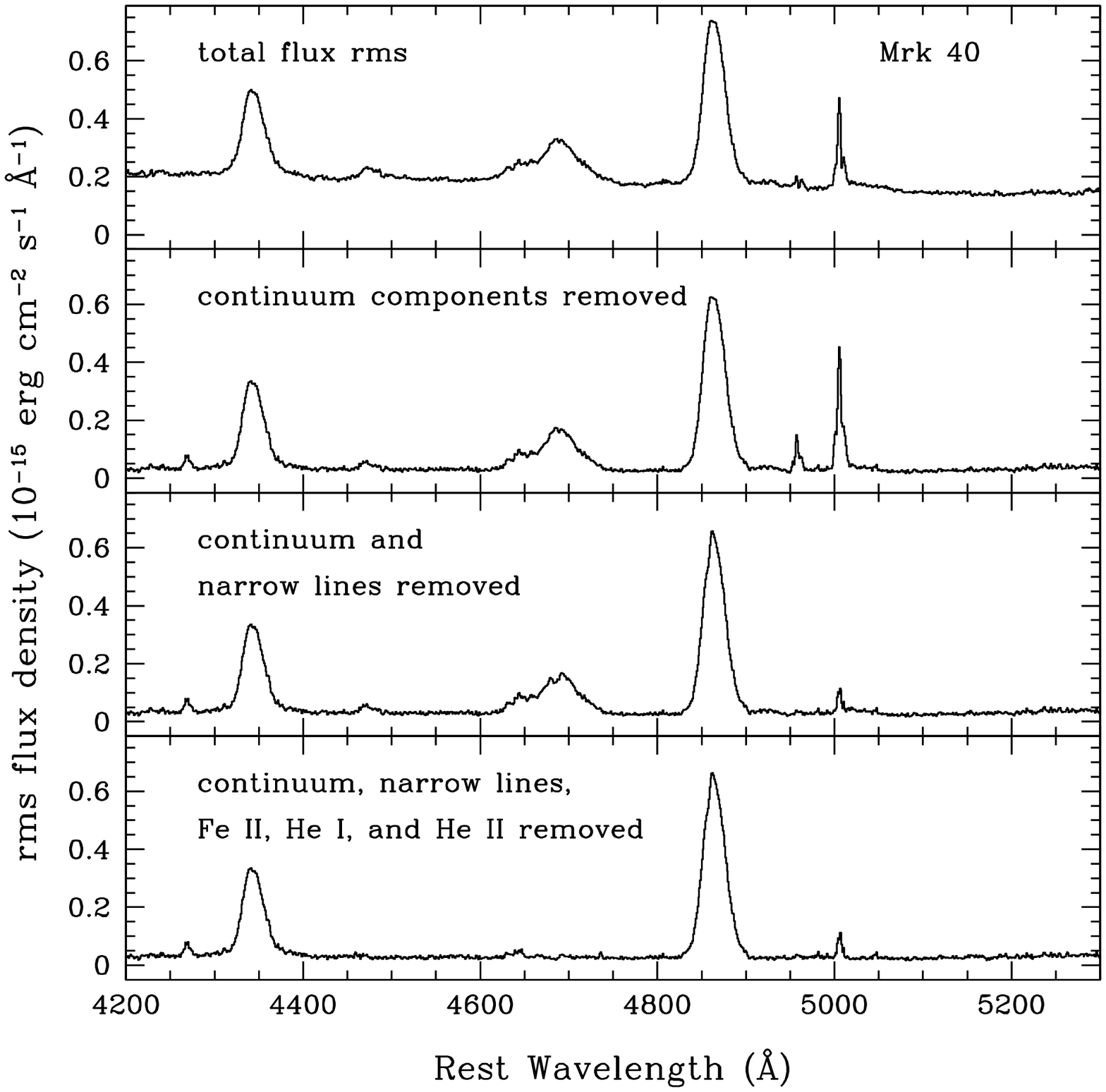}}
\caption{Different versions of the rms spectrum for Mrk 40.  The top
  panel shows the standard rms spectrum constructed using the nightly
  total flux spectra.  The second panel shows the rms after removing
  the AGN continuum and host-galaxy starlight components from each
  nightly spectrum.  In the third panel, narrow emission lines have
  also been removed ([\ion{O}{3}] $\lambda\lambda4959,5007$, \hbeta,
  and \ion{He}{2}), and in the bottom panel, the broad \ion{He}{2} and
  \ion{Fe}{2} components are also removed from the nightly spectra.
  This last version of the rms spectrum, with all model components
  removed other than broad \hbeta, is the one used to measure the
  width of broad \hbeta\ in the rms spectrum. }
\label{mrk40rmscomparison}
\end{figure}

The fitting procedure returns estimates of uncertainties on each of
the free parameters. However, for measurements of emission-line light
curves, what is needed is an estimate of the uncertainty on the flux
density at each pixel for a particular spectral component such as
\hbeta.  We employed a Monte Carlo procedure to estimate the flux
uncertainties.  For each spectrum, after finding the best model fit,
we created 100 modified versions of the spectrum by adjusting each
pixel by a random Gaussian deviate drawn from a distribution with a
Gaussian dispersion equal to the amplitude of the propagated error
spectrum at that pixel.  The spectral fit was then repeated for each
of the 100 randomly deviated spectra, and the individual fit
components were saved.  For each spectral component, the value of the
error spectrum at a given pixel was taken to be the standard deviation
of the flux values from the 100 re-fitted versions of that spectral
component.  Similarly, we produced error spectra for the decomposed
\hbeta\ profiles by taking the standard deviation of the 100 Monte
Carlo realizations of the decomposed profile (i.e., the deviated
spectrum minus all best-fit components other than \hbeta).

From this procedure, we found that the error spectrum of the
decomposed \hbeta\ profile has an amplitude at each spectral pixel
that is nearly equal to that of the original error spectrum, but
noisier as a result of the limited number of Monte Carlo iterations.
Based on the results of this procedure, we adopt the original error
spectrum as our best estimate of the pixel uncertainties on the
individual emission-line component spectra that we use to measure
light curves. The error spectra do not incorporate possible systematic
uncertainties due to differences between \ion{Fe}{2} templates or the
choice of a particular host-galaxy starlight model.

Using the fit results, we constructed revised mean and rms spectra.
Figure \ref{meanrmsspectra} shows the original mean and rms spectra
along with mean and rms spectra constructed after subtraction of the
best-fitting starlight and AGN featureless continuum components from
each nightly spectrum. In some cases such as Mrk 279, the
\hbeta\ profile in the rms spectrum changes significantly after
subtraction of the continuum components; similar results have
previously been discussed by \citet{park2012a}.  Additionally, for
some AGNs the \ion{He}{2} profiles in rms spectra are significantly
altered by removal of the continuum components; these changes are
particularly prominent in Mrk 504 and Mrk 1511.  Another version of
the rms spectrum was constructed by subtracting all model components
other than broad \hbeta\ from the nightly spectra. This version
isolates the variability of the broad \hbeta\ line itself, and was
used to measure the width of the \hbeta\ rms profile as described
in Section \ref{sec:linewidths}. Figure \ref{mrk40rmscomparison}
shows multiple versions of the rms spectrum for Mrk 40 to illustrate
the changes when different spectral components are subtracted prior to
calculating the rms flux.  One advantage of subtracting narrow
emission-line components prior to constructing the rms spectrum is
that the residual noise from these intrinsically nonvariable features
is significantly reduced.  This is clearly apparent for the
[\ion{O}{3}] lines in Mrk 40 (Figure \ref{mrk40rmscomparison}).

In future work, a variety of possible extensions and improvements to
the fitting method can be explored.  These include (a) addition of
model components for other emission lines such as \hgamma\ and
[\ion{O}{3}] $\lambda4363$ instead of simply masking them out from the
fit; (b) broadening the wavelength range of the fit blueward down to
the near-UV atmospheric cutoff, and redward to \hal; (c) inclusion of
multiple-age stellar populations; (d) addition of a Balmer continuum
component for fits extending below the Balmer break; (e) use of
higher-order models to achieve more accurate fits to strong lines such
as [\ion{O}{3}] $\lambda5007$ and for irregular or double-peaked broad
Balmer lines; (f) exploration of different methods for constraining
the decomposition in the \hbeta\ red shelf region; (g) allowing for
the possibility that different reddening values or even different
reddening laws might apply to spectral components originating at
different spatial scales.  The possible benefits of these improvements
must however be weighed against the increased complexity that comes
with adding more free parameters and the degeneracies among them,
particularly for the continuum components.  Furthermore, since the
primary goal is often to achieve the most accurate fit to the
wavelength region surrounding \hbeta, care must be taken to ensure
that fitting a broader wavelength range does not worsen the fit in
this region. The available set of \ion{Fe}{2} templates remains a
fundamental limitation and source of systematic uncertainty in the
fits, and further work to develop more flexible sets of \ion{Fe}{2}
templates, both theoretical and emipirical, covering the full
UV/optical wavelength range, would be extremely valuable for future
reverberation mapping.

\section{Emission-Line and Continuum Light Curves}
\label{sec:lightcurves}

Light curves for broad emission lines were measured by direct
integration of spectra.  The epoch of each observation is given by the
Heliocentric Julian Date (HJD) determined from the UT date saved in
the FITS image headers, using the IRAF \texttt{setjd} routine. To
measure the \hbeta\ and \hgamma\ light curves for each night's
observation of an AGN, we began with the observed spectrum and
subtracted all of the best-fitting model components except for the
broad and narrow \hbeta\ models.  This left a residual spectrum with a
continuum level of zero (modulated by the fitting residuals) and the
\hbeta\ line and \hgamma+[\ion{O}{3}] blends as the remaining
emission features.  The flux in these features was integrated over
wavelength ranges that were selected for each AGN to encompass the
full width of the line profiles.  Wavelength ranges for each light
curve are listed in Table \ref{lcstats}.  We measure the total (broad
+ narrow) flux for \hbeta\ in order to avoid introducing possible
errors into the light curves due to degeneracy in decomposing the two
components, consistent with methods typically used for reverberation
measurements \citep[e.g.,][]{kaspi2000,bentz2009lamp}.  The narrow
component adds a constant flux pedestal to the light curves.  The
\hgamma\ measurement includes the flux of the broad and narrow
\hgamma\ lines as well as [\ion{O}{3}] $\lambda4363$.

We also measured light curves for \ion{He}{2} $\lambda4686$ and for
the \ion{Fe}{2} blends integrated over the rest wavelength range
4400--4700 \AA.  The light curves for these weak features were
measured from the best-fitting model component spectra for each
night's data, and for \ion{He}{2} the light curves include both the
broad and narrow components.  The \ion{Fe}{2} emission was integrated
over the rest wavelength ranges of 4400--4700 and 5100--5400 \AA,
avoiding regions in which \ion{Fe}{2} emission is degenerate with
\hbeta\ or the possible \ion{He}{1} emission in the \hbeta\ red shelf.

For the \hal\ and \hdelta\ lines, which lie outside the wavelength
range of the spectral fits, we measured fluxes from the scaled spectra
using the traditional approach of fitting a local linear continuum to
surrounding line-free regions on either side of the line.  The
\hal\ light curves include the fluxes of the narrow [\ion{N}{2}]
$\lambda6548,6583$ lines.

\begin{figure}
\scalebox{0.4}{\includegraphics{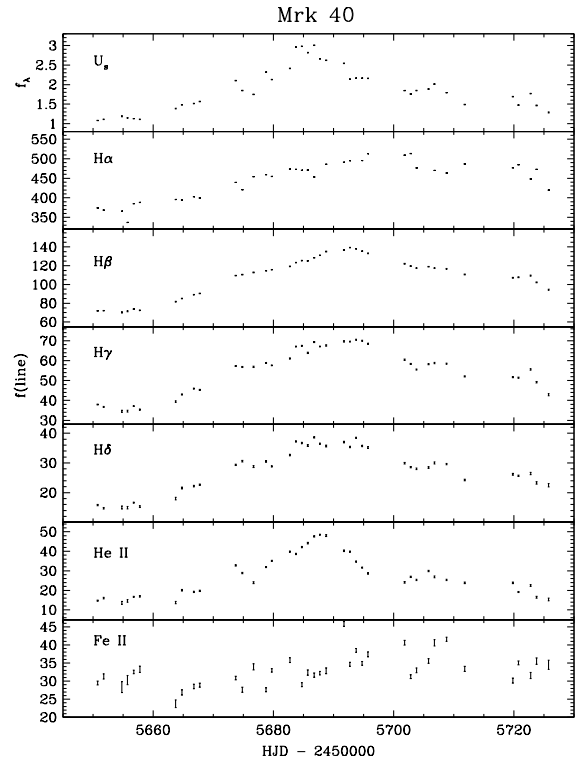}}
\caption{Light curves of Mrk 40.  The $y$-axis units for this and
  subsequent light curve plots are $10^{-15}$ erg
  cm\persq\ s\per\ \AA\per\ for the $U_s$-band continuum flux density
  and $10^{-15}$ erg cm\persq\ s\per\ for the emission-line fluxes.
  Plotted error bars represent the propagated uncertainties from the
  spectral extractions only, as listed in Table \ref{lcdata}, and do
  not incorporate the estimated residual flux normalization scatter as
  determined from the [\ion{O}{3}] emission line.}
\label{mrk40lightcurves}
\end{figure}

\begin{figure}
\scalebox{0.4}{\includegraphics{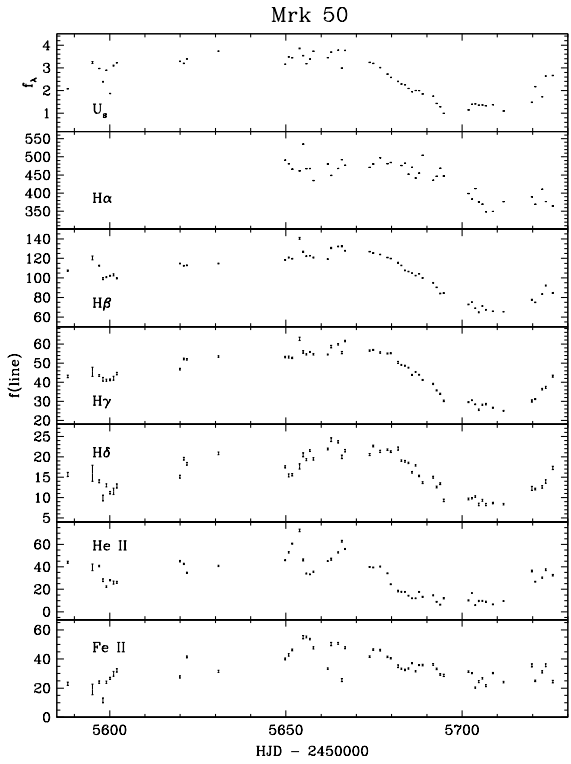}}
\caption{Light curves of Mrk 50, including observations taken during
  January--March before the start of the main campaign. }
\label{mrk50lightcurves}
\end{figure}

\begin{figure}
\scalebox{0.4}{\includegraphics{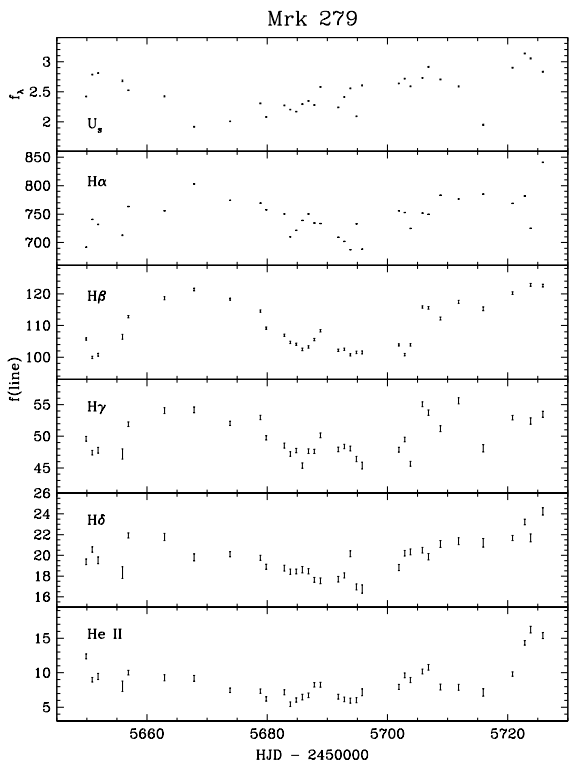}}
\caption{Light curves of Mrk 279. }
\label{mrk279lightcurves}
\end{figure}

\begin{figure}
\scalebox{0.4}{\includegraphics{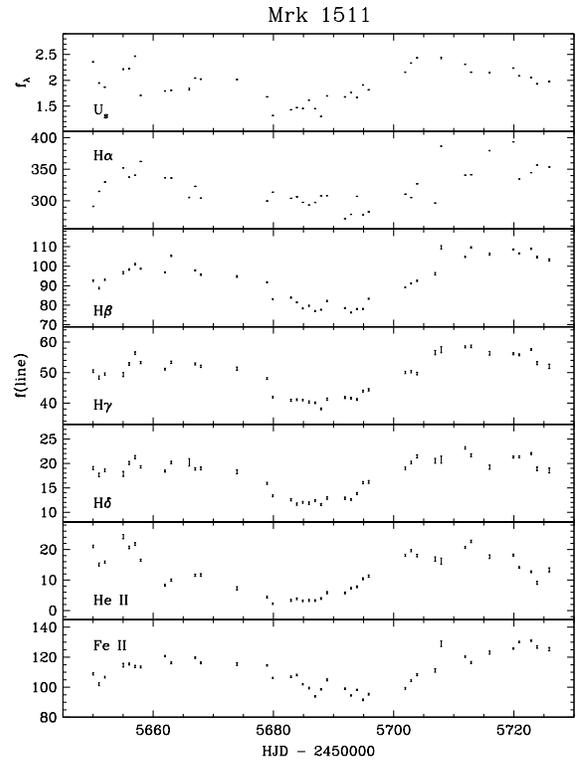}}
\caption{Light curves of Mrk 1511. }
\label{mrk1511lightcurves}
\end{figure}

\begin{figure}
\scalebox{0.4}{\includegraphics{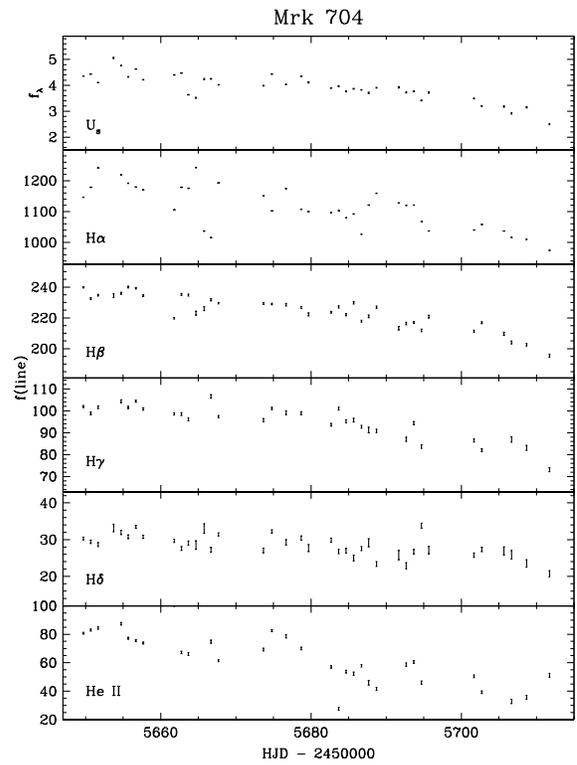}}
\caption{Light curves of Mrk 704. }
\label{mrk704lightcurves}
\end{figure}

\begin{figure}
\scalebox{0.4}{\includegraphics{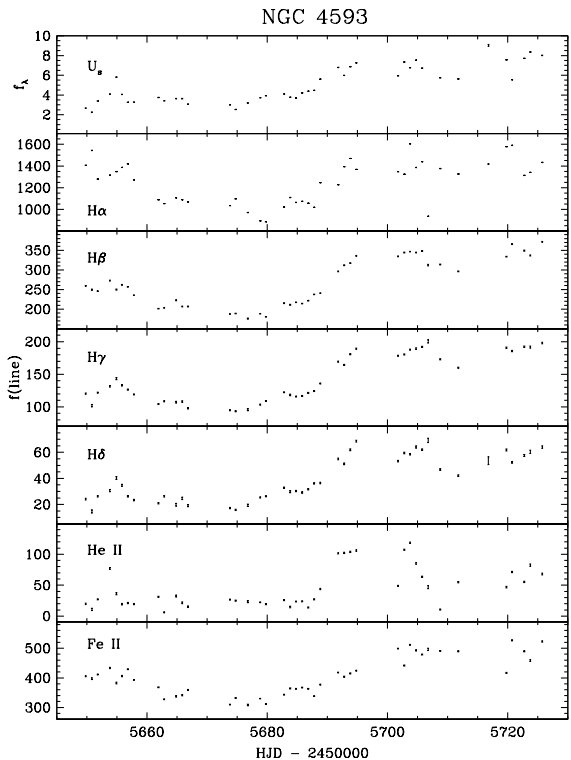}}
\caption{Light curves of NGC 4593. }
\label{ngc4593lightcurves}
\end{figure}

\begin{figure}
\scalebox{0.4}{\includegraphics{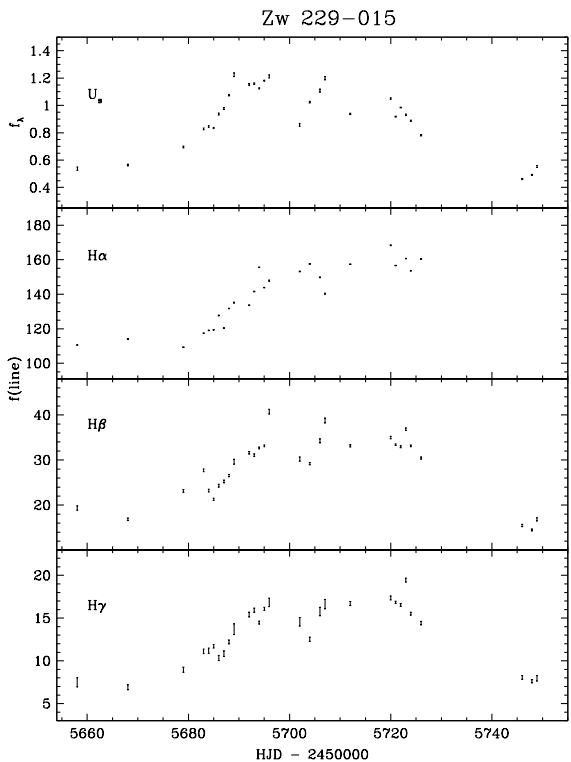}}
\caption{Light curves of Zw 229-015, including the three data points
  taken after the end of the main campaign. }
\label{zw229lightcurves}
\end{figure}

\begin{figure*}[t!]
\begin{center}
\scalebox{0.75}{\includegraphics{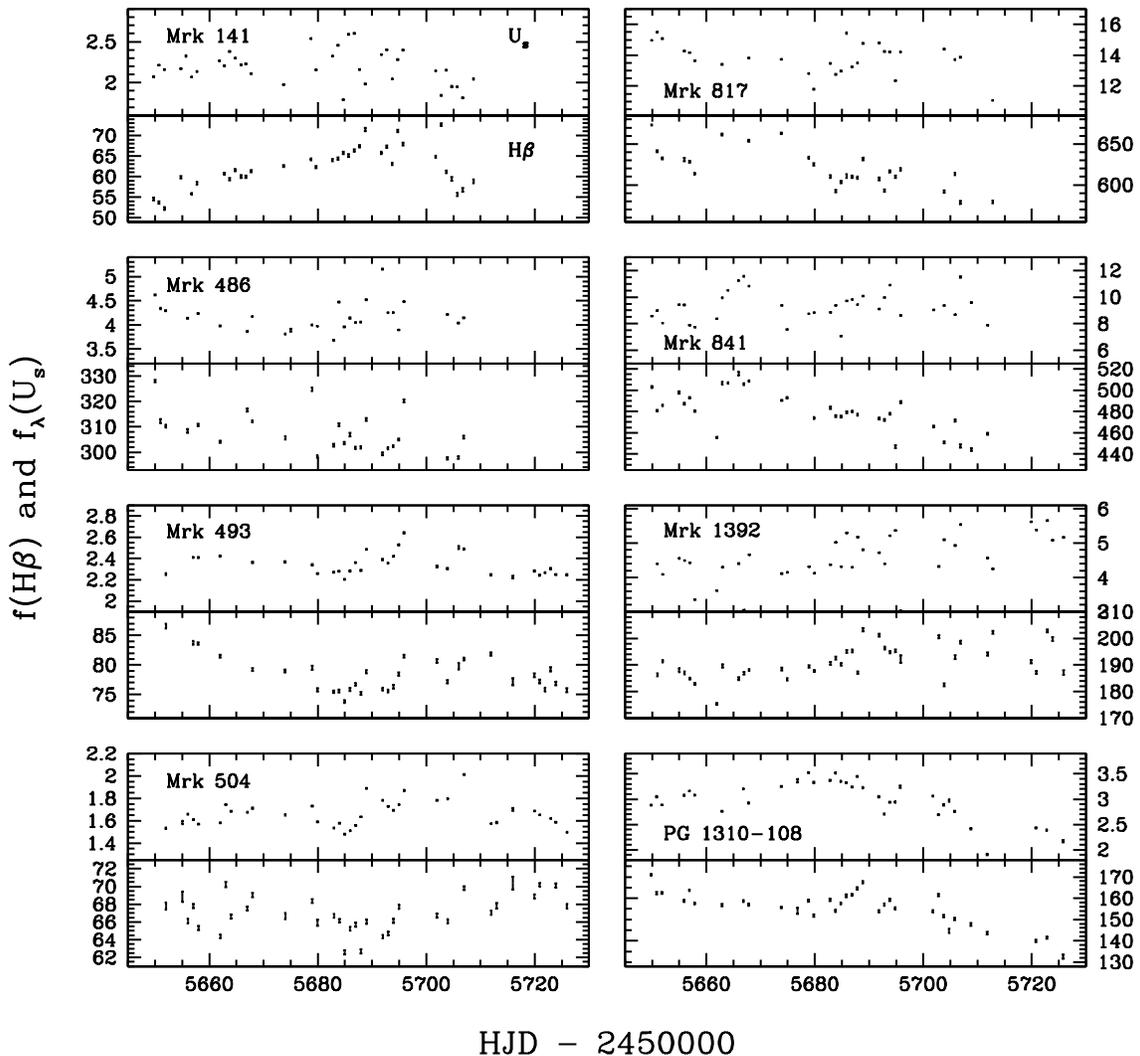}}
\end{center}
\caption{Light curves of the remaining objects in
  the sample.  In each panel, the upper and lower subpanels show the
  $U_s$ and \hbeta\ light curves, respectively.}
\label{otherlightcurves}
\end{figure*}

Light curves are displayed in Figures
\ref{mrk40lightcurves}--\ref{otherlightcurves}, and the light curve
data are listed in Table \ref{lcdata}.

Light curves were also measured for the AGN continuum in the $U$-band
spectral region, since the AGN featureless continuum tends to dominate
over starlight in this region and is more highly variable than at
redder wavelengths.  The mean continuum flux density was measured over
$\lambda_\mathrm{rest} = $ 3500--3600 \AA, and we refer to this
spectroscopic continuum flux as the $U_s$ band.  For a few objects,
this wavelength range was adjusted in order to avoid emission
lines. This measurement includes both AGN and starlight continuum.
Table \ref{lcstats} lists the wavelength range that was integrated for
each light curve.  Since this wavelength range is far from the
wavelength of [\ion{O}{3}], which was used as the flux calibration
reference for spectral scaling, the relative accuracy of the
photometric calibration is sometimes poor.  From the appearance of the
$U_s$ light curves, the flux uncertainties in portions of the light
curves are $\sim10\%$ or occasionally worse.  The scatter in the $U_s$
light curves for some objects (such as Mrk 50) is highest at times
close to the start and end of the campaign, when targets were observed
at higher airmasses, but this trend does not appear to hold for the
entire sample.

\begin{deluxetable*}{lccccc}
\tablecaption{Light Curve Wavelength Ranges and Variability Statistics}
\tablehead{
  \colhead{Galaxy} &
  \colhead{Light Curve} &
  \colhead{Rest Wavelengths (\AA)} &
  \colhead{\fvar} & 
  \colhead{\maxratio} 
}
\startdata
Mrk 40      & $U_s$       & 3500--3600 & $0.30\pm0.03$ & $2.78\pm0.04$ \\
            & \hal\       & 6493--6660 & $0.14\pm0.02$ & $2.02\pm0.02$ \\
            & \hbeta\     & 4799--4916 & $0.20\pm0.02$ & $1.98\pm0.03$ \\
            & \hgamma\    & 4309--4387 & $0.21\pm0.02$ & $2.04\pm0.04$ \\
            & \hdelta\    & 4074--4142 & $0.27\pm0.03$ & $2.61\pm0.06$ \\
            & \ion{He}{2} & 4583--4789 & $0.38\pm0.04$ & $3.57\pm0.21$ \\
            & \ion{Fe}{2} & 4400--4700,5100--5400 & $0.15\pm0.02$ & $2.05\pm0.10$ \\
Mrk 50      & $U_s$       & 3500--3600 & $0.34\pm0.03$ & $3.88\pm0.34$ \\
            & \hal\       & 6449--6693 & $0.10\pm0.01$ & $1.53\pm0.08$ \\
            & \hbeta\     & 4758--4944 & $0.20\pm0.02$ & $2.17\pm0.13$ \\
            & \hgamma\    & 4270--4417 & $0.23\pm0.02$ & $2.50\pm0.17$ \\
            & \hdelta\    & 4065--4133 & $0.29\pm0.03$ & $2.93\pm0.23$ \\
            & \ion{He}{2} & 4495--4817 & $0.55\pm0.05$ & $12.02\pm2.09$ \\
            & \ion{Fe}{2} & 4400--4700,5100--5400 & $0.35\pm0.03$ & $7.28\pm1.31$ \\
Mrk 141     & $U_s$       & 3500--3600 & $0.09\pm0.01$ & $1.45\pm0.03$ \\
            & \hbeta\     & 4782--4925 & $0.08\pm0.01$ & $1.39\pm0.03$ \\
Mrk 279     & $U_s$       & 3500--3600 & $0.13\pm0.02$ & $1.63\pm0.02$ \\
            & \hal\       & 6405--6667 & $0.04\pm0.01$ & $1.22\pm0.01$ \\
            & \hbeta\     & 4755--4920 & $0.07\pm0.01$ & $1.31\pm0.01$ \\
            & \hgamma\    & 4251--4416 & $0.07\pm0.01$ & $1.34\pm0.02$ \\
            & \hdelta\    & 4047--4115 & $0.09\pm0.01$ & $1.45\pm0.04$ \\
            & \ion{He}{2} & 4590--4765 & $0.30\pm0.04$ & $2.98\pm0.21$ \\
Mrk 486     & $U_s$       & 3500--3600 & $0.07\pm0.01$ & $1.40\pm0.04$ \\
            & \hbeta\     & 4784--4928 & $0.02\pm0.01$ & $1.10\pm0.03$ \\
Mrk 493     & $U_s$       & 3500--3600 & $0.04\pm0.01$ & $1.20\pm0.04$ \\
            & \hbeta\     & 4809--4897 & $0.03\pm0.01$ & $1.17\pm0.04$ \\
Mrk 504     & $U_s$       & 3500--3600 & $0.07\pm0.01$ & $1.36\pm0.03$ \\
            & \hbeta\     & 4798--4914 & $0.04\pm0.01$ & $1.24\pm0.03$ \\
Mrk 704     & $U_s$       & 3500--3570 & $0.13\pm0.02$ & $2.02\pm0.03$ \\
            & \hal\       & 6422--6704 & $0.07\pm0.01$ & $1.37\pm0.01$ \\
            & \hbeta\     & 4770--4936 & $0.05\pm0.01$ & $1.23\pm0.01$ \\
            & \hgamma\    & 4256--4411 & $0.08\pm0.01$ & $1.46\pm0.02$ \\
            & \hdelta\    & 4052--4149 & $0.10\pm0.01$ & $1.63\pm0.08$ \\
            & \heii\      & 4489--4848 & $0.28\pm0.04$ & $3.62\pm0.13$ \\
Mrk 817     & $U_s$       & 3500--3600 & $0.08\pm0.01$ & $1.40\pm0.03$ \\
            & \hbeta\     & 4750--4944 & $0.04\pm0.01$ & $1.16\pm0.02$ \\
Mrk 841     & $U_s$       & 3500--3570 & $0.14\pm0.02$ & $1.92\pm0.03$ \\
            & \hbeta\     & 4757--4940 & $0.05\pm0.01$ & $1.27\pm0.02$ \\
Mrk 1392    & $U_s$       & 3500--3600 & $0.14\pm0.02$ & $1.88\pm0.03$ \\
            & \hbeta\     & 4758--4932 & $0.03\pm0.01$ & $1.16\pm0.01$ \\
Mrk 1511    & $U_s$       & 3500--3600 & $0.19\pm0.02$ & $2.42\pm0.06$ \\
            & \hal\       & 6432--6693 & $0.09\pm0.01$ & $1.45\pm0.03$ \\
            & \hbeta\     & 4768--4933 & $0.12\pm0.01$ & $1.43\pm0.03$ \\
            & \hgamma\    & 4236--4430 & $0.13\pm0.01$ & $1.54\pm0.04$ \\
            & \hdelta\    & 4062--4140 & $0.20\pm0.02$ & $2.00\pm0.07$ \\
            & \heii\      & 4556--4797 & $0.54\pm0.06$ & $10.80\pm1.59$ \\
            & \feii\      & 4400--4700,5100--5400 & $0.10\pm0.01$ & $1.43\pm0.03$ \\
NGC 4593    & $U_s$       & 3500--3570 & $0.38\pm0.04$ & $4.54\pm0.25$ \\
            & \hal\       & 6442--6690 & $0.16\pm0.02$ & $1.81\pm0.07$ \\
            & \hbeta\     & 4757--4926 & $0.23\pm0.02$ & $2.11\pm0.08$ \\
            & \hgamma\    & 4242--4410 & $0.25\pm0.03$ & $2.15\pm0.09$ \\
            & \hdelta\    & 4058--4133 & $0.43\pm0.05$ & $4.72\pm0.50$ \\
            & \ion{He}{2} & 4539--4807 & $0.72\pm0.08$ & $19.28\pm4.41$ \\
            & \ion{Fe}{2} & 4400--4700,5100--5400 & $0.16\pm0.02$ & $1.71\pm0.06$ \\
PG 1310-108 & $U_s$       & 3500--3600 & $0.13\pm0.02$ & $1.85\pm0.03$ \\
            & \hbeta\     & 4787--4932 & $0.05\pm0.01$ & $1.29\pm0.02$ \\
Zw 229-015  & $U_s$       & 3500--3570 & $0.25\pm0.03$ & $2.65\pm0.06$ \\
            & \hal\       & 6470--6664 & $0.13\pm0.02$ & $1.54\pm0.02$ \\
            & \hbeta\     & 4767--4912 & $0.25\pm0.03$ & $2.81\pm0.07$ \\
            & \hgamma\    & 4281--4388 & $0.27\pm0.04$ & $2.82\pm0.13$ \\
\enddata
\tablecomments{These measurements include the constant narrow-line
  component contributions to the Balmer-line and \ion{He}{2} light
  curves, and the constant host-galaxy contribution to the $U_s$ light
  curve.  The wavelength column lists the range over which flux
  density was integrated for the light curve measurement.  As
  described in the text, the \hbeta, \hgamma, \heii, and \feii\ light
  curves were measured from individual components of the decomposed
  blue-side data, while $U_s$, \hal, and \hdelta\ were measured from the
  total-flux spectra.
}
\label{lcstats}
\end{deluxetable*}

\begin{deluxetable}{lcccc}
\tablecaption{Light Curve Data}
\tablehead{
  \colhead{Galaxy} &
  \colhead{Light Curve} &
  \colhead{HJD} &
  \colhead{$f$} & 
  \colhead{$\delta_f$} 
}
\startdata
Mrk 40 & $U_s$ &  5650.75  &    1.081  &  0.004 \\
       &       &  5651.75  &    1.112  &  0.006 \\
       &       &  5654.80  &    1.191  &  0.013 \\
       &       &  5655.73  &    1.146  &  0.010 \\
       &       &  5656.76  &    1.127  &  0.004 \\
       &       &  5657.77  &    1.111  &  0.007 \\
       &       &  5663.74  &    1.386  &  0.009 \\
       &       &  5664.78  &    1.481  &  0.006 \\
       &       &  5666.76  &    1.515  &  0.005 \\
       &       &  5667.75  &    1.567  &  0.005 \\
       &       &  5673.73  &    2.101  &  0.006 \\
       &       &  5674.82  &    1.848  &  0.006 \\
\enddata
\tablecomments{Dates are listed as HJD -- 2450000.  Units are
  $10^{-15}$ erg cm\persq\ s\per\ \AA\per\ for $U_s$, and $10^{-15}$
  erg cm\persq\ s\per\ for emission-line light curves.  The listed
  uncertainties $\delta_f$ correspond to the propagated uncertainties
  from photon counting and detector readout noise, and do not include
  the additional contribution from residual flux-scaling errors. This
  Table is published in its entirety in the electronic edition of
  \emph{The Astrophysical Journal Supplement Series}.  A
  portion is shown here for guidance regarding its form and content. }
\label{lcdata}
\end{deluxetable}

From each light curve, we calculated the statistics \fvar\ and
\maxratio\ to characterize the variability amplitude, and for
comparison with previous reverberation campaigns
\citep[e.g.,][]{bentz2009lamp}.  The quantity \fvar\ is similar to the
normalized excess scatter and is defined as
\begin{equation}
\fvar = \frac{\sqrt{{\sigma_f^2} -
    \delta_\mathrm{rms}^2}}{\langle{f}\rangle},
\end{equation}
where $\sigma_f^2$ is the variance of the flux values in the light
curve, $\delta_\mathrm{rms}$ is the rms measurement uncertainty on the
fluxes, and $\langle f \rangle$ is the mean flux
\citep{rodriguezpascual1997, edelson2002}.  \maxratio\ gives the
maximum variability amplitude and is simply the ratio of maximum to
minimum fluxes in the light curve.  To account for the contribution of
residual scatter from the spectral scaling to the error budget, we
modified the flux uncertainties by taking the product of excess
[\ion{O}{3}] flux scatter with mean flux ($\sigma_{nx} \times \langle
f \rangle$) and adding that quantity in quadrature to each individual
flux uncertainty value prior to calculating \fvar, \maxratio, and
their uncertainties.  As an overall measure of light-curve
variability, \fvar\ is preferable to \maxratio\ because \maxratio\ is
particularly sensitive to outlier values at the highest and lowest
fluxes.

Emission-line light curves having $\fvar>0.1$ correspond to strong
variability from which lags can generally be measured well.  In our
sample, the objects exceeding this threshold for \hbeta\ variations
are Mrk 40, Mrk 50, Mrk 1511, NGC 4593, and Zw 229-015.  For these
AGNs, we plot light curves of the $U_s$ flux density, the Balmer
lines, and also \ion{He}{2} and \ion{Fe}{2} in those cases where there
is some distinct variability above the level of the noise for these
components.  We also plot multi-line light curves for Mrk 279 and Mrk
704. Mrk 279 has a lower \hbeta\ variability amplitude of $\fvar =
0.07$, but its light curve exhibits distinct up-and-down variations
which provide useful diagnostics of the emission-line lags.  Mrk 704
exhibits a downward trend in the continuum and emission-line light
curves, but the nearly monotonic trend contains little short-timescale
structure and is not optimal for determination of lags. For the
remaining objects in the sample, all of which have
$\fvar$(\hbeta)$<0.1$, light curves are shown only for $U_s$ and
\hbeta\ in Figure \ref{otherlightcurves}.  Aside from \hbeta, the
light curves for other emission lines in these objects are too noisy
to be of much use.  A few of these AGNs, such as Mrk 141, Mrk 493, and
PG 1310--108, show distinct features in their light curves which may
provide some limited information on the \hbeta\ lag.  Table
\ref{lcstats} lists the light-curve statistics \fvar\ and
\maxratio\ for each light curve displayed in Figures
\ref{mrk40lightcurves}--\ref{otherlightcurves}.  For completeness, we
list statistics measured for the $U_s$ continuum light curves for each
AGN, including the low-variability objects shown in Figure
\ref{otherlightcurves}.

Table \ref{oiiitable} also presents measurements of $f_\lambda$(5100
\AA), the AGN continuum flux density at 5100 \AA\ in the AGN rest
frame.  This quantity is most often used to determine AGN luminosities
for use in the radius-luminosity relationship
\citep[e.g.,][]{bentz2009rl}.  This wavelength corresponds to a local
minimum in \ion{Fe}{2} flux, making it a good choice for measurement
of the featureless continuum level in total-flux spectra. We used the
AGN power-law component of the fit to the mean spectrum to measure
$f_\lambda$(5100 \AA) for each AGN.  The formal fitting uncertainty on
the power-law component flux for the mean spectrum is typically very
small (a few percent) and not a good measure of the actual
uncertainty.  From examination of the fitting results using all three
\ion{Fe}{2} templates, we find that the rms scatter among the
power-law component fluxes at 5100 \AA\ for the three fits ranges from
4\% to 13\% of the mean value.  For uniformity, we adopt an estimated
10\% uncertainty on the 5100 \AA\ featureless continuum flux at this
wavelength resulting from differences among the \ion{Fe}{2} templates.
We also combine with this (by addition in quadrature) an estimate of
the overall photometric uncertainty on the continuum flux density
given by $\delta$([\ion{O}{3}])/$f$([\ion{O}{3}])$\times
f_\lambda$(5100~\AA), where the [\ion{O}{3}] flux and uncertainty
values are listed in Table \ref{oiiitable}.  Improved values of
$f_\lambda$(5100 \AA) can be determined using \hst\ $V$-band images to
remove the host-galaxy contribution more accurately from the total
spectroscopic flux density.\footnote{New images of Mrk 50, Mrk 704,
  Mrk 1511, and Zw 229-015 will be obtained with WFC3 in the F547M
  filter as part of \hst\ Cycle 22 program GO-13816 (PI: Bentz) along
  with 10 other recently reverberation-mapped AGNs compiled from other
  programs.}

For the strongly variable AGNs, we generally find higher values of
\fvar\ for the higher-order Balmer lines: $\fvar($\hdelta$) >
\fvar($\hgamma$) > \fvar($\hbeta$) > \fvar($\hal$)$.  This overall
trend is consistent with results from previous work
\citep[e.g.,][]{bentz2010a}, and is in accord with expectations from
photoionization modeling \citep{korista2004}. The \ion{He}{2}
$\lambda4686$ line is by far the most strongly variable line in the
optical spectrum, following the trend first seen in NGC 5548 by
\citet{petersonferland1986}.  For the six objects having \ion{He}{2}
light curves, the ratio $\fvar($\ion{He}{2}$)/\fvar($\hbeta$)$ has a
mean value of 3.7.  The strong responsivity and prompt response of
\ion{He}{2} is easily visible in the light curves: for example, in Mrk
40, the \ion{He}{2} emission drops very steeply following the decline
in continuum luminosity at about HJD 5685, while the decline in the
Balmer-line fluxes is much shallower and more gradual (Figure
\ref{mrk40lightcurves}).

While it is possible to measure reverberation lags based on the data
presented here for the highly variable AGNs in our sample, we find
that for some objects the cross-correlation measurements show evidence
for an excess and spurious signal at zero lag, caused by correlated
flux-calibration errors in the continuum and emission-line light
curves.  Thus, we defer the measurement of lags to a later paper in
this series, in which our emission-line light curves will be combined
with $V$-band light curves which will be presented separately.  The
$V$-band light curves generally have a higher cadence and higher S/N
than the $U_s$ light curves.  Determination of virial masses for the
black holes in these AGNs will be done using lags measured against the
$V$-band light curves.

\section{Broad \hbeta\ Widths and Velocity Shifts}
\label{sec:linewidths}

\subsection{Instrumental Resolution}

To quantify the broad-line widths, it is necessary to correct
measurements for the instrumental broadening in observations with the
4\arcsec-wide slit.  The spectral resolution for observations of the
intrinsically compact BLR as observed through a 4\arcsec\ slit will be
seeing-dependent, and will vary slightly from night to night due to
variations in spectrograph focus.  However, the impact of instrumental
broadening on the broad emission lines in these AGNs is very small,
and we use a single representative value for the instrumental
broadening for simplicity.  For the blue-side data we follow the same
method used by \citet{bentz2009lamp} and \citet{barth:zw229} to
estimate the instrumental FHWM by using observations of an AGN and an
arc lamp in 2\arcsec\ and 4\arcsec\ slit widths.  Mrk 40 was observed
on the night of 2011 May 4 in both of these slit widths.  For the
spectral region around \hbeta, we use the \ion{Cd}{1} $\lambda5085$
\AA\ line as the nearest reference calibration line.  From the
2\arcsec\ observations of the AGN and the comparison lamp, we subtract
the width of the \ion{Cd}{1} line in quadrature from the observed
[\ion{O}{3}] width to derive an estimate of the intrinsic [\ion{O}{3}]
line width of FWHM$\approx4.9$ \AA.  Then, the instrumental broadening
for the 4\arcsec\ slit is taken to be the difference in quadrature
between the [\ion{O}{3}] line width as observed through the
4\arcsec\ slit (FWHM = 8.1 \AA) and the intrinsic [\ion{O}{3}] width.
This gives an instrumental FWHM of 6.45 \AA\ or 380 \kms, and an
instrumental dispersion of $\sigma_\mathrm{inst} \approx 162$ \kms.

For the red-side data, there is no strong narrow line that can be used
as a reference for comparison between the 2\arcsec\ and
4\arcsec\ observations, and instead we use the widths of comparison
lamp lines as observed through the 4\arcsec\ slit as a rough estimate
of the instrumental broadening, obtaining an instrumental FWHM of 7.7
\AA\ or $\sim330$ \kms\ for the region around the \hal\ line.  These
instrumental widths are much smaller than the broad emission-line
widths in the AGNs, and the corrections to the observed line widths
are very small.

\subsection{Broad \hbeta\ Widths in Mean and RMS Spectra}

The width of the broad \hbeta\ line is a primary quantity used for
determination of BH masses from reverberation data, and it is of
interest to measure both the FWHM and the dispersion (or second
moment) of the \hbeta\ line profile in both the mean and the rms
spectra when possible. We use the results of our spectral
decompositions to measure these width parameters for the broad
\hbeta\ line after isolating it from the contributions of other
emission-line and continuum components.  For each nightly spectrum, we
subtract all model components other than broad \hbeta\ from the data,
leaving the broad \hbeta\ profile as the residual.  From this time
series of residuals, we construct mean and rms spectra, and measure
widths empirically from the mean and rms line profiles as described
below.  The mean spectrum has a continuum level of zero (modulated by
residual noise from the spectral decompositions), while the continuum
level in the rms spectrum is nonzero due to photon-counting noise and
residual errors in the nightly spectral decompositions.  This noise
continuum level in the rms spectra is generally flat and featureless
outside the \hbeta\ profile (as seen in the bottom panel of Figure
\ref{mrk40rmscomparison}) and can be subtracted off with a simple
linear fit to regions on either side of the line.  For each AGN we
selected broad continuum windows redward and blueward of the
\hbeta\ line to fit and subtracted this continuum from the rms
spectrum, avoiding any residual spikes at the wavelengths of
[\ion{O}{3}] or other strong features.

Line width parameters were measured following methods similar to those
of \citet{peterson2004}. The profile FWHM was measured by starting at
the profile peak and moving redward until reaching a flux equal to
half the peak value (using linear interpolation between pixels on the
profile shoulder to determine the wavelength at half maximum), and
similarly moving blueward until reaching half of the peak flux, and
determining the separation in {\AA}ngstroms between the two half-peak
points. To measure the line dispersion \sigmaline\ for either the mean
or rms profile, we first determine the line centroid wavelength
$\lambda_0$:
\begin{equation}
\lambda_0 = \frac{\sum{\lambda_i f_{i}}}{\sum{f_{i}}},
\end{equation}
where $f_{i}$ is the flux density at pixel $i$ and the summation is
carried out over all pixels in the profile.  Then, the line dispersion
is calculated according to
\begin{equation}
\sigmaline^2 = \left( \frac{\sum \lambda_i^2 f_{i}}{\sum
  f_{i}} \right) - \lambda_0^2.
\end{equation}
The endpoints of the summation region are determined visually for each
AGN by identifying the location where the line profile meets the
surrounding continuum level. 

In order to assess the uncertainty in the measured width parameters,
we carry out a Monte Carlo procedure, following \citet{peterson2004}
and \citet{bentz2009lamp}, by creating $10^4$ modified realizations of
the mean and rms profiles and measuring the width parameters for each
one. Each realization is constructed from the time series of spectra
by randomly selecting $N$ individual nightly spectra (where $N$ is the
total number of spectra in the time series), allowing duplications,
calculating the mean and rms from this revised set of spectra, and
measuring the broad \hbeta\ widths from each. For measurements of the
rms profile, we allow the endpoints of the continuum fitting region in
each realization to vary randomly within a range of $\pm10$ \AA\ from
our initial selections, and for measurements of line dispersions from
either the mean or rms profiles, we allow the endpoints of the
summation region in each realization to vary randomly within $\pm10$
\AA\ from our initial selections.  The final line width and
uncertainty are taken to be the mean and standard deviation of the
distribution of measurements over these $10^4$ iterations.

Finally, the line widths are converted to velocity units, and the
instrumental FWHM or $\sigma$ are subtracted in quadrature from the
measured FWHM or $\sigma$ values (respectively) to correct for
instrumental broadening.  The results of these measurements are
reported in Table \ref{broadhbetawidths}.  Width parameters for the
rms profile are only reported for the five AGNs having
$\fvar($\hbeta$)>0.1$ and for Mrk 279.  For the remainder of the
sample, the rms spectra contain very little flux as a result of the
overall low variability amplitude, and the Monte Carlo error analysis
procedure produces many iterations in which the FWHM or
\sigmaline\ are not well-defined.  Widths measured from the rms
profiles for these low-variability AGNs have large uncertainties and
would not be particularly useful for deriving estimates of black hole
masses, despite the fact that some of these low-variability AGNs
appear to show clear signal at \hbeta\ in the rms spectra (Figure
\ref{meanrmsspectra}).

An important caveat for the line width measurements, particularly for
the line dispersions, is that the red wing of \hbeta\ is often
blended with a substantial contribution of \ion{Fe}{2} and possibly
\ion{He}{1} as well.  As described above, the exact amount of this
contamination is difficult to determine, and the endpoint of the red
wing of the line is usually not well determined.  Since the line
dispersion values are highly sensitive to the amount of flux in the
line wings, broad \hbeta\ line dispersions are particularly
susceptible to systematic errors resulting from this deblending.  We
note that in most past reverberation-mapping work, no deblending was
attempted at all, potentially leading to overestimates of \hbeta\ line
dispersions.  The FWHM may be modestly affected by this red-wing
contamination as well.

\begin{deluxetable*}{lcccc}
  \tablecaption{Broad \hbeta\ Component Widths in Mean and RMS Spectra}
  \tablehead{
    \colhead{} &
    \multicolumn{2}{c}{Mean Spectrum} & 
    \multicolumn{2}{c}{RMS Spectrum} \\
    \colhead{Galaxy} &
    \colhead{FWHM} &
    \colhead{\sigmaline} &
    \colhead{FWHM} & 
    \colhead{\sigmaline} \\
    \colhead{} &
    \colhead{(\kms)} &
    \colhead{(\kms)} &
    \colhead{(\kms)} &
    \colhead{(\kms)} 
  }
\startdata
Mrk 40      & $2021\pm17$ & $1058\pm9$  & $1688\pm26$  & $740\pm17$ \\
Mrk 50      & $4101\pm56$ & $2024\pm31$ & $3355\pm128$ & $2020\pm103$ \\
Mrk 141     & $5129\pm45$ & $2280\pm21$ & \nodata  & \nodata \\
Mrk 279     & $4099\pm43$ & $1821\pm13$ & $3306\pm338$ & $1778\pm71$ \\
Mrk 486     & $2106\pm8$  & $1332\pm14$ & \nodata & \nodata \\
Mrk 493     & $1717\pm25$ & $1128\pm41$ & \nodata  & \nodata \\
Mrk 504     & $1927\pm7$  & $1058\pm14$ & \nodata & \nodata \\
Mrk 704     & $6664\pm66$ & $2911\pm61$ & \nodata & \nodata \\
Mrk 817     & $5593\pm32$ & $2718\pm45$ & \nodata & \nodata \\
Mrk 841     & $6065\pm34$ & $2794\pm48$ & \nodata & \nodata \\
Mrk 1392    & $5421\pm31$ & $2103\pm17$  & \nodata & \nodata \\
Mrk 1511    & $4154\pm28$ & $1828\pm12$ & $3236\pm65$ & $1506\pm42$ \\
NGC 4593    & $4264\pm41$ & $1925\pm38$ & $3597\pm72$ & $1601\pm40$ \\
PG 1310-108 & $3422\pm21$ & $1823\pm20$ & \nodata & \nodata \\
Zw 229-015  & $3705\pm203$ & $1747\pm56$ & $1789\pm93$ & $1609\pm109$ \\ 
\enddata
\tablecomments{Widths are measured from the mean and rms of the broad
  \hbeta\ component spectra, as described in the text. }
\label{broadhbetawidths}
\end{deluxetable*}

\subsection{Broad \hbeta\ Width and Velocity Centroid Variations}
\label{sec:widthsandcentroids}

Width parameters (FWHM and \sigmaline) were also measured from each
nightly spectrum for the high-variability objects in order to search
for time-dependent changes in profile width, following the same
methods described above.  These widths were measured from the spectra
of the broad \hbeta\ component alone, constructed by subtracting all
other model components (including the narrow \hbeta\ component) from
each total flux spectrum.  To assess the uncertainty in FWHM and
\sigmaline, we created 1000 modified realizations of each spectrum by
adding Gaussian random noise with the noise dispersion equal to the
amplitude of the error spectrum at each wavelength, and then measuring
the FWHM and \sigmaline\ from each modified spectrum. The
uncertainties on FWHM and \sigmaline\ were taken to be the standard
deviations of the distributions of values measured from this
procedure.  In general, the line widths show an anticorrelation with
flux; \S\ref{sec:widthluminosity} discusses this trend further.

Using the same spectra of the broad \hbeta\ component, we measured the
broad-line velocity centroid shift relative to the broad
\hbeta\ component of the mean spectrum, using a cross-correlation
method based on the techniques described by \citet{shen2013bbh} and
\citet{ju2013} in the context of searching for binary black hole
orbital motion in quasars.  The wavelength centroid of the broad
\hbeta\ component is a parameter determined by our spectral fitting
procedure, but the fitted line centroid can be strongly influenced by
asymmetric wings on the line profile, particularly in the \hbeta\ red
shelf. In contrast, the cross-correlation line shift as measured by
\citet{shen2013bbh} and \cite{ju2013} is a better measure of an overall
wavelength shift of the line as a whole, and is less sensitive to
noise in the high-velocity wings.  The velocity shifts are discussed
in \S\ref{sec:velocityshifts}.

\begin{figure}
\scalebox{0.4}{\includegraphics{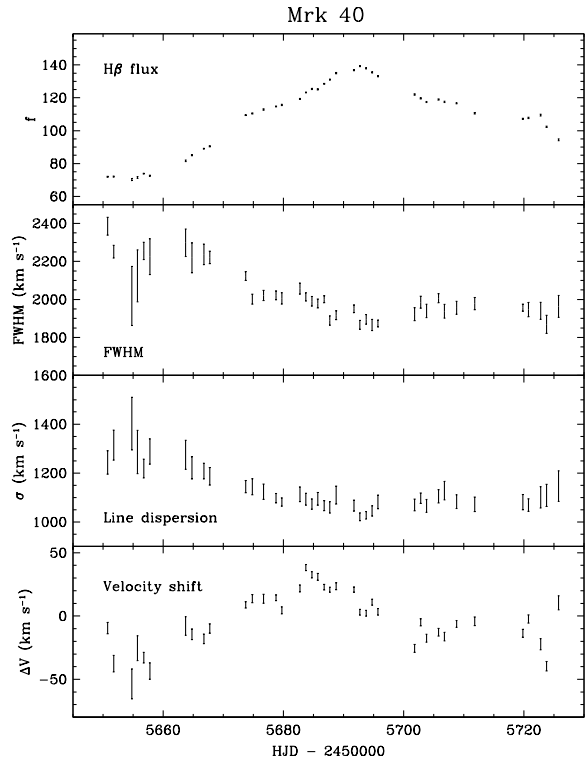}}
\caption{Broad \hbeta\ parameters as a function of time for Mrk 40.
  The top panel shows the same \hbeta\ light curve as in Figure
  \ref{mrk40lightcurves}, in units of $10^{-15}$ erg cm\persq\ s\per.
  Middle panels show the broad \hbeta\ FWHM and line dispersion
  \sigmaline, and the bottom panel shows the velocity shift $\Delta V$
  measured relative to the mean spectrum. }
\label{mrk40hbetaparams}
\end{figure}

\begin{figure}
\scalebox{0.4}{\includegraphics{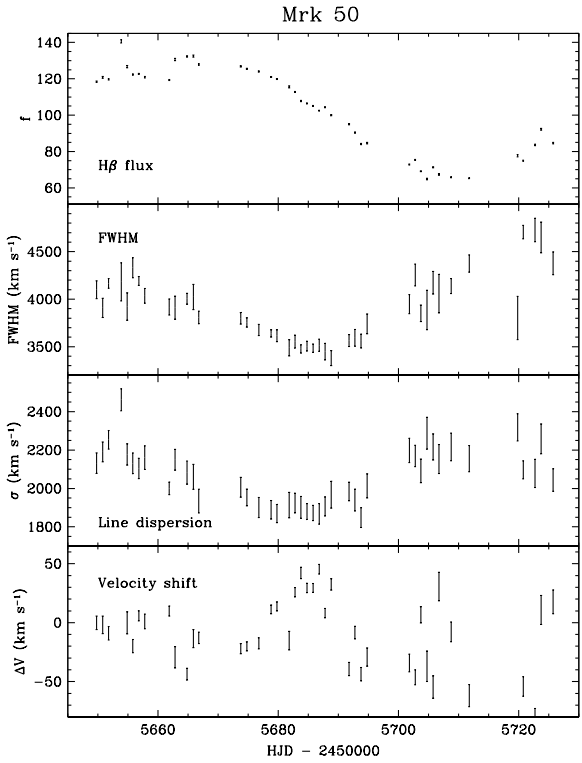}}
\caption{Broad \hbeta\ parameters for Mrk 50. Only the time span of
  the main campaign is shown, because the data points from earlier
  observations are very noisy. }
\label{mrk50hbetaparams}
\end{figure}

\begin{figure}
\scalebox{0.4}{\includegraphics{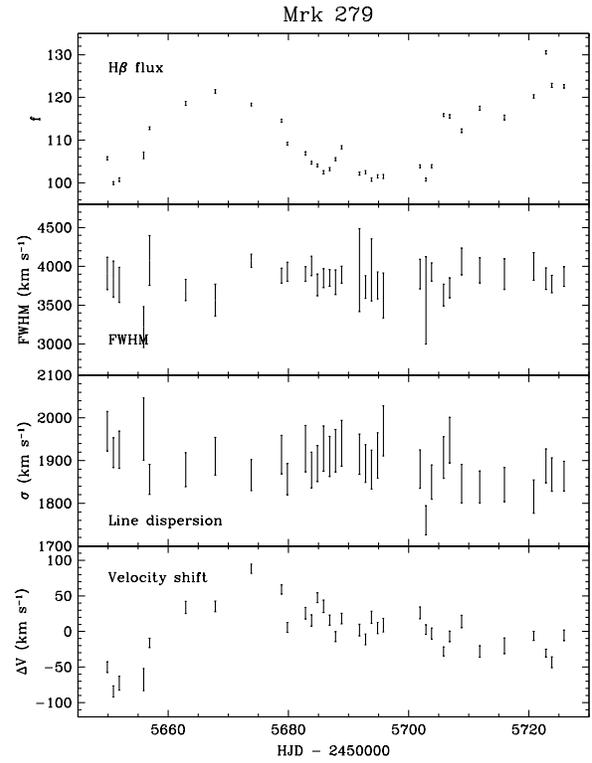}}
\caption{Broad \hbeta\ parameters for Mrk 279.  The nearly constant
  \hbeta\ width for this AGN is an artifact of the constraint that had
  to be imposed in order to prevent the line width from reaching
  unphysically large values in the fits. }
\label{mrk279hbetaparams}
\end{figure}

\begin{figure}
\scalebox{0.4}{\includegraphics{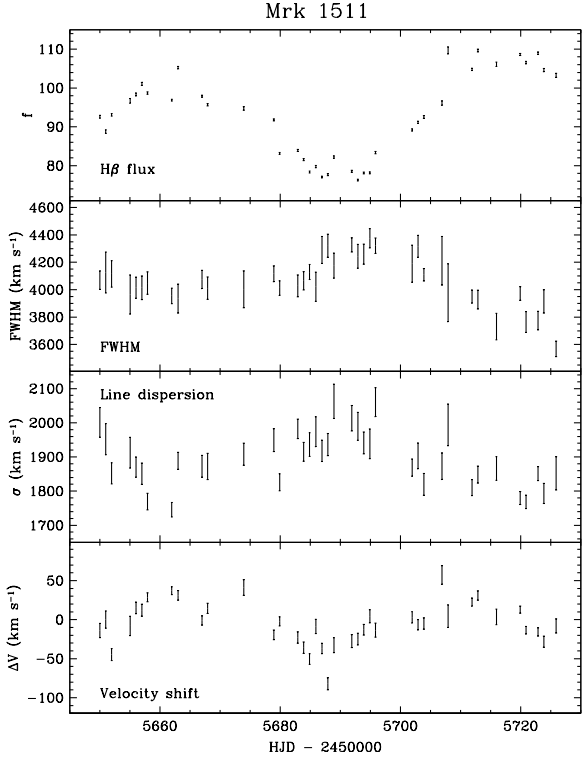}}
\caption{Broad \hbeta\ parameters for Mrk 1511. }
\label{mrk1511hbetaparams}
\end{figure}

\begin{figure}
\scalebox{0.4}{\includegraphics{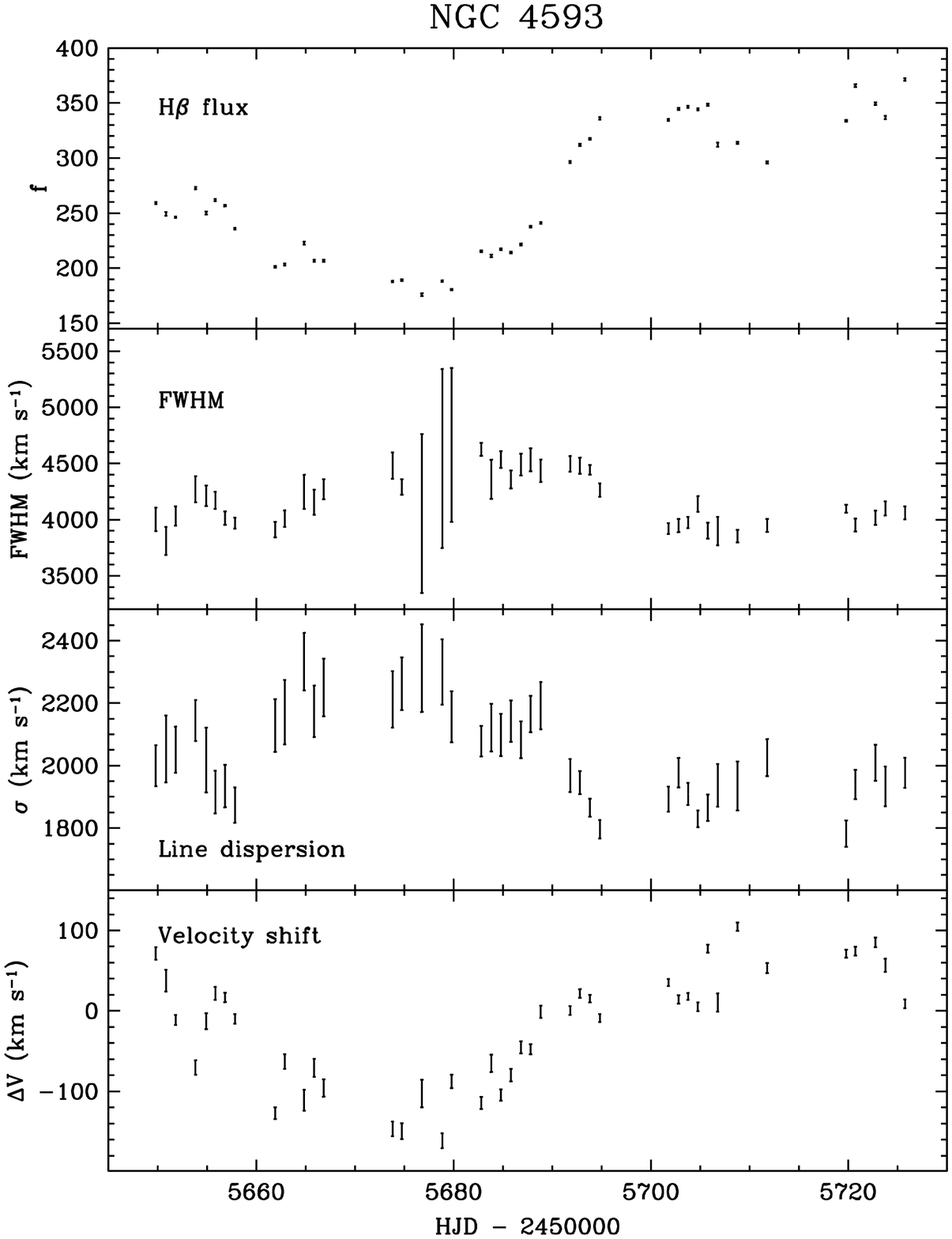}}
\caption{Broad \hbeta\ parameters for NGC 4593. }
\label{ngc4593hbetaparams}
\end{figure}

The nightly width and velocity shift parameters are displayed in
Figures \ref{mrk40hbetaparams}--\ref{ngc4593hbetaparams} for Mrk 40,
Mrk 50, Mrk 279, Mrk 1511, and NGC 4593, all of which show distinct
time-dependent trends in line profile width and/or velocity.  For the
remainder of the sample, no clear signal was detected beyond random
scatter in the measurements.

\section{Relationships between Broad-Line Width and Luminosity}
\label{sec:widthluminosity}

\begin{figure}
\scalebox{0.4}{\includegraphics{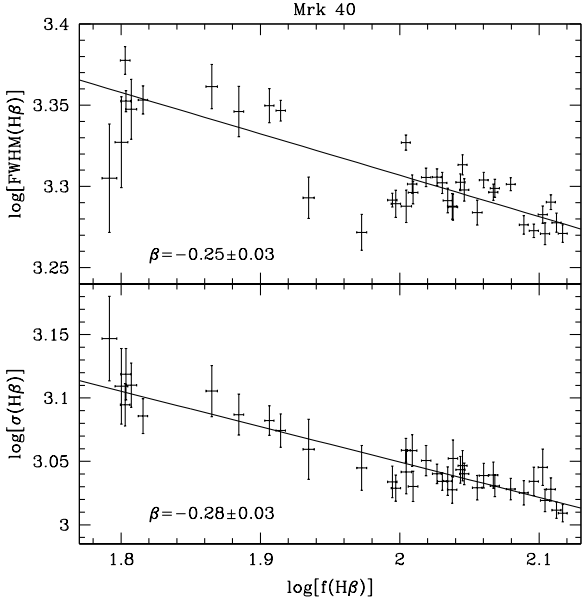}}
\caption{Relationship between broad \hbeta\ width and flux for Mrk 40.
  In each panel, the solid line shows the best fit to the relationship
  with slope $\beta$.  Line widths (FWHM and \sigmaline) are in
  \kms, and flux units are $10^{-15}$ erg cm\persq\ s\per.}
\label{mrk40virialfig}
\end{figure}

\begin{figure}
\scalebox{0.4}{\includegraphics{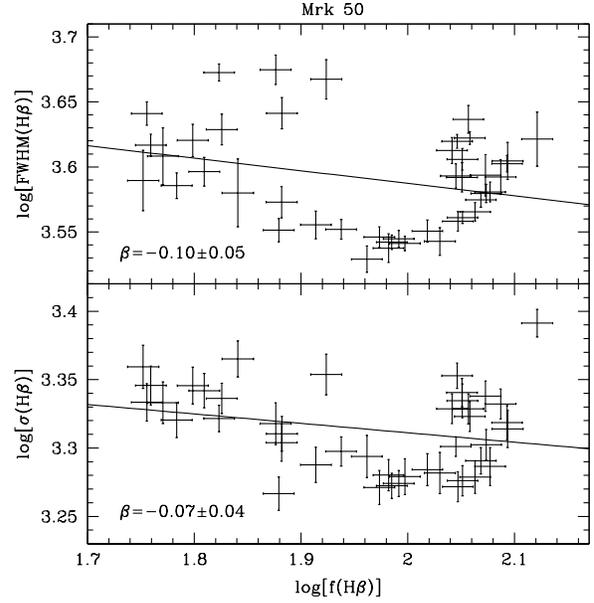}}
\caption{Relationship between broad \hbeta\ width and flux
  for Mrk 50.  Only data points from the main campaign are included.}
\label{mrk50virialfig}
\end{figure}

\begin{figure}
\scalebox{0.4}{\includegraphics{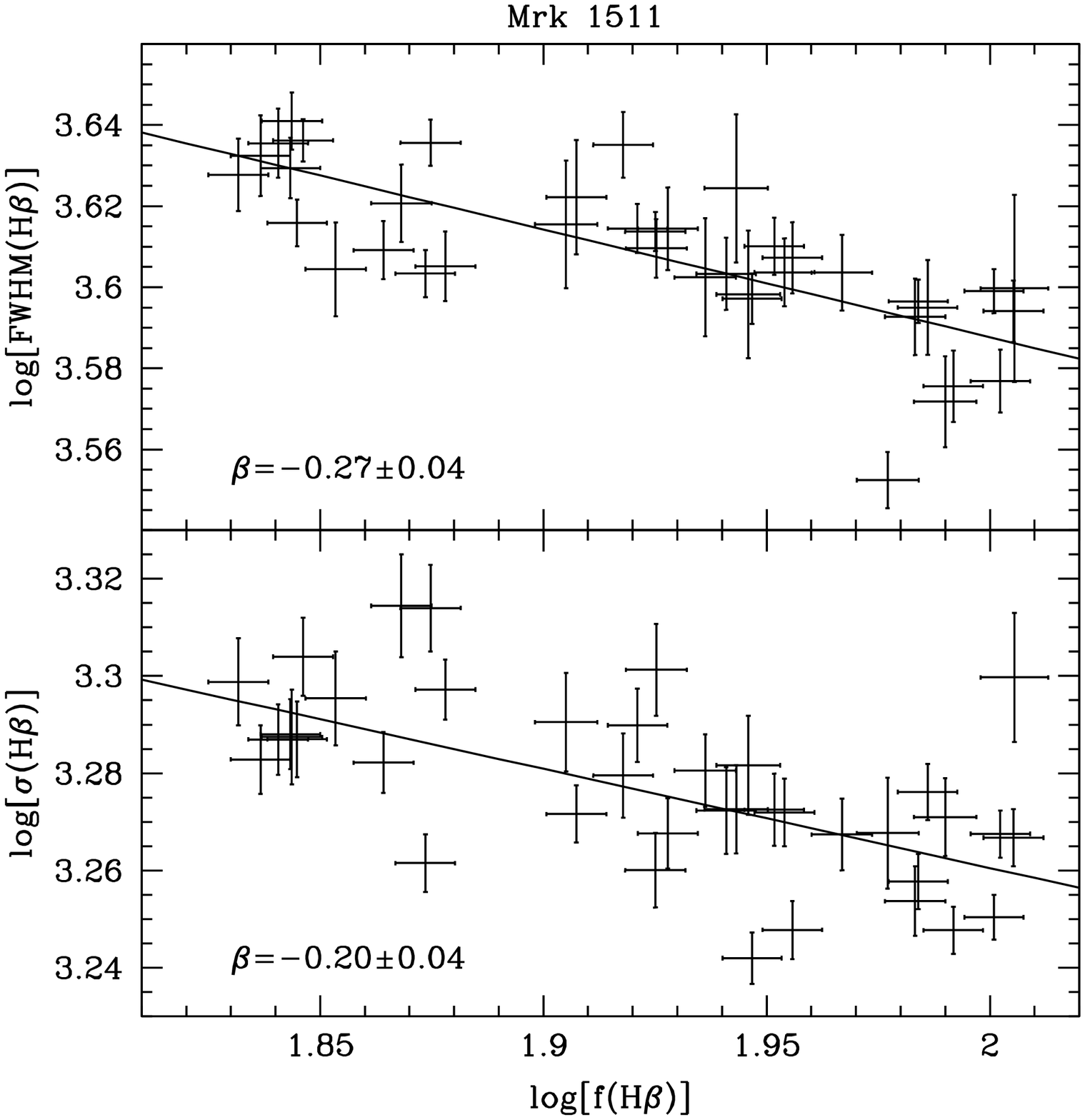}}
\caption{Relationship between broad \hbeta\ width and flux
  for Mrk 1511.}
\label{mrk1511virialfig}
\end{figure}

\begin{figure}
\scalebox{0.4}{\includegraphics{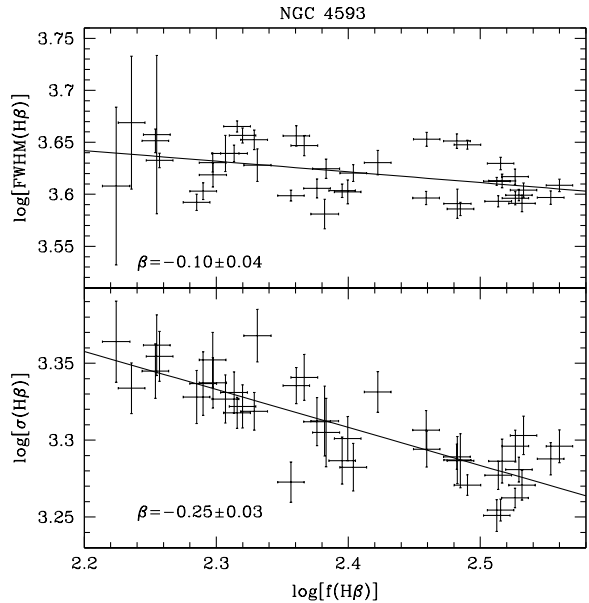}}
\caption{Relationship between broad \hbeta\ width and flux
  for NGC 4593. }
\label{ngc4593virialfig}
\end{figure}

Figures \ref{mrk40virialfig}--\ref{ngc4593virialfig} display the
relationship between broad \hbeta\ width (FWHM or \sigmaline) and
broad \hbeta\ flux.  The narrow \hbeta\ component flux from the mean
spectrum decomposition has been subtracted from the total
\hbeta\ light curve in each case, to obtain a light curve of the broad
\hbeta\ line alone.  The flux uncertainties have been expanded by
including the estimated uncertainty in spectral scaling determined
from the [\ion{O}{3}] excess scatter, using the results listed in
Table \ref{oiiitable}.  In the case of Mrk 50 (Figure
\ref{mrk50virialfig}) we include only data points from the main
observing campaign, since the observations taken prior to the start of
the campaign are much noisier.  Mrk 279 is excluded since the broad
\hbeta\ width could not be measured accurately, as described
previously.

Although there is substantial scatter in some of the plots, in each
case we find an overall anticorrelation between the broad
\hbeta\ width and flux.  To quantify this relationship, we use the
Bayesian \texttt{linmix\_err} regression method of \citet{kelly2007},
fitting a model of the form $\log[\Delta V($\hbeta$)] = \beta
\log[f($\hbeta$)] + \alpha$, where $\Delta V$ represents either FWHM
or \sigmaline.  The slopes determined by the \texttt{linmix\_err}
regression are shown in each figure and range from $\beta=-0.07$ to
$-0.28$.

Similar trends between broad \hbeta\ width and luminosity have been
seen previously in high-quality reverberation mapping datasets.
Examples include NGC 5548 \citep{denney2009se} and Mrk 40 \citep[Arp
  151;][]{park2012a}.  The anticorrelation between broad \hbeta\ width
and luminosity arises as a result of the relationship between the
ionizing flux and the local responsivity (or reprocessing efficiency)
of BLR gas: the \hbeta\ responsivity is highest in the outer portions
of the BLR where incident continuum fluxes are lower
\citep{korista2004, goadkorista2014}.  An increase in ionizing
continuum luminosity will therefore lead to an increase in the
emissivity-weighted BLR radius; this effect is known as ``breathing''.
For a BLR having a radial velocity gradient with the highest-velocity
gas at small radii close to the black hole, breathing will
preferentially boost the flux of the low-velocity line core relative
to the high-velocity wings of the broad line, making the line profile
narrower.

BLR breathing has usually been considered in the context of large
luminosity changes over sufficiently long timescales that
emission-line lags can be seen to vary between widely separated epochs
of monitoring \citep[e.g.,][]{obrien1995,korista2004,cackett2006,
  bentz2007}.  The \hbeta\ width--luminosity anticorrelation seen in
our data \citep[and in recent work by][]{park2012a} further
illustrates that BLR breathing occurs even on the shortest observed
timescales (days to weeks) in response to modest variations in the AGN
continuum.  While our data do not directly reveal short-timescale
changes in lag or BLR radius, the relationship between \hbeta\ width
and luminosity implies that the BLR radius responds very rapidly to
continuum flux changes.  A further consequence of this finding is that
the reverberation transfer function will be nonstationary even on
timescales shorter than a single typical monitoring campaign.
Measured transfer functions \citep{bentz2010b, grier2013a} will thus
represent temporally smeared averages of intrinsically time-varying
functions.

Since broad-line flux variations are driven by continuum variations,
it is also of interest to examine the relationship between broad-line
width and continuum flux.  However, the time delay between continuum
and line variations introduces substantial additional scatter to the
relationship, and in our data, the broad \hbeta\ widths show a much
clearer relationship with \hbeta\ flux than with continuum flux.  The
temporal mismatch between line width and continuum flux measurements
can be partially alleviated by selecting continuum and emission-line
measurements that are separated by the measured emission-line lag time
\citep{park2012a}. Unfortunately, the frequent weather-related gaps in
our light curves do not leave enough matched pairs of data points to
obtain clear results. Another possibility would be to interpolate the
data to obtain matched pairs of line widths and continuum fluxes after
correcting for the broad-line lag time, but we will defer that
investigation in order to use the higher-cadence $V$-band photometric
light curves instead of the spectroscopic continuum measurements
presented here.

The behavior of broad-line widths in response to luminosity variations
has important ramifications for the accuracy of single-epoch
spectroscopic determinations of black hole masses. These are derived
by using the AGN continuum luminosity as a proxy for the BLR radius
and computing a virial product based on VP $= L^{\gamma} (\Delta V)^2
/G$, where $\gamma$ represents the slope of the radius-luminosity
relationship.  Recent work finds $\rblr \propto
L_{5100}^{0.53\pm0.04}$ for the full sample of reverberation-mapped
AGNs, where $L_{5100}$ is the AGN continuum luminosity at
$\lambda=5100$ \AA\ \citep{bentz2013}.  The luminosity of the broad
Balmer emission lines can also be used as a proxy for continuum
luminosity \citep{greeneho2005}.  The accuracy of these single-epoch
methods for mass estimation relies implicitly on the assumption that
line width and luminosity must be anticorrelated in a given AGN, such
that the derived masses will not change dramatically as the luminosity
varies.  Our results provide additional reassuring evidence that this
is generally the case; see \citet{denney2009se} and \citet{park2012a}
for detailed discussion of the impact of variability on single-epoch
mass determinations.  If $\rblr \propto L_\mathrm{5100}^{0.5}$ and the
\hbeta\ luminosity scales linearly with continuum luminosity, then an
AGN maintaining a constant virial product would exhibit a line
width-luminosity relationship having slope $\beta=-0.25$, similar to
the slopes observed in Mrk 40, Mrk 1511, and NGC 4593.  However, it is
important to note that this naive prediction for the slope of the
\hbeta\ width-luminosity relationship is an oversimplification; for
example, observations of NGC 5548 show that it exhibits more complex
behavior.  The broad \hbeta\ flux in NGC 5548 does not scale linearly
with continuum flux \citep[the ``intrinsic Baldwin effect'';
][]{gilbertpeterson2003}, and the relationship between BLR radius and
continuum luminosity measured from long-term monitoring of NGC 5548
exhibits a steeper slope than the radius-luminosity relationship for
the reverberation-mapped sample as a whole \citep{kilerci2015}.

Recently, \citet{guo2014} carried out line width measurements for a
sample of 60 quasars having at least six epochs of SDSS spectroscopy,
and they examined the correlation between line FWHM and luminosity for
\ion{C}{4}, \ion{Mg}{2}, \hbeta, and \hal.  Interestingly, 75\% of
their sources show a positive correlation between broad-line width and
luminosity, and only three sources exhibited an anticorrelation at the
$\geq90\%$ confidence level.  Another example of a positive
width-luminosity correlation for the Balmer lines was recently found
in PG 1613+658 \citep{zhang2014}, using data from the Wise Observatory
quasar monitoring campaign \citep{kaspi2000}. This behavior is very
different from that seen in our sample, and is inconsistent with
expectations from the scenario of BLR breathing determined by radial
gradients in responsivity (with higher responsivity at larger radius)
as described above.  It will be interesting to examine this trend for
larger samples of quasars over long time baselines as more multi-epoch
spectra become available for quasars over longer time baselines.  If
broad-line widths in high-luminosity quasars respond to continuum
variations in a manner strongly different from what we observe in our
sample of Seyfert galaxies, this would potentially compromise the
reliability of single-epoch black hole mass estimates derived from
quasar spectra.

\section{Implications of Broad-Line Velocity Shifts for Binary
  Black Hole Searches}
\label{sec:velocityshifts}

The spectral decompositions make it possible to measure night-to-night
changes in the broad \hbeta\ velocity centroids.  Our results for Mrk
40, Mrk 50, Mrk 279, Mrk 1511, and NGC 4593 illustrate that systematic
shifts of $\sim100$ \kms, correlated with flux changes, are common and
occur over timescales similar to the continuum variability timescale.
The largest shifts are seen for NGC 4593, in which the broad
\hbeta\ centroid changed by $266\pm11$ \kms\ over a 30-day interval,
as seen in Figure \ref{ngc4593hbetaparams}.

Broad-line velocity shifts are expected in any AGN for which the
reverberation transfer function is asymmetric about the line center,
which will occur when radial motions are present in the BLR
\citep[e.g.,][]{blandfordmckee, capriotti82, antonucci1983, ulrich84,
  stirpe1988, rosenblatt1990}.  The reverberation process can be
described in terms of a convolution of the continuum light curve
$C(t)$ by a velocity-dependent transfer function,

\begin{equation}
L(v_r, t) = \int_{0}^{\infty} \Psi(v_r, \tau) C(t-\tau)
d\tau, 
\end{equation}
where the transfer function $\Psi(v_r,\tau)$ is a function of
line-of-sight velocity $v_r$ and of time delay $\tau$, and $L(v_r, t)$
gives the resulting time-dependent emission-line profile
\citep{blandfordmckee,peterson2001}.  If BLR clouds are in circular
orbits around the black hole, as in a circular Keplerian disk, then
the transfer function will show short lags at high velocities on both
the blueshifted and redshifted sides of the line profile, and at small
line-of-sight velocities there will be a broad distribution of lag
time.  If the BLR gas is dominated by radial motions, however, the
transfer function will be strongly asymmetric about the line center.
For example, in a BLR dominated by radial inflow, the redshifted gas
on the near side of the BLR will have near-zero lag, while the
blueshifted gas on the far side of the BLR will have the longest lags.
Representative transfer functions for simple kinematic models are
shown by \citet{welshhorne}, \citet{bentz2010b}, \citet{grier2013a},
and \citet{pancoast2014b}.

In addition to the imprint of BLR kinematics on the transfer function,
the transverse Doppler shift and gravitational redshift in the BLR
will produce redward asymmetries in the transfer function,
particularly in the high-velocity wings \citep{goad2012}.  The
combination of transverse Doppler shift and gravitational redshift,
together with any radial motions, should ensure that all transfer
functions are at least modestly asymmetric in structure even if the
BLR cloud motions are predominantly rotational.  Absorption or
scattering within the BLR may also alter the shape of $\Psi(v_r,\tau)$
in an asymmetric fashion.

With the best-quality reverberation-mapping data obtained over the
past several years, it has become possible to examine the
velocity-dependent structure of the transfer function in several
low-redshift AGNs by measuring the lag as a function of velocity
across the line profile \citep{kollatschny2003, bentz2008, denney2010,
  barth2011:mrk50}, and by reconstructing the full two-dimensional
shape of the transfer function \citep{bentz2010b, grier2013a}.
Dynamical modeling of reverberation data has also begun to yield
constraints on the transfer function shape \citep{pancoast2012,
  pancoast2014b}.  A primary result that has emerged from these
investigations is that the \hbeta\ transfer functions for nearby
Seyfert galaxies exhibit a broad diversity of structures: some objects
have nearly symmetric transfer functions consistent with expectations
for a Keplerian disk, while others have strong asymmetries suggestive
of inflowing or outflowing motions.

As an illustration of why an asymmetric transfer function will lead to
broad-line velocity centroid changes, consider a radially inflowing
BLR.  If the continuum luminosity increases, the prompt red-wing
response will shift the line centroid redward initially until the blue
wing response from gas on the far side of the BLR is seen at a later
date.  Similarly, a decrease in continuum luminosity will initially
shift the line centroid blueward.  The opposite behavior would be seen
in an outflowing BLR.  Line centroid variability will depend on the
continuum light curve shape and variability amplitude, the broad-line
width, and the degree of asymmetry in the transfer function.  For a
specified transfer function, the direction of line profile shifts will
depend on the recent gradient in the continuum light curve rather than
on the absolute luminosity of the continuum.  Either blueshifts or
redshifts can occur at a given continuum flux level, depending on the
recent history of continuum variations.

These considerations are relevant for binary black hole searches
because one proposed method for identification of binary black holes
involves searching for temporal variations in the line-of-sight
velocity of broad emission lines in quasars.  \citep[See][for recent
  reviews of binary black hole phenomenology, optical spectroscopic
  signatures, and search techniques]{popovic2012, dotti2012}.  If one
member of a binary black hole is active and possesses a broad-line
region, then the orbital motion of the binary will cause the broad
emission lines to oscillate in velocity over an orbital period.  A
substantial amount of effort has been invested recently in searching
for broad-line velocity changes in quasars having two or more epochs
of spectroscopic observations.  One strategy is to search for objects
having very large broad-line velocity offsets (i.e., $|\Delta V|
\gtrsim 1000$ \kms\ relative to the host galaxy), and then obtain
follow-up spectroscopy to check whether the broad-line velocity drifts
over time in a manner consistent with expectations for orbital motion
\citep{eracleous2012, decarli2013, tsalmantza2011, liu2014}.  An
alternate method is to start with the general quasar population,
rather than restricting to objects showing strong velocity offsets
initially, and to search for changes in broad-line centroid velocity
between two widely separated epochs of spectroscopic observations
\citep{shen2013bbh, ju2013}.  The discussion in this section is
primarily (but not exclusively) relevant to this latter method.

Such searches are still in their initial stages. In recent searches
using SDSS data, \citet{shen2013bbh} examined \hbeta\ velocity shifts
in a sample of 1347 quasars having two or more epochs of spectroscopy,
finding 28 objects with significant velocity shifts (of typically
$\sim100-300$ \kms) and identifying seven of these as being promising
candidate binaries.  \citet{ju2013} carried out a similar search using
\ion{Mg}{2}, finding seven candidate sub-pc binaries with velocity
changes of $>280$ \kms\ from a starting sample of 1523 quasars having
spectroscopy with S/N $>10$ at two epochs.  For these candidates,
continuing long-duration follow-up spectroscopy will be required in
order to rule out alternative explanations and conclusively
demonstrate binary motion.

In discussion of these search efforts, it is generally recognized that
broad-line velocity changes can happen as a result of structural
changes in the BLR occurring over a dynamical timescale (years to
decades) or longer \citep{shen2013bbh}, and such changes have been
seen in multi-year monitoring observations of NGC 5548, for example
\citep{wanders1996, sergeev2007}.  The appearance or fading of
hotspots in the BLR has also been suggested as a possible mechanism to
induce line profile changes and apparent velocity shifts
\citep{decarli2013}.  However, asymmetric reverberation seems to have
been overlooked as a possible cause of the detected velocity shifts in
these binary black hole searches.  Our data provide a clear
demonstration that broad-line velocity centroids can vary
significantly on short timescales solely as a result of reverberation,
and do not provide direct evidence for bulk acceleration of the BLR.

Based on our small sample of nearby, low- to moderate-luminosity AGNs,
it is difficult to extrapolate our results to make any useful
predictions for the distribution of broad-line velocity shifts
expected as a function of timescale for quasar surveys. We have
carried out simple simulations using asymmetric transfer functions
similar to those observed in Arp 151 \citep{bentz2010b} or NGC 5548
\citep{pancoast2014b} and find that velocity shifts of
$\gtrsim300-400$ \kms\ can occasionally occur for plausible AGN variability
amplitudes. However, the range of transfer function shapes and
asymmetries is completely unknown for the quasar population.  For a
given AGN variability model and transfer function model, it would be
straightforward to predict the expected distribution of velocity
changes as a function of timescale, but such an effort is beyond the
scope of this paper.

Here, our goal is simply to call attention to the fact that
reverberation does indeed produce significant velocity shifts on short
timescales as a result of the BLR response to continuum
fluctuations. This has been understood since the earliest
reverberation mapping studies, and our spectral decompositions provide
a way to examine broad \hbeta\ velocity shifts in Seyfert 1 nuclei
with high-cadence sampling over a 2.5-month baseline.  Considering
that we find one object showing a velocity change of $\gtrsim250$
\kms\ in our small sample over this brief monitoring duration, it
seems entirely plausible that, in large AGN samples observed over long
time baselines, the tail of the distribution of reverberation-induced
velocity shifts could extend to substantially larger velocities than
the values we observe here. It is difficult to exclude the possibility
that most or all of the recently identified candidate binaries might
simply be exhibiting velocity changes due to this relatively prosaic
effect, rather than exhibiting acceleration due to binary orbital
motion.  In large samples of quasars with two-epoch spectroscopy,
selection of targets with the largest velocity shifts will
preferentially select objects from the tail of the distribution of
large reverberation-induced velocity shifts, and we suggest that the
majority of candidate binaries selected via this method will most
likely be false positives. 

This reverberation effect will also contribute random velocity noise
or jitter that could impede the detection of genuine binaries
(particularly wide-separation or nearly face-on binaries), rather
analogous to the impact of stellar activity on the detectability of
exoplanets via the radial-velocity method.  Furthermore, the 
radial-velocity curve for NGC 4593 (Figure \ref{ngc4593hbetaparams})
illustrates another potential concern.  In this case, the velocity
curve could be mistaken for a near-complete cycle of sinusoidal
oscillation, when in reality the line centroid is simply responding to
a down-and-up variation in the continuum luminosity.  In order to
convincingly demonstrate binary orbital motion and rule out
reverberating impostors similar to this example, it will be necessary
to monitor broad-line velocity variations over at least two orbital
cycles, with sufficient cadence to sample the orbit well, and also to
demonstrate that the velocity variations are uncorrelated with
continuum flux variations.  The expected orbital periods for quasars
hosting supermassive black hole binaries span years to decades
\citep[e.g.,][]{shen2013bbh}, thus confirmation of black hole binaries by
this method will require a very long-term effort.

\section{Conclusions}
\label{sec:conclusions}

We present spectroscopic data, light curves, and other measurements
from a 69-night reverberation mapping campaign carried out at the Lick
3~m telescope in Spring 2011 for a sample of 15 low-redshift
Seyfert 1 galaxies.  Although our observing run was hampered by
worse-than-average weather, we obtained data of excellent quality for
a subset of our sample, suitable for measurement of light curves for
multiple emission lines.  Our primary results can be summarized as
follows.

\begin{enumerate}

\item Expanding on our team's recent work, we have carried out
  multicomponent decompositions of the nightly blue-side spectra from
  our campaign. The decompositions enable measurement of light curves
  for blended features including \ion{Fe}{2} and \ion{He}{2}, and
  measurements of \hbeta\ light curves and widths are improved by
  deblending \hbeta\ from other contaminating emission-line and
  continuum components.  Obtaining a time series of \hbeta\ line
  profiles without contamination from other overlapping features is
  also important for measurement of velocity-resolved reverberation
  signals and transfer functions, and for application of dynamical
  modeling methods to derive black hole masses \citep{pancoast2014b}.
  In Appendix \ref{appendix:biases} below, we demonstrate that fitting
  and removal of continuum components prior to constructing the rms
  spectrum is also necessary in order to avoid a potentially severe
  bias that can occur when measuring line widths from the rms
  spectrum.  Spectral decompositions should be carried out as a
  routine part of reverberation mapping data analysis, provided that
  the S/N of the data is sufficient for multicomponent fits to yield
  useful results.

\item General variability properties for our sample are consistent
  with trends seen in previous reverberation-mapping programs,
  including higher responsivity for the higher-order Balmer lines, and
  very high responsivity for \ion{He}{2}.  While cross-correlation
  lags can be measured between the emission-line and continuum light
  curves presented here, lags measured against $V$-band photometric
  light curves will have better reliability as a result of the higher
  S/N of the photometric data and the absence of correlated
  flux-calibration errors between the photometric and spectroscopic
  data.  

\item The anticorrelation between broad-line width and flux
  observed in four objects demonstrates that the BLR ``breathes''
  (i.e., the emissivity-weighted BLR radius increases as a function of
  AGN luminosity) even on short timescales of days to weeks in
  response to continuum variations. This implies that reverberation
  transfer functions will be nonstationary even during the course of
  a single monitoring campaign. This breathing effect is responsible
  for the relationship between broad-line profile widths measured in
  mean and rms spectra, as we illustrate by means of simulations in
  Appendix \ref{appendix:widthsimulations}.

\item In NGC 4593 we observe a broad \hbeta\ velocity shift of
  $\sim250$ \kms\ over the course of one month, and other AGNs in our
  sample exhibit \hbeta\ velocity shifts of order 100 \kms\ over
  similar timescales.  Such velocity shifts caused by asymmetric
  reverberation represent an important source of confusion noise for
  binary black hole searches based on multi-epoch quasar spectroscopy.
  When only two (or a few) observations of a given AGN are available,
  line-profile centroid shifts due to velocity-dependent reverberation
  can mimic the expected acceleration signature of binary orbital
  motion, producing a spurious identification of a candidate binary.
  Typical velocity changes caused by asymmetric reverberation will be
  up to a few hundred \kms, similar to the velocity shifts detected in
  many of the recently identified candidate binaries
  \citep{shen2013bbh, ju2013}.  In general, reverberation will produce
  a velocity-jitter noise floor that could limit the detectability of
  binary black holes in spectroscopic searches for radial-velocity
  oscillations.

\end{enumerate}

Additional papers in this series will present $V$-band photometric
light curves, measurements of broad-line lags and velocity-resolved
reverberation using both cross-correlation and JAVELIN \citep{zu2011}
methodology, black hole virial mass estimates, and BLR dynamical
modeling using the methods described by \citet{pancoast2011} and
\citet{pancoast2014a}.  Following the completion of this program, our
reduced and scaled spectra will be made available to the community for
archival use.\footnote{The Lick AGN Monitoring Project 2008
  spectroscopic dataset \citep{bentz2009lamp, bentz2010a} has been
  recently released for public use, and is currently available at the
  web site of A.J.B. at UC Irvine.}

\acknowledgments

We thank the Lick Observatory staff for their tireless efforts during
our observing run.  We gratefully acknowledge contributions of
observing time and data from Dawoo Park, Donghoon Son, Matthew Auger,
Alessandro Sonnenfeld, Robert da Silva, Michele Fumagalli, Jessica
Werk, Michael Gregg, Chelsea Harris, Jeffrey Lee, Liliana Lopez, Nao
Suzuki, Jonathan Trump, Hassen Yesuf, Peter Nugent, David Tytler,
Xavier Prochaska, G\'{a}bor Worseck, Melissa Graham, and Michael
Childress.  We thank Daeseong Park for many very helpful discussions
about data-analysis methods, and the anonymous referee for many
helpful suggestions that improved the presentation of this paper.

This work has been supported by National Science Foundation (NSF)
grants AST-1108835 and 1412693 (UC Irvine), 1107865 (UCSB), 1107812
and 1412315 (UCLA), and 1108665 (UC Berkeley).  A.P.\ acknowledges
support from the NSF through the Graduate Research Fellowship Program.
A.V.F.'s group at UC Berkeley received additional funding through NSF
grant AST-1211916, Gary \& Cynthia Bengier, the Richard \& Rhoda
Goldman Fund, the TABASGO Foundation, and the Christopher R. Redlich
Fund.  T.T.  acknowledges a Packard Research Fellowship. The work of
D.S.\ and R.J.A.\ was carried out at the Jet Propulsion Laboratory,
California Institute of Technology, under a contract with the National
Aeronautics and Space Administration (NASA).  R.J.A.\ was also
supported by Gemini-CONICYT grant number 32120009.  Research by
J.L.W.\ was supported by NSF grant AST-1102845.  J.H.W.\ acknowledges
support by the National Research Foundation of Korea (NRF) grant
funded by the Korea government (MEST) (No.\ 2012-006087).
V.N.B. acknowledges assistance from NSF Research at Undergraduate
Institutions (RUI) grant AST-1312296.  J.M.S. is supported by an NSF
Astronomy and Astrophysics Postdoctoral Fellowship under award
AST-1302771. A.M.D. acknowledges support from The Grainger Foundation.
Portions of the data analysis were completed during the Summer 2013
workshop ``A Universe of Black Holes'' at the Kavli Institute for
Theoretical Physics, attended by A.J.B., T.T., and J.-H.W., and we are
grateful for the hospitality of the KITP during this program.

This research has made use of the NASA/IPAC Extragalactic Database
(NED), which is operated by the Jet Propulsion Laboratory, California
Institute of Technology, under contract with NASA.  We thank Jelena
Kova{\v c}evi{\'c} and collaborators for making their new \ion{Fe}{2}
templates available to the community.

Funding for the SDSS and SDSS-II has been provided by the Alfred
P. Sloan Foundation, the Participating Institutions, the NSF, the
U.S. Department of Energy, NASA, the Japanese Monbukagakusho, the Max
Planck Society, and the Higher Education Funding Council for
England. The SDSS Web Site is http://www.sdss.org/.

The SDSS is managed by the Astrophysical Research Consortium for the
Participating Institutions. The Participating Institutions are the
American Museum of Natural History, Astrophysical Institute Potsdam,
University of Basel, University of Cambridge, Case Western Reserve
University, University of Chicago, Drexel University, Fermilab, the
Institute for Advanced Study, the Japan Participation Group, Johns
Hopkins University, the Joint Institute for Nuclear Astrophysics, the
Kavli Institute for Particle Astrophysics and Cosmology, the Korean
Scientist Group, the Chinese Academy of Sciences (LAMOST), Los Alamos
National Laboratory, the Max-Planck-Institute for Astronomy (MPIA),
the Max-Planck-Institute for Astrophysics (MPA), New Mexico State
University, Ohio State University, University of Pittsburgh,
University of Portsmouth, Princeton University, the United States
Naval Observatory, and the University of Washington.

\emph{Facilities:} \facility{Lick/Shane (Kast)}


\clearpage

\appendix

\section{A. Additional Spectra}
\label{appendix:discardedtargets}

\begin{figure}
\begin{center}
\scalebox{0.5}{\includegraphics{{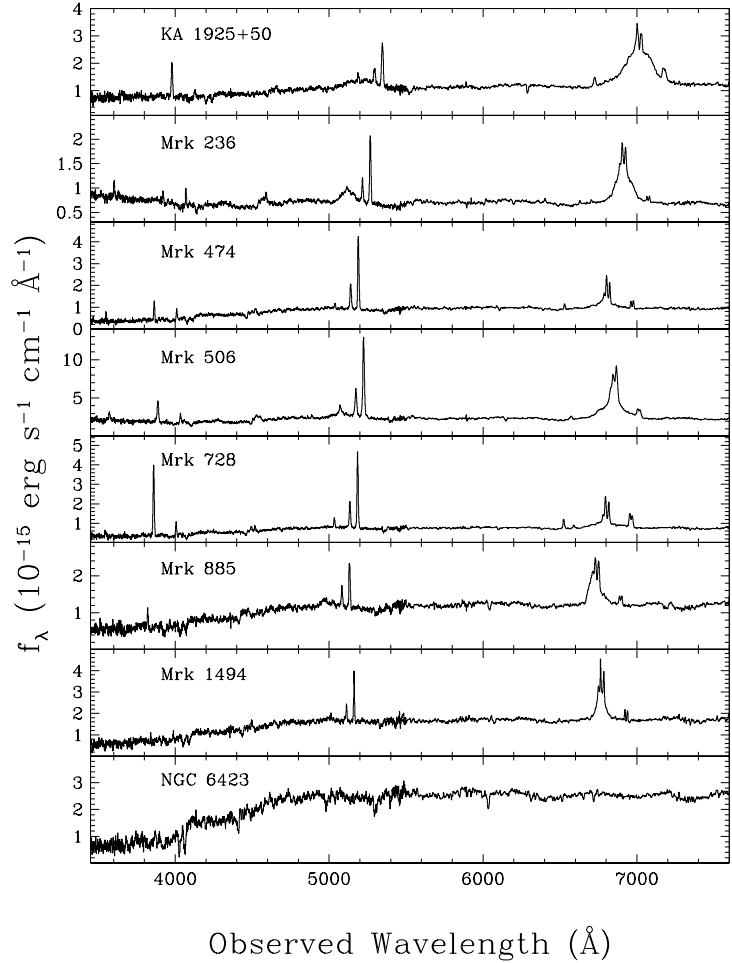}}}
\end{center}
\caption{Kast spectra of objects observed at the start of the campaign
  but not selected for continued monitoring.
\label{otheragns}}
\end{figure}

\begin{figure}
\begin{center}
\scalebox{0.4}{\includegraphics{{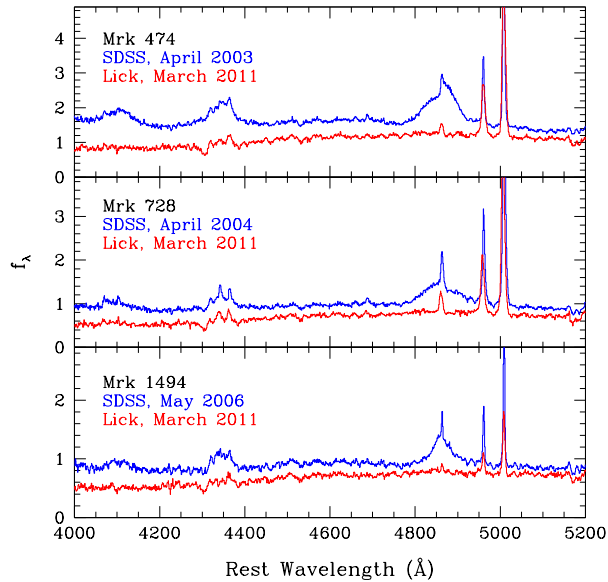}}}
\caption{Comparison between SDSS and Kast spectra for Mrk 474, Mrk
  728, and Mrk 1494, illustrating dramatic fading of the broad
  \hbeta\ emission lines over a span of 5 to 8 years. The Lick
  spectra were not taken under photometric conditions, and have been
  scaled by an arbitrary factor to facilitate comparison.
\label{sdsscomp}}
\end{center}
\end{figure}

During the first few clear nights of our Lick program (2011 March
29--31), we observed several additional galaxies that were initially
selected as possible reverberation targets based on SDSS or other
spectra or on AGN classifications from the literature, but which were
subsequently discarded from our monitoring sample.  Some of these
objects were too faint for regular monitoring, and others appeared to
be in a low flux state compared with previous observations.  In one
case we did not detect any emission lines in a galaxy previously
classified as a Seyfert 1.  Interestingly, three AGNs selected from
SDSS based on their strong broad \hbeta\ emission had significantly
different spectral properties when observed at Lick: Mrk 474, Mrk 728,
and Mrk 1494 all appeared to be excellent reverberation mapping
targets based on SDSS spectra observed in 2003--2006, but by 2011 the
previously strong broad \hbeta\ lines faded so completely that they
would be classified as Type 1.9 Seyferts.  Transitions like these are
not unprecedented \citep[e.g.,][]{tohline1976, aretxaga1999,
  shappee2014, denney2014, scott2014, lamassa2014}, but little
statistical information is available on the timescale or duty cycle of
such dramatic changes.  For completeness, we present notes on the
discarded targets below, along with plots of the spectra (Figure
\ref{otheragns}) and comparisons with the earlier SDSS spectra for the
three AGNs that exhibited a near-complete disappearance of broad
\hbeta\ (Figure \ref{sdsscomp}).

\emph{KA 1925+50:} This Seyfert 1 galaxy at $z=0.067$ is one of the
brightest AGNs in the field observed during the main \emph{Kepler}
mission, and initial \emph{Kepler} light curves were presented by
\cite{mushotzky2011}.  We confirm its Seyfert 1 classification, but
its broad \hbeta\ emission line has a low amplitude relative to the
host-galaxy continuum, so it is not an ideal target for
\hbeta\ reverberation mapping although \hal\ emission is very strong.
Its full designation is 2MASX J19250215+5043134.  See
\citet{edelson2012} for further information about this object.

\emph{Mrk 236:}  This AGN appeared to be in a fainter state than in an
earlier SDSS spectrum, and was discarded in favor of other objects in
the same RA range.

\emph{Mrk 474 (NGC 5683):} In the SDSS spectrum (observed 2003 April
4), this is a Seyfert 1 galaxy with very strong broad
\hbeta\ emission.  Our Lick spectrum shows that the AGN has faded
dramatically, and broad \hbeta\ is no longer visible.

\emph{Mrk 506:} The broad \hbeta\ emission in this galaxy is too weak
to be useful for reverberation mapping, and our spectrum appears very
similar to the SDSS spectrum.

\emph{Mrk 728:} This is another example of an object having strong
broad \hbeta\ in the SDSS spectrum (observed 2004 April 24) but
showing little or no detectable broad \hbeta\ in our Lick data.

\emph{Mrk 885:}  This object is classified in NED as a Seyfert 1.0,
but has relatively weak broad \hbeta\ and is too faint and
starlight-dominated to be a good reverberation-mapping target.

\emph{Mrk 1494:} The broad \hbeta\ line in this AGN also appears to
have faded dramatically in comparison with the SDSS spectrum, observed
on 2006 May 28.  The narrow [\ion{O}{3}] and \hbeta\ lines are also
significantly fainter in the Lick spectrum than in the earlier SDSS
data.

\emph{NGC 6423 (Ark 524):} This galaxy is listed in the
\citet{veron2006} catalog and in NED as a Type 1.5 Seyfert, but we do
not detect emission lines in its spectrum.


\section{B. On the Relationship Between Mean and RMS Profile Widths}
\label{appendix:widthsimulations}

Broad-line widths measured from rms spectra are commonly used to
derive black hole masses from reverberation data. In light of the very
widespread use of rms profile widths for determination of black hole
masses and for calibrating the overall normalization of the AGN black
hole mass scale \citep{onken2004, peterson2004, collin2006, woo2010,
  graham2011, park2012b, grier2013b, ho2014}, we include in the
following sections a description of simulations which may be helpful
in clarifying certain aspects of the behavior of the rms profile in
time-series spectroscopy of AGNs.

We first examine the relationship between broad-line profile widths in
mean and rms spectra.  The mean line profile gives a useful
representative view of the typical emission-line width during an
observational campaign. However, the mean profile incorporates the
complex and nonlinear response of the line profile to continuum
variations (including profile width variations and centroid shifts as
well as flux variations, as illustrated in Section
\ref{sec:widthsandcentroids}), so it is not identical to the
single-epoch profile that would be observed when the AGN is at its
mean flux.  The interpretation of the rms spectrum is less
straightforward.  It is sometimes stated that the rms spectrum
represents the variable portion of an emission line, because
nonvarying components do not contribute to the rms
\citep[e.g.,][]{peterson1998, petersonwandel99, fromerth2000}.
However, the BLR is not composed of distinct variable and nonvariable
components; rather, different regions of the BLR have different
responsivities to continuum variations, and the rms spectrum does not
correspond directly to the spectrum emitted by any specific physical
subcomponent of the BLR.  Essentially, the rms spectrum is useful as a
tool for visualization of the relative degree of variability as a
function of wavelength across the line profile, due to spatial
gradients in emission-line responsivity across the BLR combined with
the effects of light-travel time across the BLR, but it is not obvious
how to interpret the physical meaning of the velocity width of an rms
line profile.  Since reverberation-mapping campaigns are unable to
acquire continuously sampled observations as an AGN varies, the
detailed shape of both the mean and rms spectra depends additionally
on the observational sampling pattern used during the campaign.
Finally, continuum variability and noise in the data will have an
impact on the line profiles and widths measured from rms spectra; we
examine these effects in Appendix \ref{appendix:biases} below.

The width of broad \hbeta\ is typically found to be slightly narrower
in rms profiles than in mean profiles
\citep[e.g.,][]{sergeev1999,collin2006,bentz2010a}. Combining our
sample with the previous 2008 Lick AGN Monitoring Project sample
\citep[using measurements from][based on spectral fitting methods
  similar to those employed here]{park2012a} plus additional Lick
observations of Zw 229-015 \citep{barth:zw229} and KA1858+4850
\citep{pei2014}, we have 17 independent measurements of the ratio
$\sigma(\mathrm{rms})/\sigma(\mathrm{mean})$ for \hbeta\ in 15 AGNs,
including two measurements each for Mrk 40 and Zw 229-015.  This ratio
ranges from 0.43 to 1.06, with a mean value of 0.85 and a standard
deviation of 0.15 among these 17 measurements.  Similarly, in the
sample of 15 AGNs examined by \citet{sergeev1999}, the average value
of the ratio $\sigma(\mathrm{rms})/\sigma(\mathrm{mean})$ for the
broad \hbeta\ line is 0.87.

The narrower widths of rms profiles for \hbeta\ can be understood as a
consequence of high-ionization gas in the inner BLR having a lower
responsivity to continuum variations than low-ionization gas more
distant from the black hole; \citet{korista2004} predict a factor of
$\sim5$ change in \hbeta\ responsivity across the radial range of the
BLR.  The high-velocity wings of the line profile are emitted
primarily by low-responsivity portions of the BLR, while
higher-responsivity gas contributes mainly to the line core.  As a
result, the core of the \hbeta\ profile varies more strongly than the
wings, making the rms profile narrower than the mean profile.

Why is it that $\sigma(\mathrm{rms})/\sigma(\mathrm{mean})$ typically
takes the value $\sim0.85$?  The purpose of this Appendix is to
illustrate by means of simple numerical simulations that this ratio is
dictated by the relationship between the broad \hbeta\ single-epoch
luminosity and line width (as described in Section
\ref{sec:linewidths}), which is fundamentally a consequence of the
variation in emission-line responsivity across the BLR.  The
calculations presented here are not derived from a physical model of
BLR variations; this is merely a numerical experiment designed to
examine the behavior of rms profiles when the line width responds in a
well-defined way to flux variations.

We begin with a Gaussian line profile as the baseline profile for
steady-state emission. Using the method of \citet{timmer1995}, we
generate a random light curve of some specified duration, sampling,
power spectrum shape, and rms variability amplitude (or \fvar). As we
describe below, the details of the light curve construction have
little or no effect on the ratio of rms profile width to mean profile
width.

Then, we generate a time series of line profiles in which both the
line luminosity and its width vary in response to continuum
fluctuations.  Assuming a delta function with lag  $\tau$ for the
velocity-integrated transfer function, at each time $t$ the integrated
luminosity $L$ of the emission line is made to scale linearly with the
input continuum light curve luminosity at time $t-\tau$.  The width of
the Gaussian profile is also made to vary, such that the product
$\sigma^2 L^{0.5}$ remains constant as the line luminosity varies.
This represents the most common behavior seen in our sample [i.e.,
  $\sigma$(\hbeta)$\propto f$(\hbeta)$^{-0.25}$], and corresponds to a
situation in which the single-epoch virial product remains roughly
constant as the AGN luminosity varies.  From the time series of
resultant line profiles, we construct the mean and rms profiles and
calculate \sigmaline\ for each.

\begin{figure}
\begin{center}
\scalebox{0.4}{\includegraphics{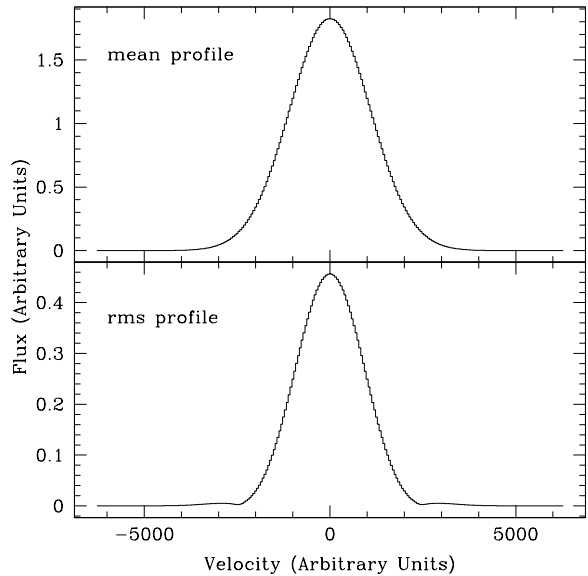}}
\caption{Example from a simulation of a mean and rms line profile,
  constructed from a time series of Gaussian profiles having variable
  luminosities and widths as described in the text, with
  $\sigma^2L^{0.5}=$~constant.  The amplitude of the rms profile is
  set by the overall variability amplitude of the input light curve in
  the simulation, but its shape is nearly independent of the light
  curve properties.
\label{rmssimulation}
}
\end{center}
\end{figure}

The result of this experiment is surprisingly simple.  In the mean
spectrum, the emission line has a profile that is very nearly
Gaussian. As expected for a broad line showing higher variability
amplitude in the core than in the wings, the rms profile is narrower
than the mean profile.  The rms profile has an approximately Gaussian
shape, but exhibits additional structure in the faint high-velocity
wings, and this profile shape is nearly independent of the input light
curve properties.  The peak height of the rms profile depends on the
variability amplitude of the light curve, while its width is only very
weakly dependent on variability amplitude.  In the limit of small but
nonzero light curve variance, the ratio
$\sigma(\mathrm{rms})/\sigma(\mathrm{mean})$ converges to 0.824.  As
the light curve variance is increased, we find that this ratio becomes
very slightly sensitive to the input light curve properties.  For
example, when the input light curve is scaled to have $\fvar=0.2$, the
simulations produce a result of
$\sigma(\mathrm{rms})/\sigma(\mathrm{mean}) = 0.83 \pm 0.01$, where
the uncertainty represents the standard deviation of values from 1000
simulations carried out with different random light curves.  Figure
\ref{rmssimulation} shows an example of a mean and rms profile
obtained from one such simulation with $\fvar=0.2$ over a light curve
having 500 equally spaced points, with a fluctuation power spectrum of
the form $P(\nu)\propto \nu^{-2.7}$ (where $\nu$ is the temporal
frequency of fluctuations).  The rms profile shape would appear
virtually identical for different choices of light curve duration or
sampling, power spectrum slope, or variability amplitude, although its
absolute normalization depends on \fvar.

These simulations produce a ratio
$\sigma(\mathrm{rms})/\sigma(\mathrm{mean})$ that is remarkably close
to the observed mean ratio of $0.85\pm0.15$ found from our Lick
Observatory monitoring campaigns.  In other words, the typical ratio
of line widths observed in rms and mean line profiles simply reflects
the fact that in most AGNs, the broad \hbeta\ line maintains a roughly
constant value of the product $L^{0.5}\sigma^2$ as the AGN luminosity
varies.  The larger dispersion of values observed in real AGNs most
likely results from the range of different (non-Gaussian) profile
shapes in different AGNs, and also differences in the scaling
relationship between single-epoch luminosity and profile width.
Changing the power-law relationship between line width and luminosity
does alter the width of the rms profile relative to the mean profile.
For example, if we calculate models for a flatter single-epoch scaling
relationship of $\sigmaline \propto L^{-0.07}$ (as observed in Mrk
50), the simulations yield $\sigma$(rms)$/\sigma$(mean) $ =
0.92\pm0.01$.

The prevailing use of rms profile widths as the basis for calibrating
the AGN black hole mass scale underscores the need for further work to
understand the rms spectrum by means of of detailed simulations of BLR
line profiles incorporating BLR dynamics and photoionization physics,
including luminosity-dependent changes in emission-line responsivity
\citep{korista2004, goadkorista2014}.  The dynamical modeling approach
for interpretation of reverberation-mapping data \citep{pancoast2012,
  li2013, pancoast2014a} avoids the problem of relying on simplistic
measures of line width from either mean or rms spectra, and can
provide more direct constraints on black hole masses.  For future
development of these models, it should be a high priority to
incorporate a spatially variable and luminosity-dependent
emission-line responsivity based on photoionization modeling of BLR
structure, in order to more accurately model an observed time series
of variable line profiles.  If the responsivity is assumed to be
spatially and/or temporally constant across the BLR, it will not
generally be possible for these models to replicate the observed
anticorrelation between line width and luminosity, or the relationship
between mean and rms profile widths.
 
\section{C. Biases in RMS Profile Widths}
\label{appendix:biases}

Since line widths measured from rms spectra are used to derive black
hole masses, it is useful to examine possible sources of systematic
error in these measurements.  The rms spectrum blends information
about emission-line variations, continuum variations, and random noise
in the data, and it can deviate substantially from the ``ideal'' rms
spectrum that would be derived if a single broad emission line could
be observed without contamination from other spectral components or
noise.  In this Appendix, we describe simulations designed to examine
the impact of continuum variability and photon-counting noise on line
widths measured from rms spectra.

\subsection{C1. Bias due to continuum variability}

This same simulation methodology described above allows us to examine
a potential source of bias in the rms profile width that can occur
when the rms spectrum is constructed from a set of spectra that have
not had the variable AGN continuum removed.  Components of constant
flux, such as narrow emission lines or starlight, do not contribute to
the rms spectrum except to the extent that they show some
night-to-night variations due to seeing differences, miscentering of
the AGN in the slit, random wavelength or flux calibration errors, or
Poisson noise.  However, it is not widely appreciated that the
variable AGN continuum underlying an emission line can affect the
shape of the rms profile of the line, and that this can potentially
introduce significant biases into black hole virial masses determined
from rms profile widths.

The reason this bias occurs is that the continuum and emission-line
variations are out of phase with one another: there will be portions
of the light curves where the continuum is rising while the
emission-line flux is falling, and vice versa.  This out-of-phase
variability alters the rms amplitude of the \emph{total} (line plus
continuum) flux, and tends to preferentially suppress the rms flux in
the line profile wings relative to the line core. Consequently, the
rms profile constructed from total-flux spectra can sometimes have a
significantly different width than it would if the emission line could
be observed in isolation without continuum contamination.  Even a
modest bias in determining the rms profile width can have a
substantial and deleterious effect on the derivation of black hole
masses, because black hole mass scales as the square of the line
width.  The typically employed procedure of fitting and subtracting
the continuum from the total-flux rms spectrum prior to measuring the
profile width does not alleviate this problem.

To quantify this bias, we carried out simulations using Gaussian line
profiles as described above.  As before, the total line luminosity
responds linearly to continuum variations, and the line luminosity and
width vary such that $\sigma^2 L^{0.5}$ remains constant.  For the
input light curves, we assumed a fluctuation power spectrum of the
form $P(\nu)\propto \nu^{-2.7}$, similar to the properties of nearby
AGNs monitored by the \emph{Kepler} mission \citep{mushotzky2011}, and
we carried out simulations with light curves of varying duration and
various values of \fvar.  The velocity-integrated transfer function
was assumed to be a delta function at a lag of 10 days.  The emission
line was set to have a mean equivalent width of 120 \AA\ relative to
the mean AGN continuum flux density (typical for the broad
\hbeta\ lines in our sample) and an initial width of $\sigmaline=2000$
\kms, and the continuum flux level was assumed to be flat over the
wavelength range of interest.  Spectra were constructed for velocity
bins of 60 km s\per, corresponding closely to the 1 \AA\ wavelength
bins of the Kast blue-side data. These simulations do not include the
effects of photon-counting noise, random wavelength or flux
calibration errors, contributions of narrow-line or host-galaxy
components to the spectra, or nonuniform sampling cadence, all of
which can further alter the rms spectrum.  (The impact of
photon-counting noise is described in the next section.)

\begin{figure}
\scalebox{0.45}{\includegraphics{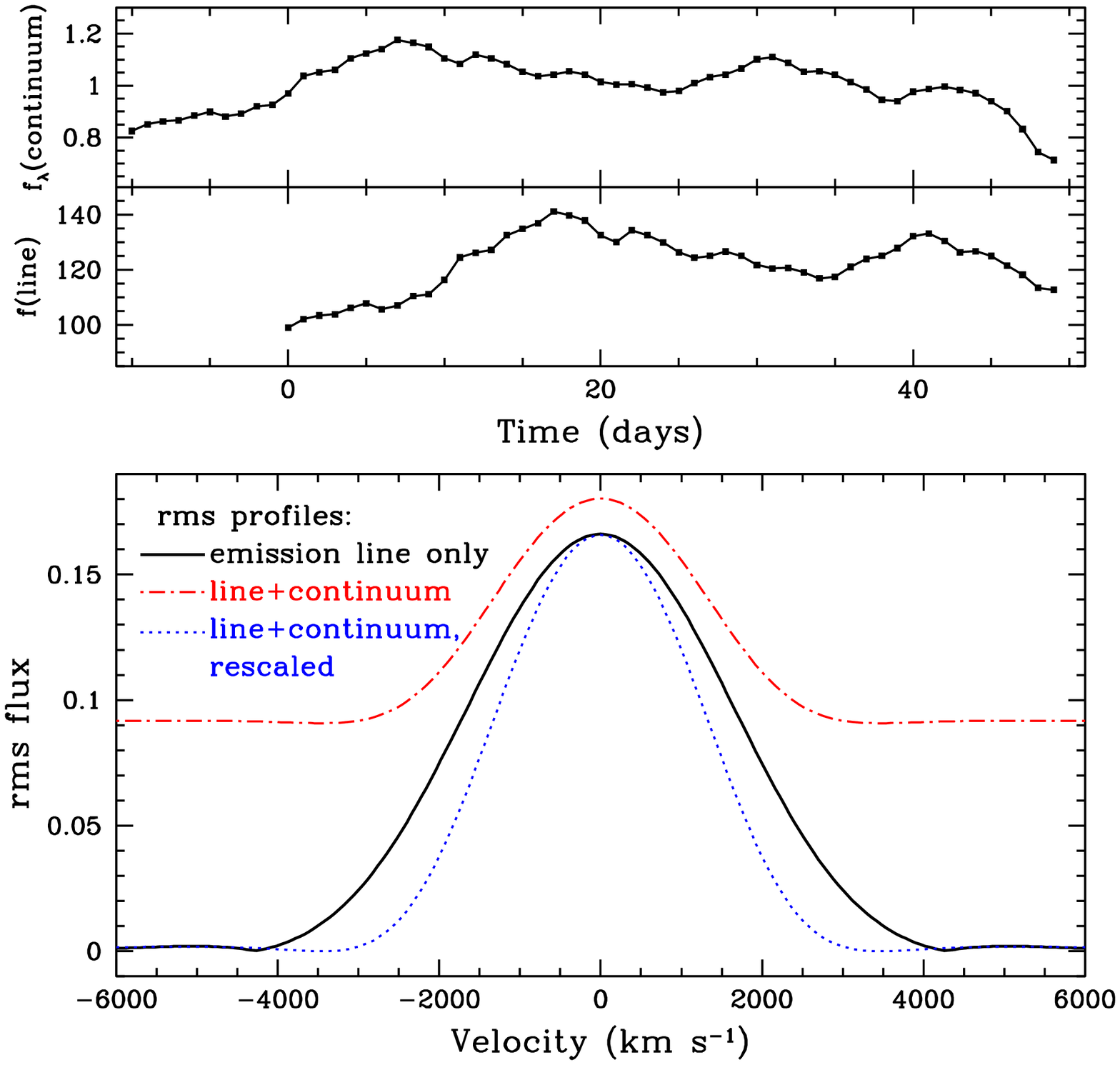}}
\scalebox{0.45}{\includegraphics{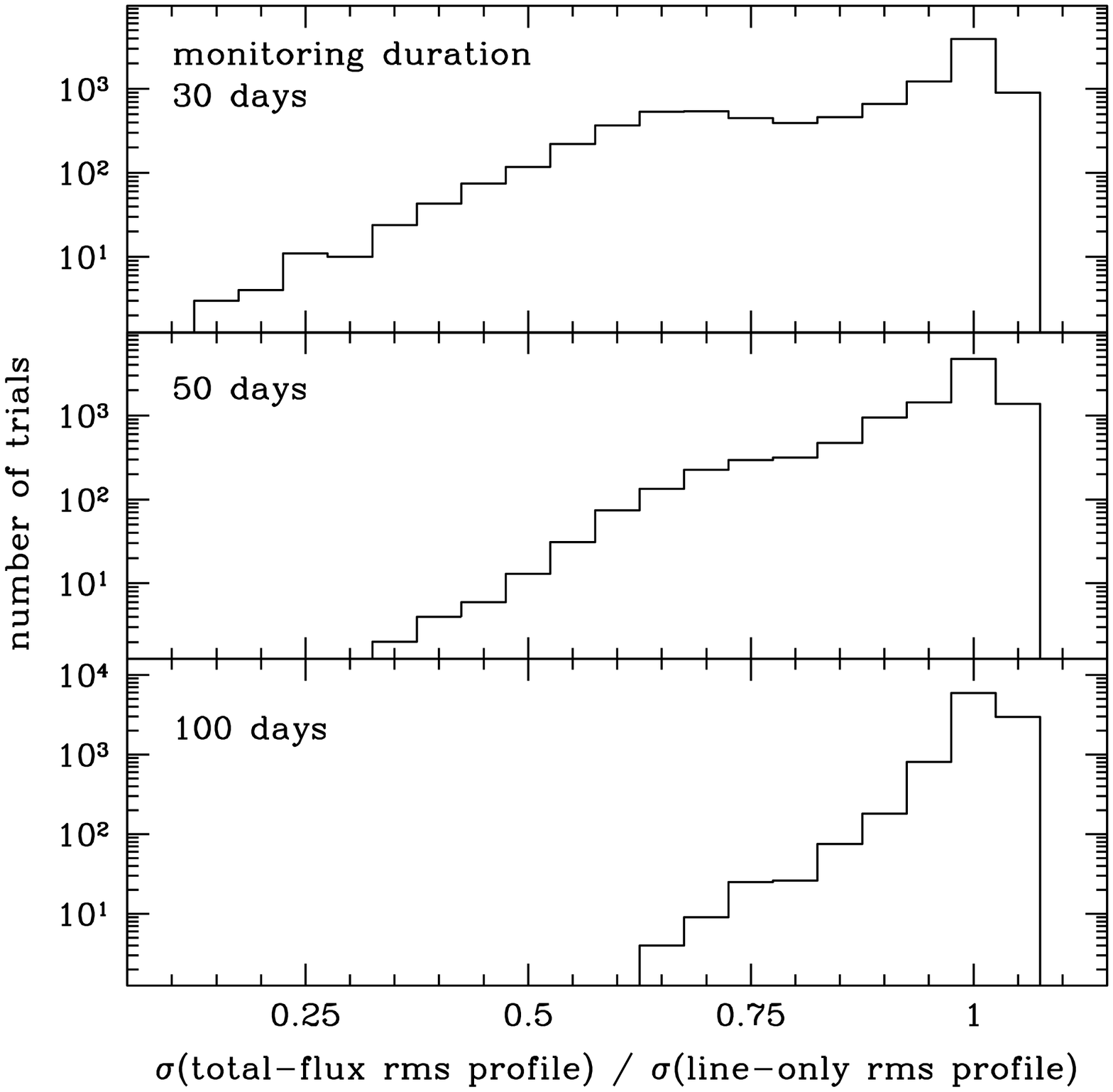}}
\caption{Results from simulations to examine the ratio of line profile
  widths in total-flux rms spectra vs.\ emission-line only rms
  spectra.  \emph{Left:} Example of a simulation exhibiting a
  significant bias in the rms profile width when using the total-flux
  rms spectrum.  The upper panels show the continuum and emission-line
  light curves; each point in the emission-line light curve
  corresponds to a simulated observation of the spectrum.  The
  continuum has $\fvar=0.1$, and the emission line has a lag of 10
  days and an equivalent width of 120 \AA.  The lower panel shows
  different versions of the rms spectrum. In black is the ``true'' rms
  spectrum determined from spectra of the emission line only (without
  continuum) over the 50 epochs of the simulation.  The red dot-dashed
  curve shows the total-flux rms spectrum which includes the
  contribution of the variable continuum.  The blue dotted curve shows
  the same total-flux rms spectrum, after subtracting the minimum
  continuum value and rescaling to match the peak flux of the true rms
  spectrum.  In this case, the dispersion $\sigmaline$ of the
  total-flux rms spectrum is 75\% of the dispersion of the true rms
  spectrum, which would lead to an inferred black hole mass that was
  only 56\% of its ``true'' value.  \emph{Right:} Simulation results
  showing the distribution of the ratio of emission-line dispersions
  \sigmaline\ measured from the total-flux and line-only spectra.
  Simulations were carried out for a lag of 10 days, monitoring
  durations of 30, 50, and 100 days with daily sampling, and
  \fvar=0.1.
\label{biasedrmsprofile}
}
\end{figure}

Starting from each simulated continuum light curve, we generated a
time series of emission-line profiles.  At each time step, we created
a spectrum of the emission line alone, and a total-flux spectrum
including both the line and the continuum.  Two versions of the rms
spectrum were then computed from the time series of line-only and
total-flux spectra.  Then, the line dispersion \sigmaline\ was
measured from each of these.  For the total-flux rms spectrum, the
surrounding continuum level was removed prior to measuring the line
dispersion by subtracting the minimum value of the rms spectrum.  In
order to minimize the impact of rms flux in the very distant
high-velocity line wings on the \sigmaline\ measurements, the
dispersion was measured only over the wavelength range between the two
minima on either side of the profile, in cases where the rms profile
exhibited local minima straddling the line profile.  Similarly, for
the line-only rms spectrum we also measured the dispersion only over
the wavelength range between the two local minima that occur in the
profile wings.  In real data, these very shallow minima would most
likely not be visible as a result of the additional contributions to
the continuum level of the rms spectra from other noise sources, and
the rms profile width would be measured over some visually defined
wavelength range corresponding roughly to the full width at zero
intensity of the rms profile.

The simulations show that the width of the total-flux rms line profile
can often deviate very substantially from the width of the line-only
rms profile, particularly when the monitoring duration is only a few
times longer than the emission-line lag.  Figure
\ref{biasedrmsprofile} (left panel) shows an example of one
simulation, with a 50-day spectroscopic monitoring duration (i.e., 5
times longer than the lag, and assuming nightly observations with no
gaps).  The figure illustrates both the line-only rms profile, which
should be considered the ``true'' rms profile for this simulation, and
the total-flux (line plus continuum) rms profile.  By subtracting the
continuum level from the total-flux rms spectrum and rescaling it to
match the peak of the true rms profile, it is immediately apparent
that the total-flux rms profile is much narrower than the true
profile.  In this instance,
$\sigma(\mathrm{total})/\sigma(\mathrm{true}) = 0.75$, which would
bias the inferred black hole mass by a factor of $0.75^2 = 0.56$.

To examine the distribution of this bias factor, we carried out a set
of these simulations for campaigns of duration 30, 50, and 100 days,
assuming nightly sampling in each case and a fixed value of
$\fvar=0.1$ for the AGN continuum.  In Figure \ref{biasedrmsprofile}
(right panel), we plot the distribution of the ratio
$\sigma(\mathrm{total})/\sigma(\mathrm{true})$ determined from a set
of $10^4$ simulations at each campaign duration.  The bias
distribution is a strong function of campaign duration.  For example,
we find that the percentage of simulations returning
$\sigma(\mathrm{total})/\sigma(\mathrm{true})<0.75$ is 22\%, 6\%, and
0.3\% for campaign durations of 30, 50, and 100 days, respectively.
In most of the recent large reverberation-mapping programs carried out
by our team and others, the monitoring duration is usually at least
$\sim5$ times longer than the measured \hbeta\ lags of the targets,
but as future programs push toward higher-luminosity AGNs at higher
redshifts there may be increasing numbers of AGNs for which the
spectroscopic time series is sampled less intensively, over a
monitoring duration that might be just barely sufficient to measure
the lag (typically about $\sim3$ times the lag).  In that situation,
the total-flux rms spectrum can frequently give a misleading measure
of the emission-line width.

In actual reverberation data sets, the magnitude of this bias will
depend on the details of the light curve shape and variability
amplitude, the transfer function shape, the profile shape and
equivalent width of the emission line, and the sampling cadence of the
observations, so the outcome will be different for every time series
of data for every AGN. Our intention is not to carry out exhaustive
simulations for a broad range of AGN parameters, but simply to
demonstrate that this bias can occur under typical circumstances of a
reverberation measurement when the total-flux rms spectrum is used.
The simulation results shown in Figure \ref{biasedrmsprofile}
illustrate that \sigmaline\ can sometimes be severely underestimated
even when the light curves have sufficient duration and sampling that
the emission-line lag could easily be measured accurately.

We propose that the best way to avoid this problem is to carry out
continuum fitting and subtraction \emph{prior to} construction of the
rms spectrum, so that line widths can be measured from an rms spectrum
that is free of continuum variability contamination.  Other advantages
of carrying out spectral decompositions for measurement of \hbeta\ rms
profile widths have been discussed by \citet{park2012a}; they focused
on the importance of removing \ion{He}{2}, \ion{Fe}{2}, narrow-line
components, and starlight prior to constructing rms spectra, but they
did not remove the nonstellar featureless continuum. Figure
\ref{mrk40rmscomparison} illustrates the construction of an rms
spectrum for the broad \hbeta\ line in isolation, with other
emission-line and continuum components removed, and in this paper we
have measured the profile widths from rms spectra constructed in this
way.  It is still necessary to carry out a local continuum subtraction
by fitting a linear continuum model to the rms flux on either side of
the \hbeta\ line in order to remove the small residual rms flux level
due to random noise in the data.  For the case of Mrk 40 illustrated
in Figure \ref{mrk40rmscomparison}, the values of \sigmaline\ for the
total-flux rms spectrum and the line-only rms spectrum differ by just
2\%.  Based on our simulations, only a very small difference would be
expected given that the campaign duration was $\sim20$ times longer
than the \hbeta\ lag of $\sim4$ days for this AGN
\citep{bentz2009lamp}.

\subsection{C2. Bias due to photon statistics}
\label{appendix:photonnoise}

The finite S/N of spectroscopic observations is the source of a
different bias that can afflict profile widths in rms spectra. The
pixel values in an rms spectrum are determined not just by variability
in the AGN, but also by the noise level in the spectra. A nonvarying
object would still be observed to have a nonzero rms spectrum due to
photon-counting noise and detector noise.  The contribution of
these noise sources can affect the profile width in the rms spectrum
simply because the flux density is largest in the core of the emission
line, hence the noise level will be highest there as well.  The larger
contribution of noise in the line core will elevate the peak of the
line profile in the rms spectrum relative to the wings, making the
line slightly narrower than it would appear in the absence of noise.
While this is not a large effect, its impact is amplified because
virial masses scale as the square of the line width.

To examine the impact of S/N on rms profile widths, we carried out
further simulations using the same methodology described above.
Continuum light curves and the resulting time series of line profiles
were created using the same prescriptions and parameters described in
the previous section.  The continuum and emission-line spectra for
each simulation epoch were added to give the total flux spectrum at
each epoch.  Then, Gaussian random noise was added to each spectrum in
the time series.  The Gaussian dispersion for this pixel-to-pixel
noise level was set by fixing the continuum S/N to a given value, and
scaling the noise level by $S_i^{1/2}$, where $S_i$ is the flux
density at pixel $i$ in the spectrum. Each simulation was run for a
total spectroscopic monitoring duration of 100 days with nightly
sampling, an emission-line lag of 10 days, and a specified value of
\fvar\ for the continuum variability amplitude.

From each simulation, we computed a ``true'' rms spectrum from the
time series of noise-free emission-line spectra only, with no
continuum contributions.  Using the noise-added data, we subtracted
the continuum flux level from each nightly spectrum and constructed a
line-only rms spectrum that included the noise contribution.  The line
dispersion of the true rms profile was measured as described
previously, by including the flux only between the two local minima on
either side of the rms profile.  To measure the width of the
noise-added rms profile, we fit and subtracted the flat noise
continuum using a straight-line fit, and then calculated the line
dispersion over the same velocity range that was used for the true rms
profile measurement.  The ratio of the noise-added rms line width to
the true rms line width gives the bias factor,
$\sigma$(noise-added)/$\sigma$(noise-free).  Since the continuum flux
was subtracted from each noise-added spectrum prior to constructing
the noise-added rms spectrum, this measurement is immune to the
continuum-variability bias described in the previous section.

\begin{figure}
\begin{center}
\scalebox{0.43}{\rotatebox{-90}{\includegraphics{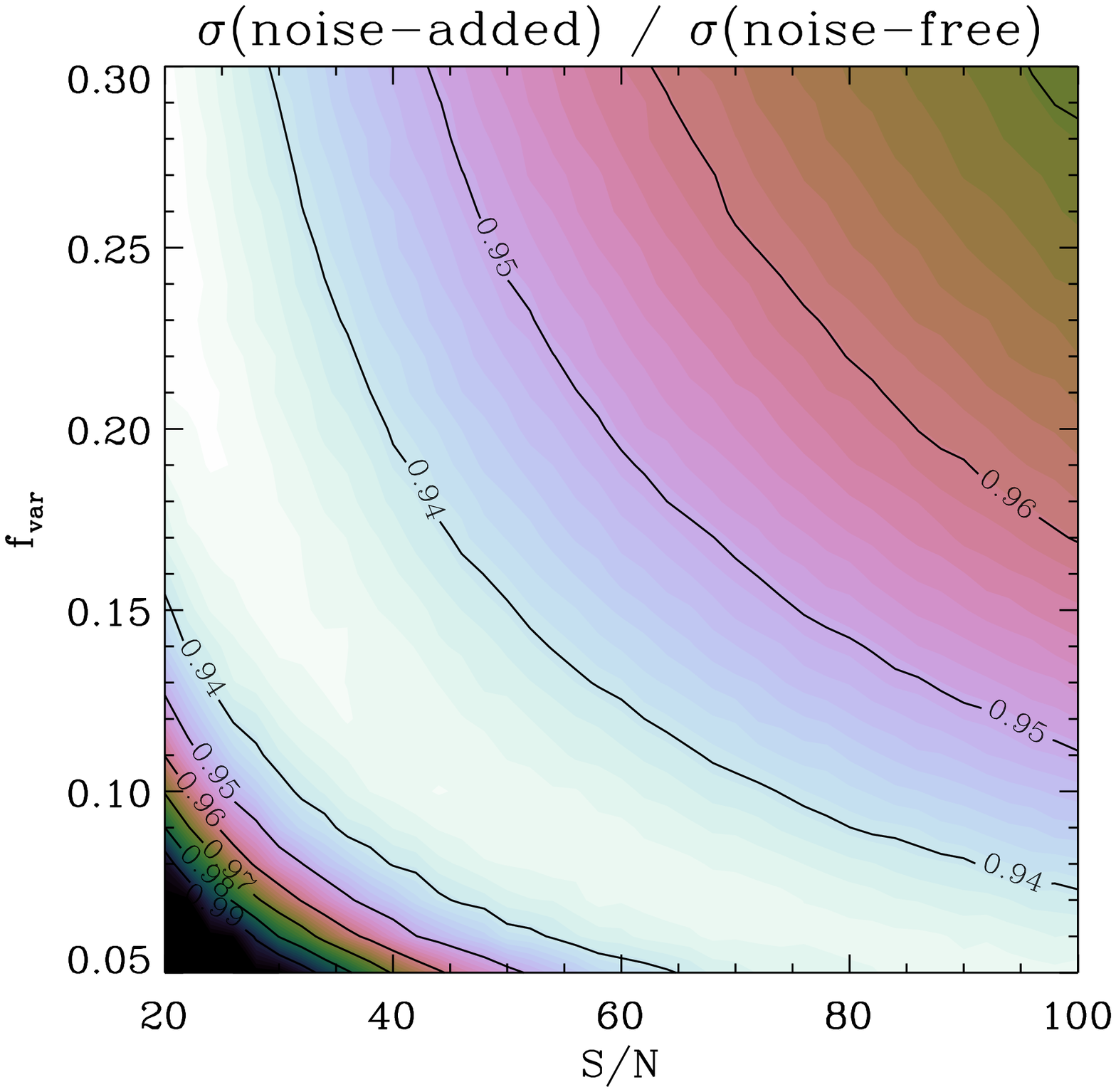}}}
\hspace*{0.1in}
\scalebox{0.43}{\rotatebox{-90}{\includegraphics{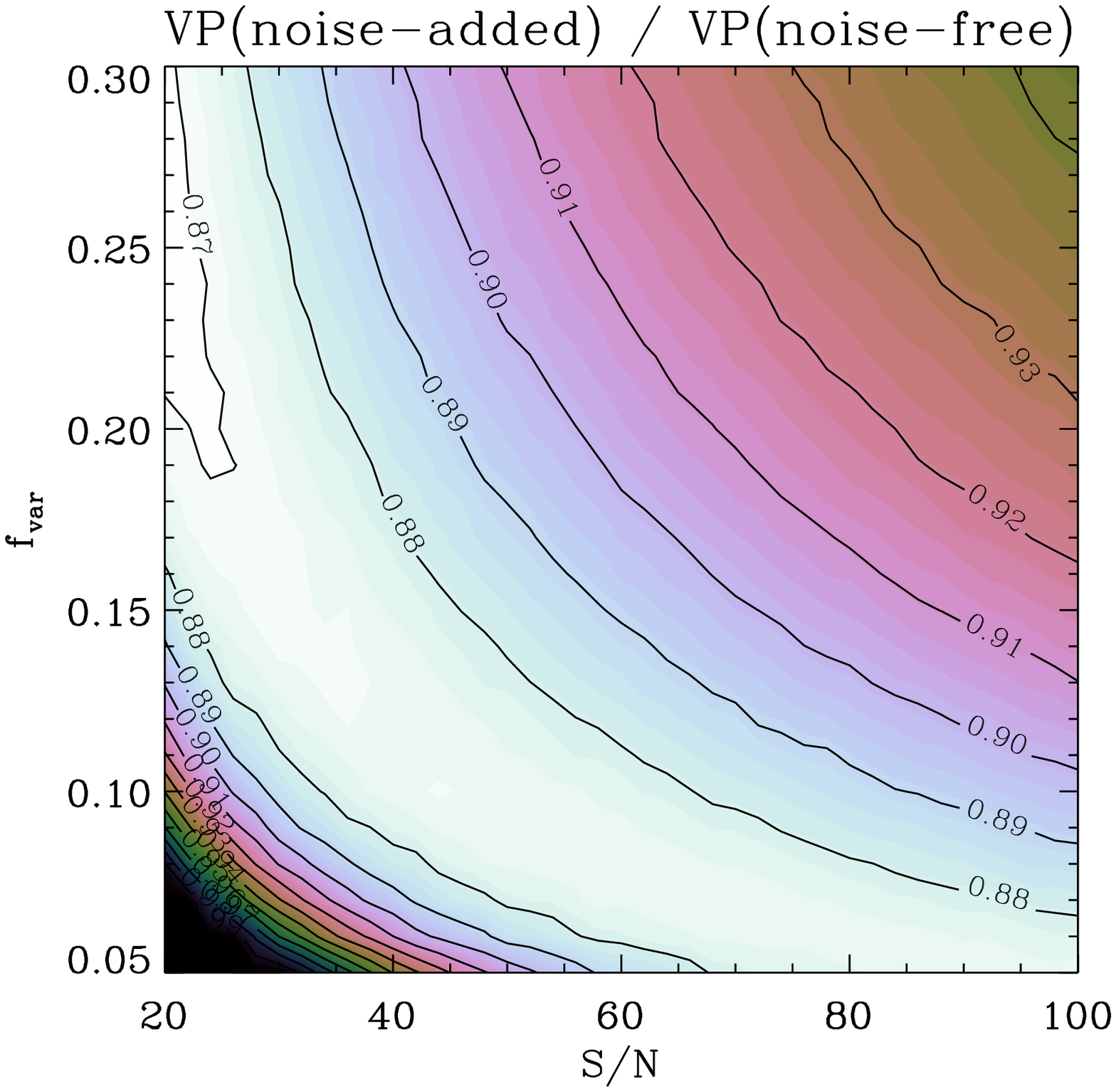}}}
\end{center}
\caption{Results of simulations testing the impact of S/N on line
  widths measured from the rms spectrum.  Tests were carried out for a
  range of values of S/N and \fvar\ as indicated on the $x$ and $y$
  axes.  Contour levels depict the median result from 1000 simulations
  done at each value of S/N and \fvar.  \emph{Left panel:} Ratio of
  the line dispersion of the rms profile including noise
  contributions, to the line dispersion of the rms profile constructed
  from noise-free spectra.  \emph{Right panel:} Ratio of the virial
  product (VP) for the noise-added and noise-free cases, where the
  virial product is proportional to the square of the line dispersion.
  The lower left corner of each panel, at $\fvar<0.1$ and low values
  of S/N, corresponds to reverberation measurements that would most
  likely fail to detect a lag.}
\label{snbiasfigure}
\end{figure}

The degree of bias in the rms profile width depends primarily on
\fvar\ and on the S/N of the data. We carried out simulations for
continuum S/N ranging from 20 to 100 (in increments of 2) and for
\fvar\ ranging from 0.05 to 0.30 (in increments of 0.01).  For each
set of these parameters, we carried out 1000 simulation trials and
determined the median bias factor.  Figure \ref{snbiasfigure} displays
the results, illustrating the ratio
$\sigma$(noise-added)$/\sigma$(noise-free) as a function of \fvar\ and
the continuum S/N.  We also display the bias in the virial product,
VP(noise-added)/VP(noise-free)
=$[\sigma$(noise-added)$/\sigma$(noise-free)$]^2$.  For typical
parameters of a successful reverberation measurement, corresponding to
$\fvar\gtrsim0.1$ and S/N of 30--100 per pixel, the rms profile width
is biased low by $\sim3-7\%$, and virial products are biased low by
$\sim6-13\%$. The bias vanishes in the limit of high S/N and high
variability amplitude, but even at S/N=100 per pixel the virial
products are biased low by $\sim5-10\%$.  Unlike the
continuum-variability bias described in the previous section, this
effect is not alleviated by having a long and well-sampled campaign
with a large number of observations.  For given values of S/N and
\fvar, increasing the number of spectra in the time series will not
remove this systematic effect of photon-counting noise on the rms
profile width, although it will make the rms spectrum less noisy.

Given the data quality for typical reverberation campaigns, these
simulations suggest that virial products calculated using rms spectrum
profile widths will be systematically low typically by $\sim10\%$.
The standard practice of normalizing reverberation-based black hole
masses to the local \msigma\ relation of inactive galaxies
\citep{onken2004, woo2010} will compensate for this offset, and
consequently the the virial $f$-factor derived from this procedure
\citep[e.g.,][]{onken2004, woo2010} will be biased high by $\sim10\%$.
The impact of this effect is much smaller than the overall uncertainty
on the mean $f$ factor for the set of reverberation-mapped AGNs, so
this does not currently represent a major problem for derivation of
AGN black hole masses, but nevertheless it is a real effect that is
present in all reverberation-mapping data.  \citet{park2012a}
presented a maximum-likelihood method to calculate rms spectra that
was designed to be immune to this problem; further tests of this or
other alternative methods to calculate rms spectra would be
worthwhile.

\begin{thebibliography}{}

\bibitem[Abazajian et al.(2009)]{abazajian2009} Abazajian, K.~N.,
  Adelman-McCarthy, J.~K., Ag{\"u}eros, M.~A., et al.\ 2009, \apjs,
  182, 543

\bibitem[Antonucci \& Cohen(1983)]{antonucci1983} Antonucci,
  R.~R.~J., \& Cohen, R.~D.\ 1983, \apj, 271, 564

\bibitem[Aretxaga et al.(1999)]{aretxaga1999} Aretxaga, I., Joguet, 
B., Kunth, D., Melnick, J., \& Terlevich, R.~J.\ 1999, \apjl, 519, L123 

\bibitem[Barth et al.(2011a)]{barth:zw229} Barth, A.~J., Nguyen, 
M.~L., Malkan, M.~A., et al.\ 2011a, \apj, 732, 121 

\bibitem[Barth et al.(2011b)]{barth2011:mrk50} Barth, A.~J., Pancoast, 
A., Thorman, S.~J., et al.\ 2011b, \apjl, 743, L4 

\bibitem[Barth et al.(2013)]{barth2013} Barth, A.~J., Pancoast, 
A., Bennert, V.~N., et al.\ 2013, \apj, 769, 128 

\bibitem[Bentz et al.(2007)]{bentz2007} Bentz, M.~C., Denney, 
K.~D., Cackett, E.~M., et al.\ 2007, \apj, 662, 205 

\bibitem[Bentz et al.(2008)]{bentz2008} Bentz, M.~C., Walsh, 
J.~L., Barth, A.~J., et al.\ 2008, \apjl, 689, L21 

\bibitem[Bentz et al.(2009a)]{bentz2009rl} Bentz, M.~C., Peterson,
  B.~M., Netzer, H., Pogge, R.~W., \& Vestergaard, M.\ 2009a, \apj,
  697, 160

\bibitem[Bentz et al.(2009b)]{bentz2009lamp} Bentz, M.~C., Walsh, 
J.~L., Barth, A.~J., et al.\ 2009b, \apj, 705, 199 

\bibitem[Bentz et al.(2010a)]{bentz2010a} Bentz, M.~C., Walsh, J.~L.,
  Barth, A.~J., et al.\ 2010a, \apj, 716, 993

\bibitem[Bentz et al.(2010b)]{bentz2010b} Bentz, M.~C., et al.\ 
2010b, \apjl, 720, L46 

\bibitem[Bentz et al.(2013)]{bentz2013} Bentz, M.~C., Denney, 
K.~D., Grier, C.~J., et al.\ 2013, \apj, 767, 149 

\bibitem[Bian et al.(2010)]{bian2010} Bian, W.-H., Huang, K., 
Hu, C., et al.\ 2010, \apj, 718, 460 

\bibitem[Blandford \& McKee(1982)]{blandfordmckee} Blandford,
  R.~D., \& McKee, C.~F.\ 1982, \apj, 255, 419

\bibitem[Boroson \& Green(1992)]{bg92} Boroson, T.~A.,
\& Green, R.~F.\ 1992, \apjs, 80, 109 (BG92)

\bibitem[Brewer et al.(2011)]{brewer2011} Brewer, B.~J., Treu, T., 
Pancoast, A., et al.\ 2011, \apjl, 733, L33 

\bibitem[Bruzual \& Charlot(2003)]{bc03} Bruzual, G., \&
Charlot, S.\ 2003, \mnras, 344, 1000

\bibitem[Cackett \& Horne(2006)]{cackett2006} Cackett, E.~M.,
  \& Horne, K.\ 2006, \mnras, 365, 1180

\bibitem[Capriotti et al.(1982)]{capriotti82} Capriotti, E.~R., Foltz,
  C.~B., \& Peterson, B.~M.\ 1982, \apj, 261, 35

\bibitem[Cardelli et al.(1989)]{ccm1989} Cardelli, J.~A., 
Clayton, G.~C., \& Mathis, J.~S.\ 1989, \apj, 345, 245 

\bibitem[Carini \& Ryle(2012)]{carini2012} Carini, M.~T., \& Ryle,
  W.~T.\ 2012, \apj, 749, 70

\bibitem[Collin et al.(2006)]{collin2006} Collin, S., Kawaguchi, T.,
  Peterson, B.~M., \& Vestergaard, M.\ 2006, \aap, 456, 75

\bibitem[Decarli et al.(2013)]{decarli2013} Decarli, R., Dotti, M., 
Fumagalli, M., et al.\ 2013, \mnras, 433, 1492 

\bibitem[Denney et al.(2006)]{denney2006} Denney, K.~D., Bentz, 
M.~C., Peterson, B.~M., et al.\ 2006, \apj, 653, 152 

\bibitem[Denney et al.(2009a)]{denney2009se} Denney, K.~D., Peterson, 
B.~M., Dietrich, M., Vestergaard, M., \& Bentz, M.~C.\ 2009a, \apj, 692, 246 

\bibitem[Denney et al.(2009b)]{denney2009vr} Denney, K.~D., Peterson, 
B.~M., Pogge, R.~W., et al.\ 2009b, \apjl, 704, L80 

\bibitem[Denney et al.(2010)]{denney2010} Denney, K.~D., Peterson, 
B.~M., Pogge, R.~W., et al.\ 2010, \apj, 721, 715 

\bibitem[Denney \etal(2014)]{denney2014} Denney, K. D., De Rosa, G.,
  Croxall, K., \etal\ 2014, ApJ, submitted (arXiv:1404.4879)

\bibitem[Dietrich et al.(1994)]{dietrich1994} Dietrich, M.,
  Kollatschny, W., Alloin, D., et al.\ 1994, \aap, 284, 33

\bibitem[Dietrich et al.(2002)]{dietrich2002} Dietrich, M., 
Appenzeller, I., Vestergaard, M., \& Wagner, S.~J.\ 2002, \apj, 564, 581 

\bibitem[Dotti et al.(2012)]{dotti2012} Dotti, M., Sesana, A., 
\& Decarli, R.\ 2012, Advances in Astronomy, 2012,  

\bibitem[Edelson et al.(2002)]{edelson2002} Edelson, R., Turner, 
T.~J., Pounds, K., et al.\ 2002, \apj, 568, 610 

\bibitem[Edelson \& Malkan(2012)]{edelson2012} Edelson, R., \& Malkan,
  M.\ 2012, \apj, 751, 52

\bibitem[Edelson et al.(2014)]{edelson2014} Edelson, R., Vaughan, 
S., Malkan, M., et al.\ 2014, arXiv:1409.1613 

\bibitem[Eracleous et al.(2012)]{eracleous2012} Eracleous, M., 
Boroson, T.~A., Halpern, J.~P., \& Liu, J.\ 2012, \apjs, 201, 23 

\bibitem[Filippenko(1982)]{filippenko1982} Filippenko, A. V. 1982,
  PASP, 94, 715

\bibitem[Foley et al.(2013)]{foley2013} Foley, R.~J., Challis, 
P.~J., Chornock, R., et al.\ 2013, \apj, 767, 57 

\bibitem[Fromerth \& Melia(2000)]{fromerth2000} Fromerth,
  M.~J., \& Melia, F.\ 2000, \apj, 533, 172

\bibitem[Gilbert \& Peterson(2003)]{gilbertpeterson2003} Gilbert,
  K.~M., \& Peterson, B.~M.\ 2003, \apj, 587, 123

\bibitem[Goad et al.(2012)]{goad2012} Goad, M.~R., Korista, 
K.~T., \& Ruff, A.~J.\ 2012, \mnras, 426, 3086 

\bibitem[Goad \& Korista(2014)]{goadkorista2014} Goad, M.~R., \&
  Korista, K.~T.\ 2014, \mnras, 444, 43

\bibitem[Graham et al.(2011)]{graham2011} Graham, A.~W., Onken, 
C.~A., Athanassoula, E., \& Combes, F.\ 2011, \mnras, 412, 2211 

\bibitem[Greene \& Ho(2005)]{greeneho2005} Greene, J.~E., \&
  Ho, L.~C.\ 2005, \apj, 630, 122

\bibitem[Greene \& Ho(2007)]{gh07} Greene, J.~E., \&
  Ho, L.~C.\ 2007, \apj, 670, 92

\bibitem[Grier et al.(2013a)]{grier2013a} Grier, C.~J., Peterson, 
B.~M., Horne, K., et al.\ 2013a, \apj, 764, 47 

\bibitem[Grier et al.(2013b)]{grier2013b} Grier, C.~J., Martini, 
P., Watson, L.~C., et al.\ 2013b, \apj, 773, 90 

\bibitem[Guo \& Gu(2014)]{guo2014} Guo, H., \& Gu,
  M.\ 2014, \apj, 792, 33

\bibitem[Hillenbrand et al.(2013)]{hillenbrand2013} Hillenbrand,
  L.~A., Miller, A.~A., Covey, K.~R., et al.\ 2013, \aj, 145, 59

\bibitem[Hinshaw et al.(2013)]{hinshaw:wmap} Hinshaw, G., Larson, 
D., Komatsu, E., et al.\ 2013, \apjs, 208, 19 

\bibitem[Ho \& Kim(2014)]{ho2014} Ho, L.~C., \& Kim,
  M.\ 2014, \apj, 789, 17

\bibitem[Horne(1986)]{horne1986} Horne, K.\ 1986, \pasp, 98, 609 

\bibitem[Horne et al.(2004)]{horne2004} Horne, K., Peterson, B.~M.,
  Collier, S.~J., \& Netzer, H.\ 2004, \pasp, 116, 465

\bibitem[Ju et al.(2013)]{ju2013} Ju, W., Greene, J.~E., 
Rafikov, R.~R., Bickerton, S.~J., \& Badenes, C.\ 2013, \apj, 777, 44 

\bibitem[Kaspi et al.(2000)]{kaspi2000} Kaspi, S., Smith, P.~S.,
  Netzer, H., et al.\ 2000, \apj, 533, 631

\bibitem[Kelly(2007)]{kelly2007} Kelly, B.~C.\ 2007, \apj, 665, 
1489 

\bibitem[Kilerci Eser et al.(2015)]{kilerci2015} Kilerci Eser, E., 
Vestergaard, M., Peterson, B.~M., Denney, K.~D., 
\& Bentz, M.~C.\ 2015, \apj, in press (arXiv:1411.2977)

\bibitem[Kollatschny(2003)]{kollatschny2003} Kollatschny, W.\ 2003,
  \aap, 407, 461

\bibitem[Korista \& Goad(2004)]{korista2004} Korista, K.~T.,
  \& Goad, M.~R.\ 2004, \apj, 606, 749

\bibitem[Kova{\v c}evi{\'c} et al.(2010)]{kovacevic2010} Kova{\v 
c}evi{\'c}, J., Popovi{\'c}, L.~{\v C}., 
\& Dimitrijevi{\'c}, M.~S.\ 2010, \apjs, 189, 15 

\bibitem[Krolik(2001)]{krolik2001} Krolik, J.~H.\ 2001, \apj, 551, 
72 

\bibitem[LaMassa et al.(2014)]{lamassa2014} LaMassa, S.~M., Cales, S.,
  Moran, E.~C., et al.\ 2014, arXiv:1412.2136

\bibitem[Li et al.(2013)]{li2013} Li, Y.-R., Wang, J.-M., Ho, 
L.~C., Du, P., \& Bai, J.-M.\ 2013, \apj, 779, 110 

\bibitem[Liu et al.(2014)]{liu2014} Liu, X., Shen, Y., Bian, 
F., Loeb, A., \& Tremaine, S.\ 2014, \apj, 789, 140 

\bibitem[Maoz et al.(1990)]{maoz1990} Maoz, D., Netzer, H., 
Leibowitz, E., et al.\ 1990, \apj, 351, 75 

\bibitem[Markwardt(2009)]{markwardt2009} Markwardt, C.~B.\ 2009, 
Astronomical Data Analysis Software and Systems XVIII, 411, 251 

\bibitem[Matheson et al.(2000)]{matheson2000} Matheson, T., 
Filippenko, A.~V., Ho, L.~C., Barth, A.~J., 
\& Leonard, D.~C.\ 2000, \aj, 120, 1499 

\bibitem[Miller \& Stone(1993)]{millerstone} Miller, J.~S., \& Stone, 
R.~P.~S. 1993, Lick Obs. Tech. Rep. 66 (Santa Cruz, CA: Lick 
Observatory)

\bibitem[Mushotzky et al.(2011)]{mushotzky2011} Mushotzky, R.~F., 
Edelson, R., Baumgartner, W., \& Gandhi, P.\ 2011, \apjl, 743, L12 

\bibitem[O'Brien et al.(1995)]{obrien1995} O'Brien, P.~T., Goad, 
M.~R., \& Gondhalekar, P.~M.\ 1995, \mnras, 275, 1125 

\bibitem[Onken et al.(2003)]{onken2003} Onken, C.~A., Peterson, B.~M.,
  Dietrich, M., Robinson, A., \& Salamanca, I.~M.\ 2003, \apj, 585,
  121

\bibitem[Onken et al.(2004)]{onken2004} Onken, C.~A., Ferrarese, 
L., Merritt, D., et al.\ 2004, \apj, 615, 645 

\bibitem[Pancoast et al.(2011)]{pancoast2011} Pancoast, A., Brewer, 
B.~J., \& Treu, T.\ 2011, \apj, 730, 139 

\bibitem[Pancoast et al.(2012)]{pancoast2012} Pancoast, A., Brewer, 
B.~J., Treu, T., et al.\ 2012, \apj, 754, 49 

\bibitem[Pancoast et al.(2014a)]{pancoast2014a} Pancoast, A., Brewer, 
B.~J., \& Treu, T.\ 2014a, arXiv:1407.2941 

\bibitem[Pancoast et al.(2014b)]{pancoast2014b} Pancoast, A., Brewer, 
B.~J., Treu, T., et al.\ 2014b, MNRAS, in press (arXiv:1311.6475)

\bibitem[Park et al.(2012a)]{park2012a} Park, D., Woo, J.-H., 
Treu, T., et al.\ 2012a, \apj, 747, 30 

\bibitem[Park et al.(2012b)]{park2012b} Park, D., Kelly, B.~C., 
Woo, J.-H., \& Treu, T.\ 2012b, \apjs, 203, 6 

\bibitem[Pei et al.(2014)]{pei2014} Pei, L., Barth, A.~J., 
Aldering, G.~S., et al.\ 2014, \apj, 795, 38 

\bibitem[Peterson \& Ferland(1986)]{petersonferland1986} Peterson,
  B.~M., \& Ferland, G.~J.\ 1986, \nat, 324, 345

\bibitem[Peterson et al.(1998)]{peterson1998} Peterson, B.~M., 
Wanders, I., Bertram, R., et al.\ 1998, \apj, 501, 82 

\bibitem[Peterson \& Wandel(1999)]{petersonwandel99} Peterson,
  B.~M., \& Wandel, A.\ 1999, \apjl, 521, L95

\bibitem[Peterson(2001)]{peterson2001} Peterson, B.~M.\ 2001,
  Advanced Lectures on the Starburst-AGN Connection, ed. I. Aretxaga,
  D. Kunth, \& R. M\'ujica (Singapore: World Scientific), 3
  (arXiv:astro-ph/0109495)

\bibitem[Peterson et al.(2004)]{peterson2004} Peterson, B.~M., 
Ferrarese, L., Gilbert, K.~M., et al.\ 2004, \apj, 613, 682 

\bibitem[Peterson et al.(2013)]{peterson2013} Peterson, B.~M., 
Denney, K.~D., De Rosa, G., et al.\ 2013, \apj, 779, 109 

\bibitem[Popovi{\'c}(2012)]{popovic2012} Popovi{\'c}, L.~{\v C}.\ 
2012, New Astronomy Reviews, 56, 74 

\bibitem[Richards et al.(2011)]{richards2011} Richards, G.~T., 
Kruczek, N.~E., Gallagher, S.~C., et al.\ 2011, \aj, 141, 167 

\bibitem[Rodr{\'{\i}}guez-Pascual et al.(1997)]{rodriguezpascual1997} 
Rodr{\'{\i}}guez-Pascual, P.~M., Alloin, D., Clavel, J., et al.\ 1997, 
\apjs, 110, 9 

\bibitem[Rosenblatt \& Malkan(1990)]{rosenblatt1990} Rosenblatt,
  E.~I., \& Malkan, M.~A.\ 1990, \apj, 350, 132

\bibitem[Santos-Lle{\'o} et al.(2001)]{santos2001} Santos-Lle{\'o},
  M., Clavel, J., Schulz, B., et al.\ 2001, \aap, 369, 57

\bibitem[Schlafly \& Finkbeiner(2011)]{schlafly2011} Schlafly,
  E.~F., \& Finkbeiner, D.~P.\ 2011, \apj, 737, 103

\bibitem[Scott et al.(2014)]{scott2014} Scott, B., Bennert, V.,
  Komossa, S., et al.\ 2014, BAAS, 223, \#250.16

\bibitem[Sergeev et al.(1999)]{sergeev1999} Sergeev, S.~G., Pronik, 
V.~I., Sergeeva, E.~A., \& Malkov, Y.~F.\ 1999, \aj, 118, 2658 

\bibitem[Sergeev et al.(2007)]{sergeev2007} Sergeev, S.~G., 
Doroshenko, V.~T., Dzyuba, S.~A., et al.\ 2007, \apj, 668, 708 

\bibitem[Seyfert(1943)]{seyfert1943} Seyfert, C.~K.\ 1943, \apj, 
97, 28 

\bibitem[Shapovalova et al.(2012)]{shapovalova2012} Shapovalova, A.~I., 
Popovi{\'c}, L.~{\v C}., Burenkov, A.~N., et al.\ 2012, \apjs, 202, 10 

\bibitem[Shappee et al.(2014)]{shappee2014} Shappee, B.~J., Prieto, 
J.~L., Grupe, D., et al.\ 2014, \apj, 788, 48 

\bibitem[Shen et al.(2008)]{shen2008} Shen, Y., Greene, J.~E., 
Strauss, M.~A., Richards, G.~T., \& Schneider, D.~P.\ 2008, \apj, 680, 169 

\bibitem[Shen(2013)]{shen2013review} Shen, Y.\ 2013, Bulletin of the
  Astronomical Society of India, 41, 61

\bibitem[Shen et al.(2013)]{shen2013bbh} Shen, Y., Liu, X., Loeb, 
A., \& Tremaine, S.\ 2013, \apj, 775, 49 

\bibitem[Silverman et al.(2013)]{silverman:latetime} Silverman, J.~M.,
  Ganeshalingam, M., \& Filippenko, A.~V.\ 2013, \mnras, 430, 1030

\bibitem[Smith et al.(2012)]{smith2012:sn2010jl} Smith, N., Silverman,
  J.~M., Filippenko, A.~V., et al.\ 2012, \aj, 143, 17

\bibitem[Stirpe et al.(1988)]{stirpe1988} Stirpe, G.~M., de
  Bruyn, A.~G., \& van Groningen, E.\ 1988, \aap, 200, 9

\bibitem[Timmer \& K\"{o}nig(1995)]{timmer1995} Timmer, J., \&
  K\"{o}nig, M.\ 1995, \aap, 300, 707

\bibitem[Tohline \& Osterbrock(1976)]{tohline1976} Tohline,
  J.~E., \& Osterbrock, D.~E.\ 1976, \apjl, 210, L117

\bibitem[Tsalmantza et al.(2011)]{tsalmantza2011} Tsalmantza, P., 
Decarli, R., Dotti, M., \& Hogg, D.~W.\ 2011, \apj, 738, 20 

\bibitem[Ulrich et al.(1984)]{ulrich84} Ulrich, M.~H., 
Boksenberg, A., Bromage, G.~E., et al.\ 1984, \mnras, 206, 221 

\bibitem[van der Marel \& Franx(1993)]{vdm93} van der
  Marel, R.~P., \& Franx, M.\ 1993, \apj, 407, 525

\bibitem[van Dokkum(2001)]{vandokkum2001} van Dokkum, P.~G.\ 2001, 
\pasp, 113, 1420 

\bibitem[van Groningen \& Wanders(1992)]{vgw1992} van
Groningen, E., \& Wanders, I.\ 1992, \pasp, 104, 700

\bibitem[V{\'e}ron et al.(2002)]{veron2002} V{\'e}ron, P.,
  Gon{\c c}alves, A.~C., \& V{\'e}ron-Cetty, M.-P.\ 2002, \aap, 384,
  826

\bibitem[V{\'e}ron-Cetty et al.(2004)]{veron2004}
  V{\'e}ron-Cetty, M.-P., Joly, M., \& V{\'e}ron, P.\ 2004, \aap, 417,
  515 (V04)

\bibitem[V{\'e}ron-Cetty \& V{\'e}ron(2006)]{veron2006}
  V{\'e}ron-Cetty, M.-P., \& V{\'e}ron, P.\ 2006, \aap, 455, 773

\bibitem[Vestergaard \& Peterson(2005)]{vp05}
  Vestergaard, M., \& Peterson, B.~M.\ 2005, \apj, 625, 688

\bibitem[Wade \& Horne(1988)]{wadehorne} Wade, R.~A., \&
  Horne, K.\ 1988, \apj, 324, 411

\bibitem[Wanders \& Peterson(1996)]{wanders1996} Wanders, I.,
  \& Peterson, B.~M.\ 1996, \apj, 466, 174

\bibitem[Wang et al.(2014)]{wang2014} Wang, J.-M., Du, P., Hu, 
C., et al.\ 2014, arXiv:1408.2337 

\bibitem[Welsh \& Horne(1991)]{welshhorne} Welsh, W.~F., \& Horne,
  K.\ 1991, \apj, 379, 586

\bibitem[Wills et al.(1985)]{wills1985} Wills, B.~J., Netzer, H., \&
  Wills, D.\ 1985, \apj, 288, 94

\bibitem[Woo et al.(2006)]{woo2006} Woo, J.-H., Treu, T., 
Malkan, M.~A., \& Blandford, R.~D.\ 2006, \apj, 645, 900 

\bibitem[Woo et al.(2010)]{woo2010} Woo, J.-H., Treu, T., 
Barth, A.~J., et al.\ 2010, \apj, 716, 269 

\bibitem[Wright(2006)]{wright:cosmo} Wright, E.~L.\ 2006, \pasp, 
118, 1711 

\bibitem[Zhang(2014)]{zhang2014} Zhang, X.\ 2014, arXiv:1410.7537 

\bibitem[Zu et al.(2011)]{zu2011} Zu, Y., Kochanek, C.~S., \&
  Peterson, B.~M.\ 2011, \apj, 735, 80




\end{thebibliography}
\end{document}